\documentclass[journal]{IEEEtran}

\usepackage{cite}

\ifCLASSINFOpdf
  \usepackage[pdftex]{graphicx}
  \graphicspath{{../pdf/}{../jpeg/}}
  \DeclareGraphicsExtensions{.pdf,.jpeg,.png}
\else
  \usepackage[dvips]{graphicx}
  \graphicspath{{../eps/}}
  \DeclareGraphicsExtensions{.eps}
\fi

\usepackage{amsmath}
\usepackage{enumitem}

\usepackage{array}

\ifCLASSOPTIONcompsoc
 \usepackage[caption=false,font=normalsize,labelfont=sf,textfont=sf]{subfig}
\else
 \usepackage[caption=false,font=footnotesize]{subfig}
\fi

\usepackage{dblfloatfix}

\usepackage{url}

\usepackage{amssymb}
\usepackage{amsthm}
\usepackage{bm}
\usepackage{mathtools}
\usepackage{dsfont}

\usepackage{url}
\usepackage{color,soul}

\newtheorem{theorem}{Theorem}
\newtheorem{remark}[theorem]{Remark}

\DeclareMathOperator*{\argmin}{arg\,min}

\usepackage[percent]{overpic}
\usepackage{siunitx}

\usepackage{float}

\usepackage{url}
\usepackage[utf8]{inputenc}

\newcommand{\Rd}{\mathbf{R}}                         
\newcommand{\XX}{\mathbf{x}}                         
\newcommand{\ZZ}{\mathbf{z}}                         
\newcommand{\YY}{\mathbf{y}}                         
\newcommand{\Ad}{\mathbf A}                         
\newcommand{\Fd}{\mathbf F}                         
\newcommand{\Ed}{\mathbf E} 
\newcommand{\Cd}{\mathbf C} 
\newcommand{\Id}{\mathbf I} 
\newcommand{\Hd}{\mathbf H}

\newcommand{\Eu}{\mathbf{E}_I}  
\newcommand{\Su}{\mathbf{S}_I}                         
\newcommand{\Xu}{\mathbf{x}_I}                         
\newcommand{\Yu}{\mathbf{y}_I}                         
\newcommand{\Yn}{\mathbf{y}_{\eta}}
\newcommand{\Xn}{\mathbf{x}_{\eta}} 
\newcommand{\Xcnn}{\mathbf{x}_{\mathrm{CNN}}} 
\usepackage{graphicx}

\usepackage{xcolor}

\usepackage[percent]{overpic}

\usepackage{algpseudocode}
\usepackage{algorithm}

\usepackage{siunitx}

\newcommand{\herm}{{\scriptstyle \boldsymbol{\mathsf{H}}}}
 \newcommand{\trans}{{\scriptstyle \boldsymbol{\mathsf{T}}}}

\usepackage{booktabs}

\begin{document}

\title{Neural Networks-based Regularization for Large-Scale  Medical Image Reconstruction}

\author{Andreas~Kofler,
        Markus~Haltmeier,
        Tobias~Schaeffter,
        Marc~Kachelrieß,
        Marc~Dewey,
        Christian~Wald,
        and Christoph~Kolbitsch
\thanks{A. Kofler, C. Wald and Marc Dewey are with the Department of Radiology, Charit\'{e} - Universit\"{a}tsmedizin Berlin, Berlin,
Germany (e-mail: $\{$andreas.kofler, christian.wald, marc.dewey$\}$@charite.de)}
\thanks{M. Haltmeier is with the Department of Mathematics, University of Innsbruck, Innsbruck, Austria (e-mail: markus.haltmeier@uibk.ac.at)}
\thanks{T. Schaeffter is with the Physikalisch-Technische Bundesanstalt (PTB), Braunschweig and Berlin, Germany, King’s College London, London, UK and the Department of Medical Engineering, Technical University of Berlin, Berlin, Germany 
(e-mail: tobias.schaeffter@ptb.de)}
\thanks{M. Kachelrieß is with the Division of X-Ray Imaging and CT, German Cancer Research Center, Heidelberg, Germany (e-mail: marc.kachelriess@dkfz.de)}

\thanks{C. Kolbitsch is with the Physikalisch-Technische Bundesanstalt (PTB), Braunschweig and Berlin, Germany and King’s College London, London, UK
(e-mail: christoph.kolbitsch@ptb.de)}
}

\maketitle

\begin{abstract}
In this paper we present a generalized Deep Learning-based approach for solving ill-posed large-scale inverse problems occuring in medical image reconstruction. Recently, Deep Learning methods using iterative neural networks and cascaded neural networks have been reported to achieve state-of-the-art results with respect to various quantitative quality measures as PSNR, NRMSE and SSIM across different imaging modalities. However, the fact that these approaches employ the forward and adjoint operators repeatedly in the network architecture requires the network to process the whole images or volumes at once, which for some applications is computationally infeasible. In this work, we follow a different reconstruction strategy by decoupling the regularization of the solution from ensuring consistency with the measured data. The regularization is given in the form of an image prior obtained by the output of a previously trained neural network which is used in a Tikhonov regularization framework. By doing so, more complex and sophisticated network architectures can be used for the removal of the artefacts or  noise than it is usually the case in iterative networks. Due to the large scale of the considered problems and the resulting computational complexity of the employed  networks, the priors are obtained by processing the images or volumes as patches or slices. 
We evaluated the method for the cases of 3D cone-beam low dose  CT and undersampled 2D radial cine MRI and compared it to a total variation-minimization-based reconstruction algorithm as well as to a method with regularization based on learned overcomplete dictionaries. The proposed method outperformed all the reported methods with respect to all chosen quantitative measures and further accelerates the regularization step in the reconstruction by several orders of magnitude.
\end{abstract}


\IEEEpeerreviewmaketitle
\section{Introduction}
\IEEEPARstart{I}{n} inverse problems, the goal is to recover an object of interest from a set of indirect and possibly incomplete observations. In medical imaging, for example, a classical inverse problem is given by the task of reconstructing a diagnostic image from a certain number of measurements, e.g.\ X-ray projections in computed tomography (CT) or the spatial frequency information ($k$-space data) in magnetic resonance imaging (MRI). 
The reconstruction from the measured data can be an ill-posed inverse problem for different reasons. In low-dose CT, for example, the reconstruction from noisy data is ill-posed because of the ill-posedeness of the inversion of the Radon transform. In accelerated MRI, on the other hand, the reconstruction from incomplete data is ill-posed since the underlying problem is underdetermined and therefore no unique solution exists without integrating prior information.

In order to constrain the space of possible solutions, a typical approach is to impose specific a-priori chosen properties on the solution by adding a regularization (or penalty) term to the problem. 
Well known choices for the regularization are for example given  by the popular total variation-minimization and sparse regularization approaches, where the  solution is transformed using a sparsifying transform such as the Wavelet-transform or the Fourier-transform \cite{Lustig2008} or a finite-differences filter \cite{block2007} and the $L_1$-norm of the latter is minimized.
While the aforementioned methods use hand-crafted priors, other methods learn the regularization directly within the reconstruction of the images  where the regularization is imposed patch-wise by the sparse approximation using a dictionary which is learned in an unsupervised manner during the reconstruction \cite{wang2014compressed}, \cite{xu2012low}.
However, these learning-based methods are usually time consuming since the regularization is adaptive and learned during an iterative reconstruction scheme. Further, in the specific dictionary learning framework, the regularization requires training of a dictionary and sparse coding of all patches of the current image estimate at each iteration. This is computationally demanding and makes the application in the clinical routine challenging.

Recently, Convolutional Neural Networks (CNNs) have been applied in the field of inverse problems, either as direct full inversion methods \cite{zhu2018image}, as post processing methods \cite{jin2017deep}, \cite{schwab2018deep}, \cite{han2018framing}, as learned iterative schemes \cite{adler2017solving}, \cite{adler2018learned}, or as learned regularizers \cite{schlemper2017deep}, \cite{kofler2018u}, \cite{li2018nett}, \cite{aggarwal2018modl}, \cite{qin2018convolutional}.
When used as post-processing methods, the networks are trained to denoise or remove artefacts from images obtained by the direct reconstruction of the noisy or incomplete data. Although a wide range of different network architecture has been proposed, e.g.\ \cite{han2018framing}, \cite{yang2018low},  a major concern is that the estimated output of the CNN might lack data-consistency.
In order to ensure the obtained image is consistent with the acquired raw data, methods have been proposed where the constructed networks define unrolled iterative schemes which employ the forward and the adjoint operators. These methods can be interpreted as learned iterative schemes and have been successfully applied to different imaging modalities \cite{adler2017solving}, \cite{adler2018learned}, \cite{hammernik2018learning}, \cite{hauptmann2018model}, \cite{schlemper2017deep}, \cite{kofler2018u}, \cite{aggarwal2018modl}, \cite{qin2018convolutional}. Thereby, the subnetworks containing trainable parameters can be thought of regularizers which are learned by end-to-end training of the whole network cascade. Due to the integration of the forward and the adjoint operators, iterative or cascaded networks seem to be a choice for various image reconstruction task.
However, the main advantage of these methods at the same time represents  the computational bottleneck of the approaches. The fact that the forward and the adjoint operators are integrated as layers in the networks requires that the whole object of interest has to be processed at once. Since CNNs typically increase the input size by extracting several feature maps per layer, end-to-end training might be infeasible for
some high-dimensional problems, including high-resolution 3D CT volumes or non-Cartesian MR acquisitions. \\
In order to overcome these limitations, we propose to decouple the regularization of the solution from ensuring consistency with the measured data.  We present a general framework to use CNNs as learned regularizers and still ensure data-consistency of the obtained solution. In particular, we consider high-dimensional problems where either the object of interest or the measured data are high-dimensional  (high-resolution 3D CT) or the evaluation of the forward or the adjoint operators is computationally expensive (dynamic 2D non-Cartesian radial MR acquisition).\\
This paper is organized as follows. In Section \ref{proposed_approach_section}, we formally introduce the inverse problem of image reconstruction and motivate our proposed approach for the solution of large-scale  ill-posed inverse problems. We demonstrate the feasibility of our method by applying it to 3D low-dose cone beam CT and 2D radial cine MRI in Section \ref{sec_experiments}. We further compare the proposed approach to an iterative reconstruction method given by total variation-minimization (TV) and a learning-based method (DIC) using Dictionary Learning-based priors in Section \ref{sec_results}. We then conclude the work with a discussion and conclusion in Section \ref{discussion_section} and Section \ref{conclusion_section}.

\section{Iterative Image Reconstruction with CNN-Priors}\label{proposed_approach_section}
In this Section, we present the proposed deep learning scheme for solving large-scale, possibly non-linear, inverse problems. For the sake of clarity, we do not focus on a functional analytical setting but consider discretized problems of the form  
\begin{equation}\label{general_inv_problem}
 \YY = \Ad \XX + \ZZ,
\end{equation}
where $\Ad\colon X \rightarrow Y$ is a discrete  possibly non-linear forward operator  between finite dimensional Hilbert spaces, $\YY \in Y$ is the measured  data, $\ZZ \in Y$ the noise and $\XX \in X$ the unknown object to be recovered.   
The operator $\Ad$ could for example model the measurement process in various imaging modalities such as the X-ray projection in CT or the Fourier encoding in MRI. Depending on the nature of the underlying imaging modality one is considering, problem (\ref{general_inv_problem}) can be ill-posed for different reasons. For example, in low-dose CT, the measurement data is inherently contaminated by noise. In cardiac MRI, $k$-space data is often undersampled in order to speed up the acquisition process.  This leads to incomplete data and therefore to an undetermined  problem with an infinite number of theoretically possible solutions.\\
In order to constrain the space of solutions of interest, a typical approach is to impose specific a-priori chosen properties on the solution $\XX$ by adding a regularization (or penalty) term $\mathcal{R}(\XX)$ and using Lagrange multipliers. Then,  we solve the relaxed problem
\begin{equation}\label{reg_inv_problem}
D(\Ad \XX, \YY) + \lambda\, \mathcal{R}(\XX) \rightarrow \mathrm{min},
\end{equation}
where $D(\, \cdot\,,\,\cdot\,)$ is an appropriately chosen data-discrepancy measure and $\lambda>0$ controls the strength of the regularization. The choice of $D(\, \cdot\,,\,\cdot\,)$ depends on the considered problem. 
 For  the examples presented  in Sections~\ref{sec_experiments} and \ref{sec_results} we choose the discrepancy measure as the squared norm distance in the case of  radial cine MRI  and  the  Kullback-Leibler divergence in the case of  low dose CT, respectively.
 
\subsection{CNN-based Regularization}
Clearly, the regularization term $\mathcal{R}(\XX)$ significantly affects the quality and the characteristics of the solution $\XX$. 
Here, we propose a generalized approach for solving high-dimensional inverse problems by the following three steps: First, an initial guess of the solution is provided by a direct reconstruction from the measured data, i.e. $\XX_{\mathrm{ini}} = \Ad^{\dagger}\YY$, where $\Ad^{\dagger}\colon Y \to X$ denotes some reconstruction operator. Then, a CNN is used to remove the noise or the artefacts from the direct reconstruction $\XX_{\mathrm{ini}}$ in order to obtain another intermediate reconstruction  $\Xcnn$ which is  used  as a CNN-prior in a Tikhonov functional
\begin{equation}\label{NN_reg_inv_problem}
F_{\YY,\Xcnn,\lambda} (\XX) := D(\Ad \XX,\YY) + \lambda \| \XX - \Xcnn\|_2^2\rightarrow \mathrm{min}.
\end{equation}
As a third and final step, the CNN-Tikhonov functional (\ref{NN_reg_inv_problem}) is minimized resulting in the proposed CNN-based reconstruction. \\ Note that the regularization of the problem, i.e.\ obtaining the CNN-prior, is decoupled from the step of ensuring data-consistency of the solution via minimization of (\ref{NN_reg_inv_problem}). This allows to use deeper and more sophisticated CNNs as the ones typically used in iterative networks. Given the high-dimensionality of the considered problems, network training is further carried out on sub-portions of the image samples, i.e.\ on patches or slices which are previously extracted from the images or volumes. This is motivated by the fact that in most medical imaging applications, one has typically access to datasets with only a relatively small number of subjects. The images or volumes of these subjects, on the other hand, are elements of a  high-dimensional space. Therefore, one is concerned with the problem of having topologically sparse training data with only very few data points in the original high-dimensional image space. Working with sub-portions of the image samples increases the number of available data points and at the same time decreases its ambient dimensionality.

\subsection{Large-Scale CNN-Prior}

Suppose we have access to a finite set of $N$ ground truth samples $(\XX_{k})_{k=1}^N$ and corresponding initial estimates $(\XX_{\mathrm{ini},k})_{k=1}^N$. 
 We  are in particular  interested in  the case where $N$ is relatively small and the considered samples $\XX_{k}$ have a relatively large size,
which is the case for most medical imaging applications.
For any sample $\XX \in X$ we consider its 
decomposition in $N_{\mathbf{p},\mathbf{s}}$ 
possibly overlapping patches  
\begin{equation}\label{xest_repr}
\XX = \mathbf{W}_{\mathbf{p},\mathbf{s}}\ \sum_{j=1}^{N_{\mathbf{p},\mathbf{s}}} (\Rd_j^{\mathbf{p},\mathbf{s}})^{\trans} \, \Rd_j^{\mathbf{p},\mathbf{s}} \, \XX,
\end{equation}
where $\Rd_j^{\mathbf{p},\mathbf{s}}$ and $(\Rd_j^{\mathbf{p},\mathbf{s}})^{\trans}$ extract and reposition the patches at the original position, respectively, and the diagonal operator $\mathbf{W}_{\mathbf{p},\mathbf{s}}$ accounts for weighting of regions containing overlaps. The entries of the tuples $\mathbf{p}$ and $\mathbf{s}$ specify the size of the patches and the strides in each dimension and therefore the number  of patches $N_{\mathbf{p},\mathbf{s}}$ which are extracted from a single image.

We aim  for improved estimates $\XX_{\mathrm{CNN},k} = f_\theta( \XX_{\mathrm{ini},k}) \approx \XX_{k}$ via a trained network  function $f_\theta$ to be constructed. Since the operator norm of  $\mathbf{W}_{\mathbf{p},\mathbf{s}}$ is less or equal to one, by the triangle inequality, we can estimate the average error  
\begin{multline}\label{error_estimate} 
e_N:=  \sum_{k=1}^N \|\XX_{k} - \XX_{\mathrm{CNN},k} \|_2 
\\  \leq    \sum_{k=1}^N \sum_{j=1}^{N_{\mathbf{p},\mathbf{s}}} 
\bigl\|  
 \Rd_j^{\mathbf{p},\mathbf{s}} \XX_{k} - \Rd_j^{\mathbf{p},\mathbf{s}} \XX_{\mathrm{CNN},k}   \bigr\|_2  =: e_{N,N_{\mathbf{p},\mathbf{s}}}.
\end{multline}
 Inequality  \eqref{error_estimate} suggests that it is  beneficial   estimating each  patch of the sample $\XX_{k}$ by a neural network $u_{\theta}$  applied  to  $\Rd_j^{\mathbf{p},\mathbf{s}} \XX_{\mathrm{ini},k}$ rather than estimating the whole sample at once. 
The neural network $u_{\theta}$
is trained on a subset of pairs 
\begin{eqnarray}\label{dataset}
\mathcal{D} = \Big\{\Big(\Rd_j^{\mathbf{p},\mathbf{s}} (\XX_{\mathrm{ini},k}), \Rd_j^{\mathbf{p},\mathbf{s}} (\XX_{k} ) \Big) : (k,j) \in \mathcal{I}_{N,N_{\mathbf{p},\mathbf{s}}} \Big\},
\end{eqnarray}
of all possible patches extracted from the $N$ samples in the dataset, where $ \mathcal{I}_{N,N_{\mathbf{p},\mathbf{s}}}:= \{1,\ldots,N\} \times \{1,\ldots,N_{\mathbf{p},\mathbf{s}} \}$.
During training, we optimize the set of parameters $\theta$ to minimize the $L_2$-error between the estimated output of the patches and the corresponding ground truth patch by minimizing  
\begin{equation}\label{loss_fct}
\mathcal{L}(\theta) =\frac{1}{N_{\mathrm{train}}} \sum_{ (\mathbf{z}_{\mathrm{ini}}, \mathbf{z}) \in \mathcal{D} }\| u_{\theta}(\mathbf{z}_{\mathrm{ini}}) - \mathbf{z}\|_2^2 \,,
\end{equation}
where $N_{\mathrm{train}}$ is the number of  training patches.

Denote by $f_{\theta}$ the composite function which decomposes a sample image or volume $\XX$ into patches, applies a neural network $u_{\theta}$ to each patch, and reassembles the sample from them. 
This results in the proposed CNN-prior  $\Xcnn$ given by 
\begin{multline}\label{xcnn}
\Xcnn := f_{\theta}(\XX_{\mathrm{ini}}) \\
= \mathbf{W}_{\mathbf{p},\mathbf{s}}\sum_j (\Rd_j^{\mathbf{p},\mathbf{s}})^{\trans} ( u_{\theta}(\Rd_j^{\mathbf{p},\mathbf{s}}(\XX_{\mathrm{ini}}) ) ),
\end{multline}
where $\XX_{\mathrm{ini}} = \Ad^\dag \YY$ is the initial reconstruction obtained from the measured data.   

\begin{figure}
\centering
\includegraphics[width=0.9\linewidth]{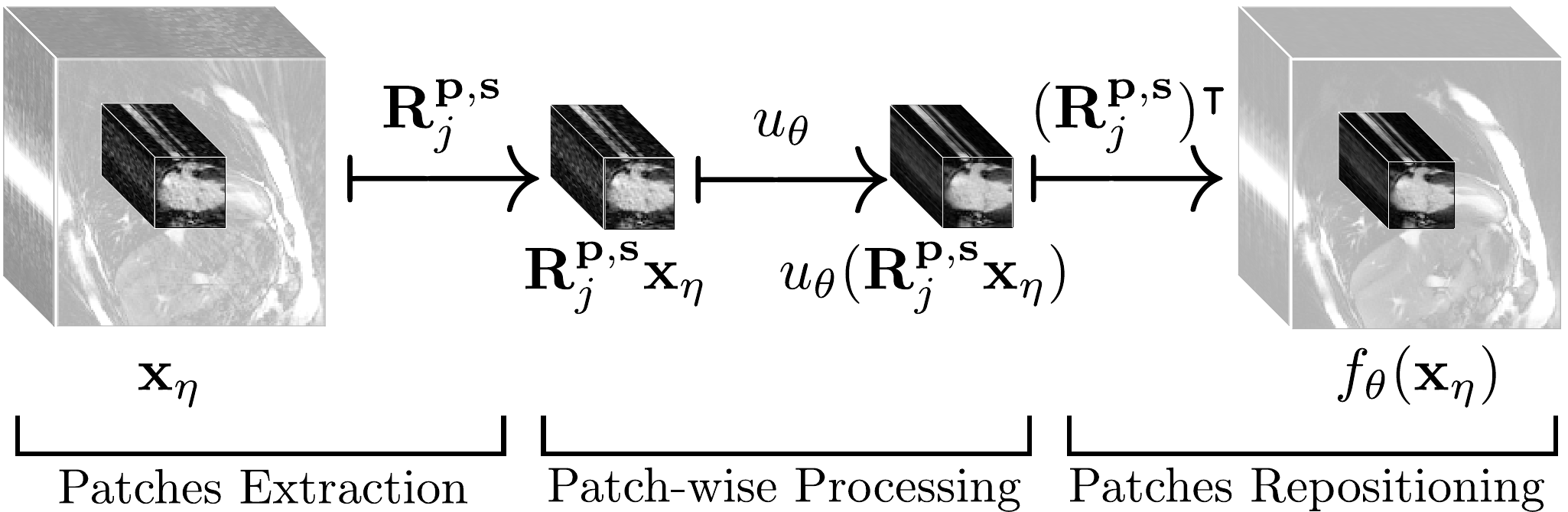}
\caption{Workflow for obtaining a CNN-prior by patch-based processing: First, the initial reconstruction is divided into patches, then the network $u_{\theta}$ is applied to all patches. Reassembling all processed patches results in the CNN-prior which is then used for regularization of the inverse problem.}\label{patch_based_processing}
\end{figure}

\begin{remark}
The inequality in (\ref{error_estimate}) guarantees that the set of parameters  found by minimizing (\ref{loss_fct}) is also suitable for obtaining the prior $\Xcnn$.   Therefore, $u_{\theta}$ is powerful enough to deliver a CNN-prior to regularize the solution of (\ref{NN_reg_inv_problem}).  Figure \ref{patch_based_processing} illustrates the process of extracting patches from a volume using the operator $\Rd_j^{\mathbf{p},\mathbf{s}}$, processing it with a neural network $u_{\theta}$ and repositioning it at the original position using the transposed operator $(\Rd_j^{\mathbf{p},\mathbf{s}})^{\trans}$. The example is shown for a 2D cine MR image sequence.
\end{remark}

\subsection{Reconstruction Algorithm}
After having found the CNN prior \eqref{xcnn}, as a final reconstruction step, the optimality condition for problem (\ref{NN_reg_inv_problem}) is solved with an iterative method  dependent on the specific application. The solution of (\ref{NN_reg_inv_problem}) is then the final CNN-based reconstruction. 
 Algorithm \ref{alg} summarizes the proposed three-step reconstruction scheme.
\begin{algorithm}
\begin{algorithmic}
\State \textbf{Data:} trained network $u_{\theta}$, function $f_{\theta}$, noisy or incomplete measured data $\YY$, regularization parameter $\lambda > 0$
\State \textbf{Output:} reconstruction $\XX_{\mathrm{REC}}$
\State 1) $\XX_{\mathrm{ini}} \gets \Ad^{\dagger} \YY$
\State 2) $\Xcnn \gets f_{\theta}(\XX_{\mathrm{ini}})$
\State 3) $\XX_{\mathrm{REC}} \gets \argmin_{\XX}  D(\Ad \XX, \YY) + \lambda \|  \XX  - \Xcnn \|_2^2$
\State \textbf{Return} $\XX_{\mathrm{REC}}$
\end{algorithmic}
\caption{ Proposed CNNs-based large scale image reconstruction algorithm.\label{alg}}
\end{algorithm}

Note that  the regularizer   $  \mathcal{R}(\XX) = 
\|  \XX  - \Xcnn \|_2^2$ is strongly convex.  Therefore,  if the  
discrepancy term $D(\Ad \XX, \YY)$  is convex, 
then  the Tikhonov functional \eqref{NN_reg_inv_problem} is strongly 
convex and can be efficiently minimized by most gradient based iterative  schemes including Landweber's iteration and Conjugate Gradient type methods.
The specific strategy for minimizing (\ref{NN_reg_inv_problem}) depends on the  considered application.
In the case of an  ill-conditioned inverse problem with noisy measurements,     it might be beneficial that (\ref{NN_reg_inv_problem}) is  only  approximately minimized. For example, for the case of low-dose CT, early stopping of the Landweber iteration is applied  as additional  regularization method due to the semi-convergence property of the Landweber iteration \cite{strand1974theory}.  Such a strategy is  used for the numerical results presented in Section~\ref{sec_results}. 
\subsection{Convergence Analysis}
\label{sec:theory}

Another benefit of our approach is that minimization  of the Tikhonov  functional  \eqref{NN_reg_inv_problem}  corresponds  to
 convex  variational regularization with a  quadratic regularizer.
 Therefore, one can use well known stability, convergence and 
 convergence rates results 
\cite{engl1996regularization,scherzer2009variational,grasmair2010generalized,li2018nett}. Consequently,   opposed to most existing  
neural network based reconstruction algorithms, the  proposed
framework is build on the  solid theoretical fundament 
for regularizing  inverse problems.  As example of such results we  
have the following theorem.            

\begin{theorem}[Convergence of CNN-based regularization]\label{thm}
Let  $\Ad \colon X \to Y$  be  linear,  
$\Xcnn \in X$, $\YY_0 \in \Ad(X)$, and $\YY_\delta  \in Y$ satisfy 
 $\| \YY_\delta  -\YY_0 \| \leq \delta$. Then the following hold:
\begin{enumerate}[label=(\alph*)]
\item \label{tik1}
For all $\delta, \lambda >0$, the 
quadratic Tikhonov functional  
\begin{equation}\label{reg-q}
	F_{\YY_\delta, \Xcnn,\lambda} (\XX) 
	:=  \| \Ad \XX-\YY_\delta \|_2^2  + \lambda \| \XX - \Xcnn\|_2^2
\end{equation}
has a  unique minimizer  $\XX_{\delta,\lambda}$.  
  
\item \label{tik2}
The equation $\Ad \XX = \YY_0$ has a  
unique $\Xcnn$-minimizing solution         
$\XX_0 \in \argmin \{ \| \XX - \Xcnn\|^2 :   \Ad \XX = \YY_0 \}$.

\item \label{tik3}
If  the parameter choice $\lambda = \lambda(\delta)$ satisfies 
$\lambda, \delta^2/\lambda \to 0$
as $\delta \to 0$, then $\lim_{\delta \to 0}
\|  \XX_0 -  \XX_{\delta,\lambda} \| =0$. 
\end{enumerate}\label{thm:tik}
\end{theorem}

\begin{IEEEproof}
The change of variables   
\begin{itemize}
\item $\bar \XX := \XX - \Xcnn$
\item $ \bar \YY_0 := \YY_0 - \Ad \Xcnn$
\item $ \bar \YY_\delta :=  \YY_\delta - \Ad \Xcnn$ 
\end{itemize}
 reduces \eqref{reg-q} to  standard  
 Tikhonov regularization $ \| \Ad \bar \XX - \bar \YY_\delta \|^2  + \lambda \| \bar \XX \|_2^2 \to \min_\XX$  for the inverse problem $\Ad \bar \XX = \bar \YY_0$.
Therefore, Items~\ref{tik1} - \ref{tik3}  follow from standard results that can be found for example in~\cite[Section 5]{engl1996regularization}.
\end{IEEEproof}
 
 Theorem \ref{thm:tik} also holds in the infinite-dimensional 
 setting \cite[Section 5]{engl1996regularization} reflecting the 
 stability of the proposed CNN regularization.  
 Similar results hold for  nonlinear problems and  general 
 discrepancy measures \cite{grasmair2010generalized}. Moreover, one  can derive quantitative error 
 estimates  similar to \cite{scherzer2009variational,grasmair2010generalized,li2018nett}.
Such  theoretical investigations, however, are beyond the scope of this paper.  

\section{Experiments}\label{sec_experiments}
In the following, we evaluated our proposed method on two different examples of large-scale inverse problems given by 3D low-dose CT and 2D undersampled radial cine MRI. We compared our proposed method to the well-known TV-minimization-based and dictionary learning-based approaches presented in \cite{block2007}, \cite{wang2014compressed} and \cite{tian2011low}, \cite{xu2012low}, which we abbreviate by TV and DIC, respectively. Further details about the comparison  methods are discussed later in the paper.

\subsection{2D Radial Cine MRI}

Here we applied our method to image reconstruction in undersampled 2D radial cine MRI. Typically, MRI is performed using multiple receiver coils and therefore, the inverse problem is given by 
\begin{equation}\label{IP_MRI_multiple}
\Eu \XX = \Yu,
\end{equation}
where $\XX \in \mathbb{C}^N$ with $N=N_x\cdot N_y\cdot N_t$ is an unknown complex-valued image sequence. The encoding operator $\Eu$ is given by $\Eu = \mathbf{S}\circ \Ed \circ \Cd$ where 
\begin{eqnarray}\label{enc_op}
\Cd &= [ \Cd_1,\ldots, \Cd_{n_c}]^{\trans}, \\
\Ed &= \mathrm{diag}( \Fd,\ldots, \Fd ), \\ 
\mathbf{S}  &= \mathrm{diag}( \Su,\ldots, \Su).  
\end{eqnarray}
Here, $\Cd_i$ denotes the $i$-th coil sensitivity map, $n_c$ is the number of coil-sensitivity maps, $\Fd$ the 2D frame-wise operator and $\Su$ with $I \subset J = \{1,\ldots,N_{\mathrm{rad}}\}$, $|I|:=m \leq N_{\mathrm{rad}}$, a binary mask which models the undersampling process of the $N_{\mathrm{rad}}$ Fourier coefficients sampled on a radial grid. The vector  $\Yu \in \mathbb{C}^M$ with $M=m\cdot n_c$ corresponds to the measured data. 
Here, we sampled the $k$-space data along radial trajectories chosen according to the golden-angle method \cite{winkelmann2006optimal}. 
Note that problem (\ref{IP_MRI_multiple}) is mainly ill-posed not due to the presence of noise in the acquisition, but because the data acquisition is accelerated and hence only a fraction of the required measurements is acquired. 

If we assume a radial data-acquisition grid, problem (\ref{IP_MRI_multiple}) is a large-scale  inverse problem mainly because of two reasons. First, the measurement vector $\Yu$ corresponds to $n_c$ copies of the Fourier encoded image data multiplied by the corresponding coil sensitivity map. Second, the adjoint operator $\Eu^\herm$ consists of two computationally demanding steps. The radially acquired $k$-space data is first properly re-weighted and interpolated to a Cartesian grid, for example by using Kaiser-Bessel functions \cite{rasche1999resampling}. Then, a 2D inverse Fourier operation is applied to the image of each cardiac phase and the final image sequence is obtained by weighting the images from each estimated coil-sensitivity map and combining them to a single image sequence. We refer to the reconstruction obtained by $\Xu=\Eu^\herm\Yu$ as the non-uniform fast Fourier-transform (NUFFT) reconstruction.
Therefore, in radial multi-coil MRI, the measured $k$-space data is high-dimensional and the application of the encoding operators $\Eu$ and $\Eu^\herm$ is further more computationally demanding than sampling on a Cartesian grid, see e.g\ \cite{smith2019trajectory}. This makes the construction of cascaded networks which also process the $k$-space data \cite{han2019k} or repeatedly employ the forward and adjoint operators \cite{schlemper2017deep}, \cite{qin2018convolutional} computationally challenging. Therefore, decoupling the regularization given by the CNNs from the data-consistency step is necessary in this case.

As proposed in Section \ref{proposed_approach_section}, we solve a regularized version of problem (\ref{IP_MRI_multiple}) by minimizing 
\begin{equation}\label{DC_eq_MRI}
F_{ \Yu, \Xcnn, \lambda} (\XX) = \| \Eu \XX - \Yu  \|_2^2 + \lambda \|  \XX  - \Xcnn \|_2^2,
\end{equation}
where $\Xcnn$ is obtained a-priori by using an already trained network.
For this example, for obtaining the CNN-prior $\Xcnn$, we adopted the XT,YT approach presented in \cite{kofler2019}, where a modified version of the 2D U-net is used to process spatio-temporal slices which can be extracted from the image sequence. Since the XT,YT method was previously introduced to only process real-valued data (i.e.\ the magnitude images), we followed a similar strategy by processing the real and imaginary parts of the image sequences separately but using the same real-valued network $u_{\theta}$. This further increases the amount of training data by a factor of two.
More precisely, let $ \Rd^{xt}_j$ and $\Rd^{yt}_j$ denote the operators which extract the $j$-th two-dimensional spatio-temporal slices in $xt$- and $yt$-direction from a 3D volume $ (\Rd^{xt}_j)^\trans$ and $(\Rd^{yt}_j)^\trans$ their respective transposed operations which reposition the spatio-temporal slices at their original position. 

By $u_{\theta}$ we denote a 2D U-net as the one described in \cite{kofler2019} which is trained on spatio-temporal slices, i.e.\ on a dataset of pairs which consist of the spatio-temporal slices in $xt$- and $yt$-direction of both the real and imaginary parts of the complex-valued images. The network $u_{\theta}$ was trained to minimize the $L_2$-error between the ground truth image and the estimated output of the CNN. 
Our dataset consists of radially acquired 2D cine MR images from $n=19$ subjects (15 healthy volunteers and 4 patients with known cardiac dysfunction) with 30 images covering the cardiac cycle. The ground truth images were obtained by $kt$-SENSE reconstruction using $N_{\theta} = 3400$ radial lines. We retrospectively generated the radial $k$-space data $\Yu$ by sampling the $k$-space data along $N_{\theta} = 1130$ radial spokes using $n_c=12$ coils. Note that sampling $N_{\theta}=3400$ already corresponds to an acceleration factor of approximately $\sim 3$ and therefore, $N_{\theta}=1130$ corresponds to an accelerated data-acquisition by an approximate factor of $\sim 9$. The forward and the adjoint operators $\Ed_I$ and $\Ed_I^\herm$ were implemented using the \texttt{ODL} library \cite{adler_github}.
The complex-valued CNN-regularized  image sequence $\Xcnn$ was obtained by
\begin{align}\label{xcnn_MRI}
\Xcnn = f_{\theta}(\Xu) \nonumber \\ = \frac{1}{2} \Big[ \sum_j &(\Rd^{xt}_j)^{\trans}  \big(u_{\theta}( \Rd^{xt}_j (\mathrm{Re}\,\Xu)) \big) \nonumber \\ +
&(\Rd^{yt}_j)^{\trans}  \big(u_{\theta}( \Rd^{yt}_j (\mathrm{Re}\,\Xu)) \big) \nonumber \\ +
\mathrm{i}\, \Big(&(\Rd^{xt}_j)^{\trans}  \big(u_{\theta}( \Rd^{xt}_j (\mathrm{Im}\, \Xu))\big)\Big)  \nonumber \\ +
\mathrm{i}\, \Big(&(\Rd^{yt}_j)^{\trans}  \big(u_{\theta}( \Rd^{yt}_j (\mathrm{Im}\, \Xu))\big)\Big)\Big ] 
\end{align}
Given $\Xcnn$, functional  (\ref{DC_eq_MRI}) was minimized by setting its derivative with respect to $\XX$ to zero and applying the pre-conditioned conjugate gradient (PCG) method to iteratively solve the resulting system. PCG was used to solve the system $\Hd \XX = \mathbf{b}$ with
\begin{eqnarray}\label{lin_system_MRI}
\Hd = \Eu^\herm\Eu + \lambda\, \Id, \nonumber \\
\mathbf{b} = \Xu + \lambda\, \Xcnn.  
\end{eqnarray} 
Since the XT,YT method gives access to a large number of training samples, training the network $u_{\Theta}$ for 12 epochs was sufficient. The CNN was trained by minimizing the $L_2$-norm of the error between labels and output by using the Adam optimizer \cite{Adam}. We split our dataset in 12/3/4 subjects for training, validation and testing and performed a 4-fold cross-validation. For the experiment, we performed $n_{\mathrm{iter}} = 16$ subsequent iterations of PCG and empirically set $\lambda=0.1$.   Note that due to strong convexity, \eqref{DC_eq_MRI} has a unique minimizer and solving system \eqref{lin_system_MRI} yields the desired minimizer.  The obtained results can be found in Subsection \ref{subsec_MRI_results}.

\subsection{3D Low-Dose Computed Tomography}

The current generation of CT scanners performs the data-acquisition by emitting X-rays along trajectories in the form of a cone-beam for each angular position of the scanner. Therefore, for each angle $\phi$ of the rotation, one obtains an X-ray image which is measured by the detector array and thus, the complete sinogram data can be identified with a 3D array of shape $(N_{\phi},N_{r_x},N_{r_y})$. Thereby, $N_{\phi}$ corresponds to the number of angles the rotation of the scanner is discretized by and $N_{r_x}$ and $N_{r_y}$ denote the number of elements of the detector array. The values of these parameters vary from scanner to scanner but are in the order of $N_{\phi}\approx 1000$ for a full rotation of the scanner and $N_{r_x} \times N_{r_y} \approx 320 \times 800$ for a 320-row detector array, which is for example used for cardiac CT scans \cite{dewey2009noninvasive}. The volumes obtained from the reconstructions are typically given by an in-plane number of pixels of $N_x \times N_y = 512 \times 512$ and varying number of slices $N_z$, dependent on the specific application. 
For this example, we consider a similar set-up as in \cite{adler2017solving}. The non-linear problem is given by 
\begin{equation}\label{IP_CT}
\Yn = \mathbf{T} \XX + \mathbf{\eta} = p \, \mathrm{exp} \{ - \mu\, \Rd \XX\} + \mathbf{\eta},
\end{equation}
where $p$ denotes the average number of photons per pixel, $\mu$ is the linear attenuation coefficient of water, $\Rd$ corresponds to the discretized version of a ray-transform with cone-beam geometry and the vector $\mathbf{\eta}$ denotes the Poisson-distributed noise in the measurements. 
Following our approach, we are interested in solving
\begin{equation}\label{DC_eq_CT}
F_{ \Yn, \Xcnn, \lambda} (\XX) = D_{\mathrm{KL}}(\mathbf{T}\XX, \Yn) + \lambda \|  \XX  - \Xcnn \|_2^2 \rightarrow \mathrm{min},
\end{equation} 
where $D_{\mathrm{KL}}$ denotes the Kullback-Leibler divergence which corresponds to the log-likelihood function for Poisson-distributed noise. According to the previously introduced notation, the prior $\Xcnn$ is given by $\Xcnn = f_{\theta}(\Xn)$, where $f_{\theta}$ denotes a CNN-based processing method with trainable parameters $\theta$ and $\Xn = \Rd^{\dagger}(-\mu^{-1}\mathrm{ln}(p^{-1}\Yn))$ with $\Rd^{\dagger}$ being the filtered back-projection (FBP) reconstruction. 

Since our object of interest $\XX$ is a volume, it is intuitive to choose a NN which involves 3D convolutions in order to learn the filters by exploiting the spatial correlation of adjacent voxels in $x$-, $y$- and $z$-direction. In this particular case, $u_{\theta}$ denotes a 3D U-net similar to the one presented in \cite{Hauptmann2019}. Due to the large dimensionality of the volumes $\XX$, the network $u_{\theta}$ cannot be applied to the whole volume. Instead, following our approach, the volume was divided into patches to which the network $u_{\theta}$ is applied. Therefore, the output $\Xcnn$ was obtained as described in (\ref{xcnn}), where $u_{\theta}$ operates on 3D patches given by the vector $\mathbf{p}=(128,128,16)$, which denotes the maximal size of 3D patches which we were able to process by a 3D U-net. The strides used for the extraction and the reassembling of the volumes used in (\ref{xcnn}) is empirically chosen to be $\mathbf{s}=(16,16,8)$. 

Training of the network $u_{\theta}$ was performed on a dataset of pairs according to (\ref{dataset}), where we retrospectively generated the measurements $\Yn$ by simulating a low-dose scan on the ground truth volumes. For the experiment, we used 16 CT volumes from the randomized DISCHARGE trial \cite{napp2017computed} which we cropped to a fixed size of $512\times 512 \times 128$.  The simulation of the low-dose scan was performed as described in  \cite{adler2017solving} by setting  $p=10\,000$ and $\mu=0.02$. The operator $\Rd$ is assumed to  perform $N_{\phi}=1000$ projections which are measured by a detector array of shape $N_{r_x} \times N_{r_y} = 320 \times 800$. For the implementation of the operators, we used the  \texttt{ODL} library \cite{adler_github}. The source-to-axis  and source-to-detector distances were chosen according to the DICOM files.
Since the dataset is relatively small, we performed a 7-fold cross-validation where for each fold we split the dataset in 12 patients for training, 2 for validation and 2 for testing. The number of training samples $N_{\mathrm{train}}$ results from the number of patches times the number of volumes contained in the training set.  We trained the network $u_{\theta}$ for 115 epochs by minimizing the $L_2$-norm of the error between labels and outputs. For training, we used the Adam optimizer \cite{Adam}. With the described configuration of $\mathbf{p}$ and $\mathbf{s}$, the resulting number of patches to be processed in order to obtain the prior $\Xcnn$ is therefore given by $N_{\mathbf{p},\mathbf{s}}=9\,375$.
In this example, the solution $\XX_{\mathrm{REC}}$ to problem (\ref{DC_eq_CT}) was then obtained by performing $n_{\mathrm{iter}}=4$ iterations of Landweber's method where we further used the filtered-back projection $\Rd^{\dagger}$ as a left-preconditioner to accelerate the convergence of the scheme. For the derivation of the gradient of (\ref{DC_eq_CT}) with respect to $\XX$, we refer to \cite{adler2017solving}. The regularization parameter  was empirically set to $\lambda=1$. The results can be found in Subsection \ref{subsec_CT_results}. 

\subsection{Reference Methods}

Here we discuss the methods of comparison in more detail and report the times needed to process and reconstruct the images or volumes. The data-discrepancy term $D(\,\cdot\, , \,\cdot\,)$ was again chosen according to the considered examples as previously discussed. The TV-minimization approach used for comparison is given by solving
\begin{equation}\label{TV_min_problem}
 D(\Ad \XX , \YY) + \lambda \| \mathbf{G}\XX \|_1  \to \min_{\XX}
\end{equation}
where $\mathbf{G}$ denotes the discretized version of the isotropic first order finite differences filter in all three dimensions.
The solution of problem (\ref{TV_min_problem}) was obtained by introducing an auxiliary variable $\mathbf{z}$ and alternating between solving for $\XX$ and $\mathbf{z}$. For the solution of one of the sub-problems, an iterative shrinkage method was used, see \cite{chambolle2005total} for more details. The second resulting sub-problem was solved by iteratively solving a system of linear equations, either by Landweber for the CT example or by PCG for the MRI example, as  mentioned before.

The dictionary learning-based method used for comparison is given by the solution of the problem
\begin{equation}
D( \Ad \XX, \YY) + \lambda \| \XX - \XX_{\mathrm{DIC}}\|_2^2 \to \min_{\XX}
\end{equation}
where, in contrast to our proposed method, $\XX_{\mathrm{DIC}}$ was obtained by the patch-wise sparse approximation  of the initial image estimate using an already trained dictionary $\mathbf{D}$. Therefore, using a similar notation as in (\ref{xcnn}), the prior $\XX_{\mathrm{DIC}}$ is given by
\begin{equation}\label{xDL}
\XX_{\mathrm{DIC}} = \mathbf{W}_{\mathbf{p},\mathbf{s}}\sum_j (\Rd_j^{\mathbf{p},\mathbf{s}})^{\trans}\, \mathbf{D} \boldsymbol{\gamma}_j,
\end{equation}
where the dictionary $\mathbf{D}$ was previously trained by 15 iterations of the iterative thresholding and $K$ residual means algorithm (ITKRM) \cite{schnass2018convergence} on a set of ground truth images which were given by the high-dose images for the CT example and the $kt$-SENSE reconstructions from $N_{\theta} = 3400$ radial lines for the MRI example. Note that for each fold, for  training the dictionary $\mathbf{D}$, we only used the data which we included in the training set for our method. This means we trained a total of seven dictionaries for the CT example and four dictionaries for the MRI example. For each iteration of ITKRM, we randomly selected a subject to extract $10\ 000$  3D training patches. The corresponding sparse codes $\boldsymbol{\gamma}_j$ were then obtained by solving 
\begin{equation}\label{sparse_coding}
\mathrm{min}_{\{\boldsymbol{\gamma}_j\}_j} \sum_j \big( \|  (\Rd_j^ {\mathbf{p},\mathbf{s}}) \XX_{\mathrm{ini}} - \mathbf{D} \boldsymbol{\gamma}_j\|_2^2 + \| \boldsymbol{\gamma}_j\|_0\big),
\end{equation}
which is a sparse coding problem and was solved using orthogonal matching pursuit (OMP) \cite{tropp2007signal}. Thereby, the image $\XX_{\mathrm{ini}}$ corresponds to either the FBP-reconstruction $\XX_{\eta}$ for the CT example or to the NUFFT-reconstruction $\Xu$ for the MRI example. In both cases, we used patches of shape given by $\mathbf{p} = (4,4,4)$ and strides given by $\mathbf{s} = (2,2,2)$. The number of atoms $K$ and the sparsity levels were set to $K=4\cdot d$, with $d=4\cdot 4 \cdot 4$ and $S=16$.
Note that, in contrast to \cite{xu2012low} and \cite{wang2004image}, \cite{caballero2014dictionary}, the dictionary and the sparse codes were not learned during the reconstruction, as the sparse coding step of all patches would be too time consuming for very large-scale  inverse problems, such as  the CT example. Instead, the dictionary and the sparse codes were used to generate the prior $\XX_{\mathrm{DIC}}$ which makes the method also more similar and comparable to ours. The parameter $\lambda$ is set as previously stated in the manuscript, depending on the considered example.

\subsection{Quantitative Measures}
For the evaluation of the reconstructions we report the normalized root mean squared error (NRMSE) and the peak signal-to-noise ratio (PSNR) as error-based measures and the structural similarity index measure (SSIM) \cite{wang2004image} and the Haar Wavelet-based perceptual similarity index measure (HPSI) \cite{reisenhofer2018haar} as image-similarity-based measures. The reported statistics were obtained by calculating the measures of the images in the $xy$-plane and averaging them over the different folds.

\section{Results}\label{sec_results}

\subsection{Results for 2D Radial Cine MRI}\label{subsec_MRI_results}

\begin{figure}[!h]
\centering
\begin{minipage}{0.1\linewidth}\vspace{-3.2cm}
\includegraphics[width=0.55\linewidth]{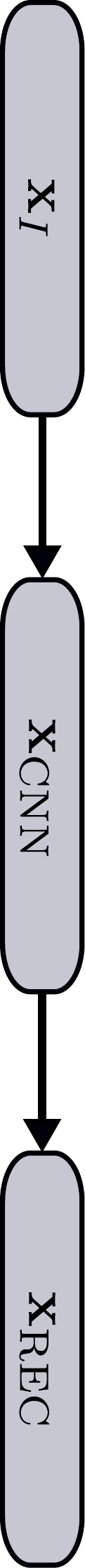}
\end{minipage} \hspace{-0.6cm} 
\begin{minipage}{0.9\linewidth}
\resizebox{\linewidth}{!}{
\includegraphics[height=3.4cm]{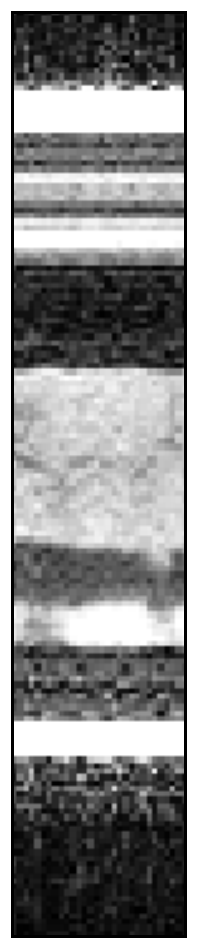}\hspace{-0.2cm}
\begin{overpic}[height=3.4cm,tics=10]{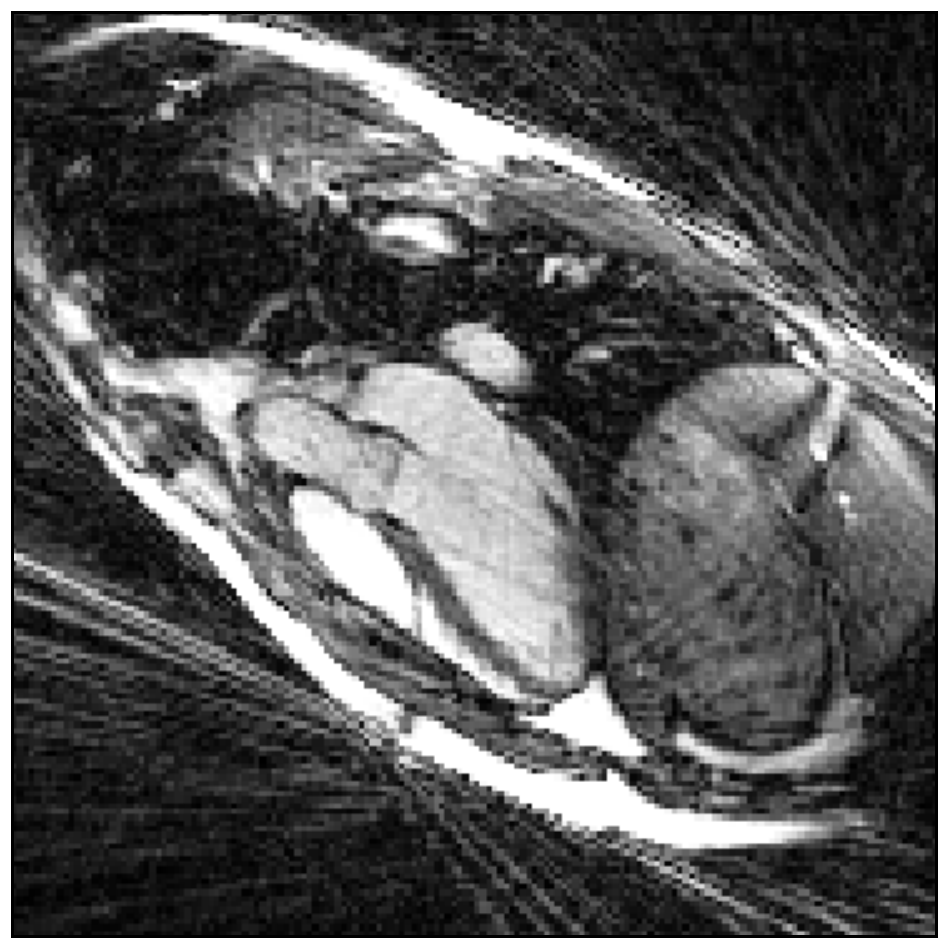}
 \put (82,82) {\Large\textcolor{white}{A}}\end{overpic}\hspace{-0.1cm}
\includegraphics[height=3.4cm]{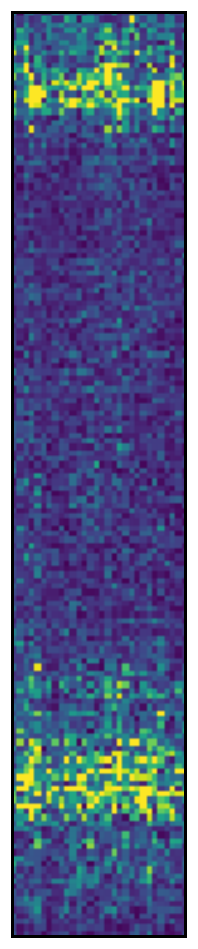}\hspace{-0.2cm}
\begin{overpic}[height=3.4cm,tics=10]{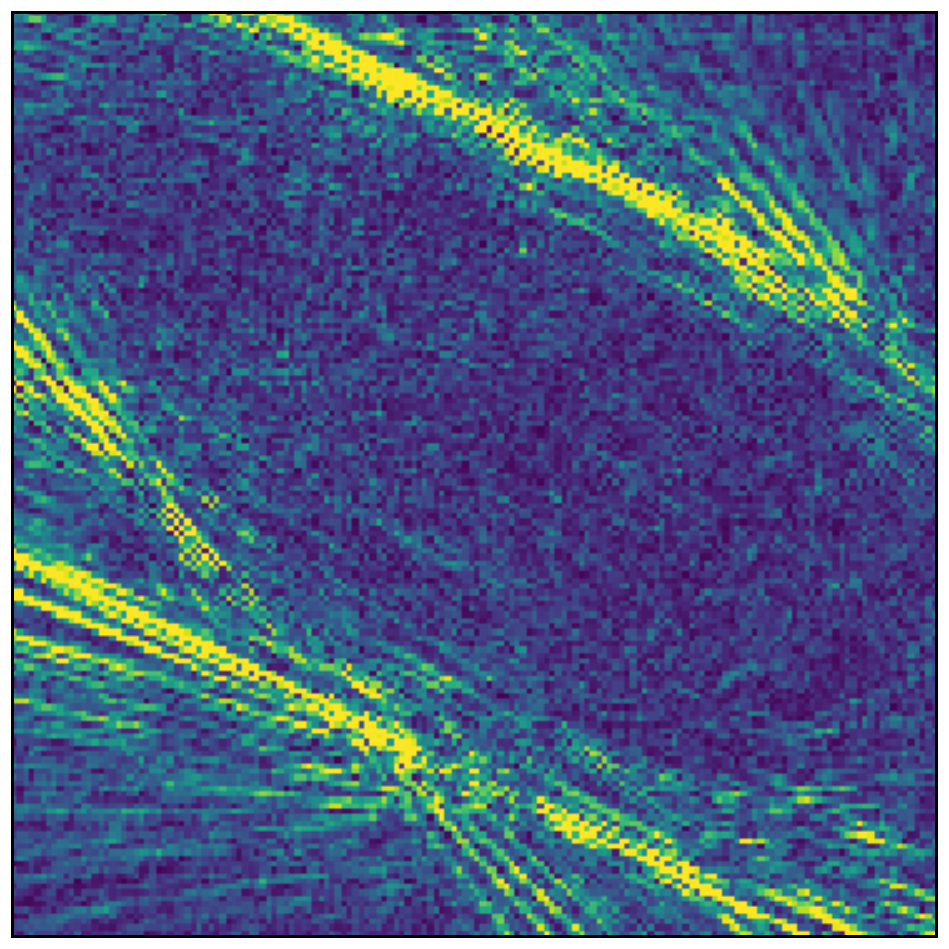}
\end{overpic}\hspace{-0.1cm}
}
\resizebox{\linewidth}{!}{
\includegraphics[height=3.4cm]{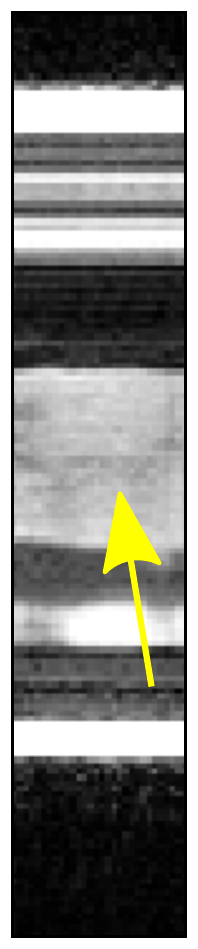}\hspace{-0.2cm}
\begin{overpic}[height=3.4cm,tics=10]{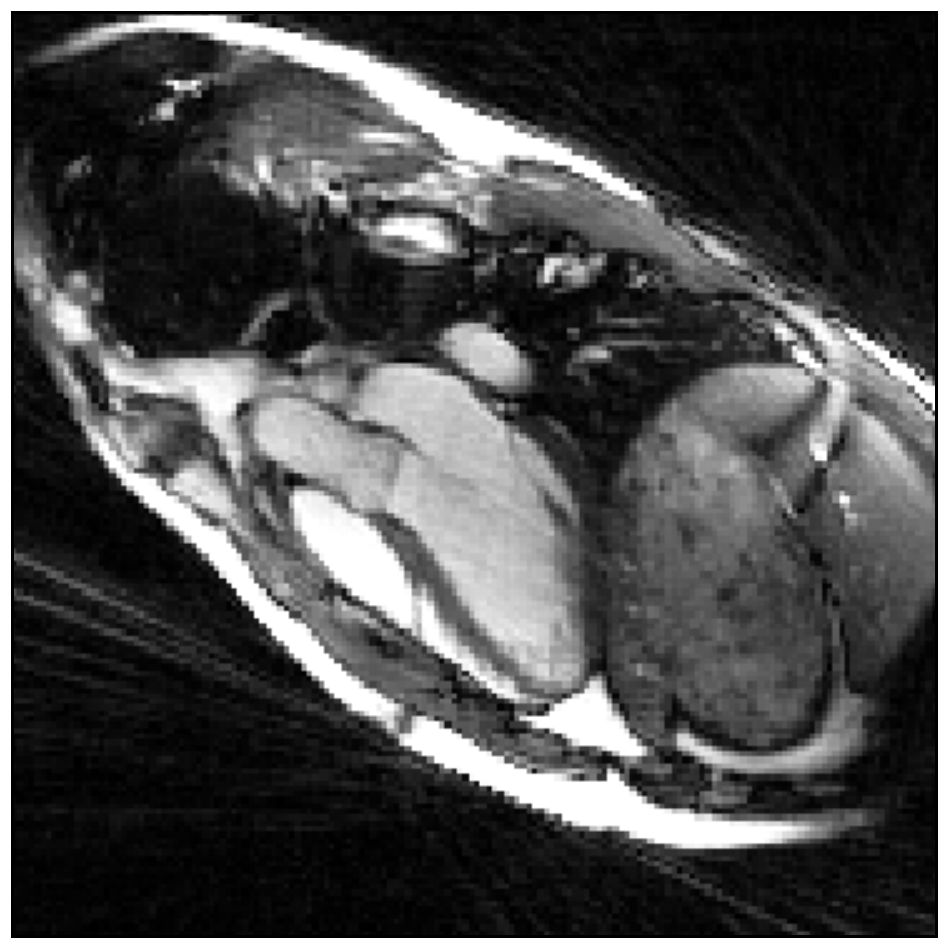}
  \put (82,82) {\Large\textcolor{white}{B}}
\end{overpic}\hspace{-0.1cm}
\includegraphics[height=3.4cm]{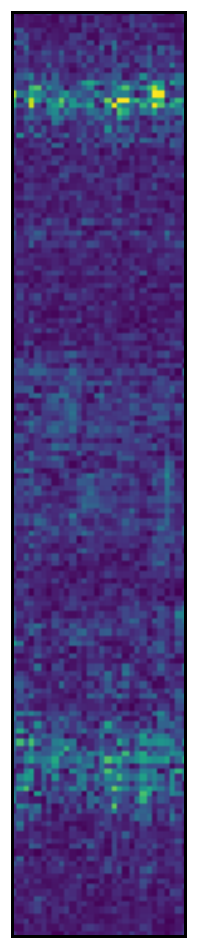}\hspace{-0.2cm}
\begin{overpic}[height=3.4cm,tics=10]{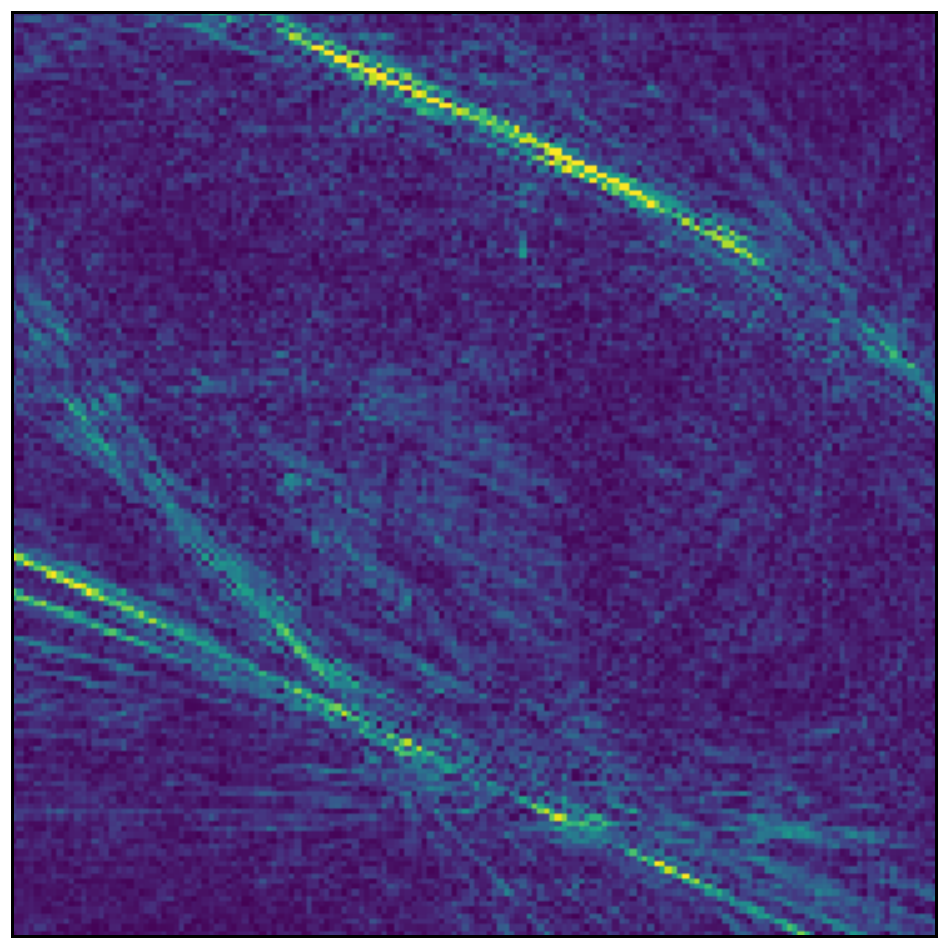}
\end{overpic}\hspace{-0.1cm}
}
\resizebox{\linewidth}{!}{
\includegraphics[height=3.4cm]{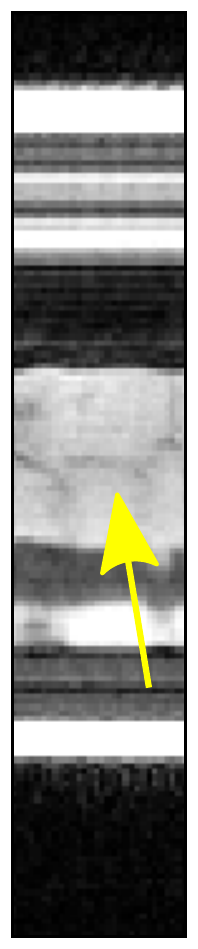}\hspace{-0.2cm}
\begin{overpic}[height=3.4cm,tics=10]{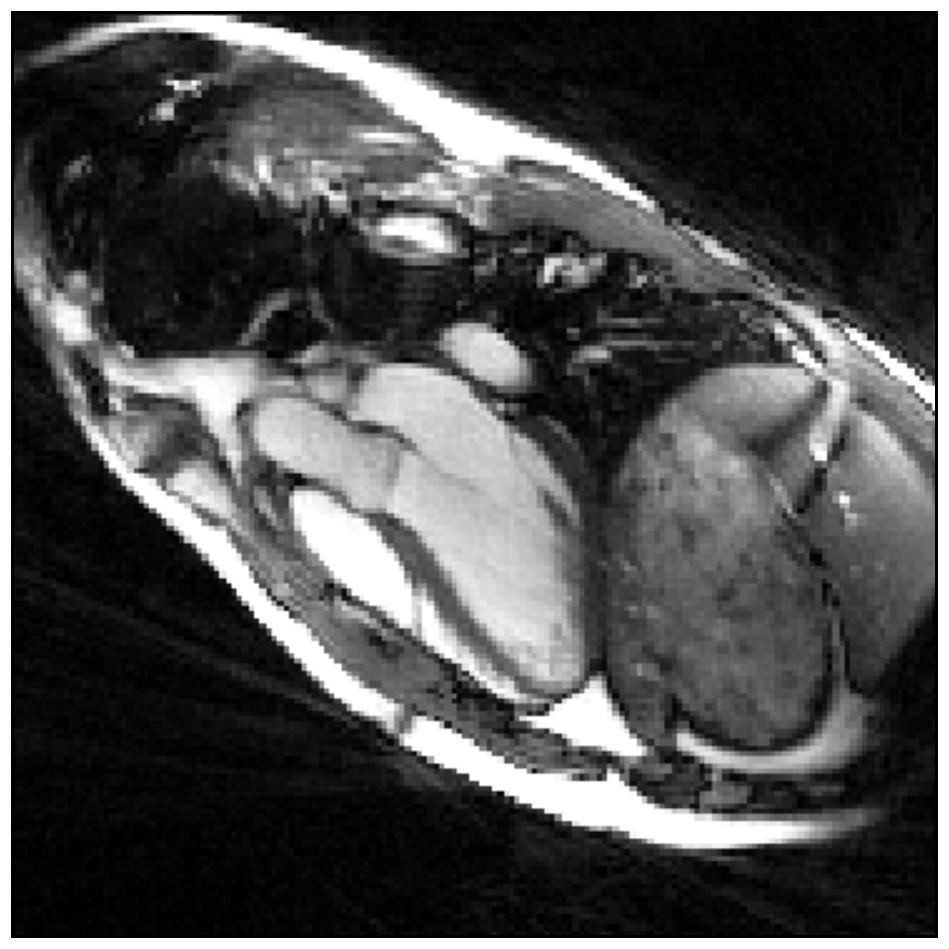}
 \put (82,82) {\Large\textcolor{white}{C}}
\end{overpic}\hspace{-0.1cm}
\includegraphics[height=3.4cm]{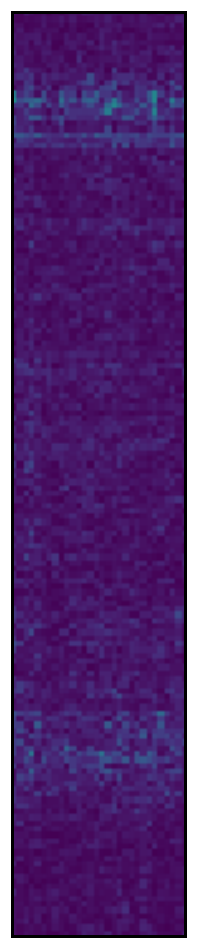}\hspace{-0.2cm}
\begin{overpic}[height=3.4cm,tics=10]{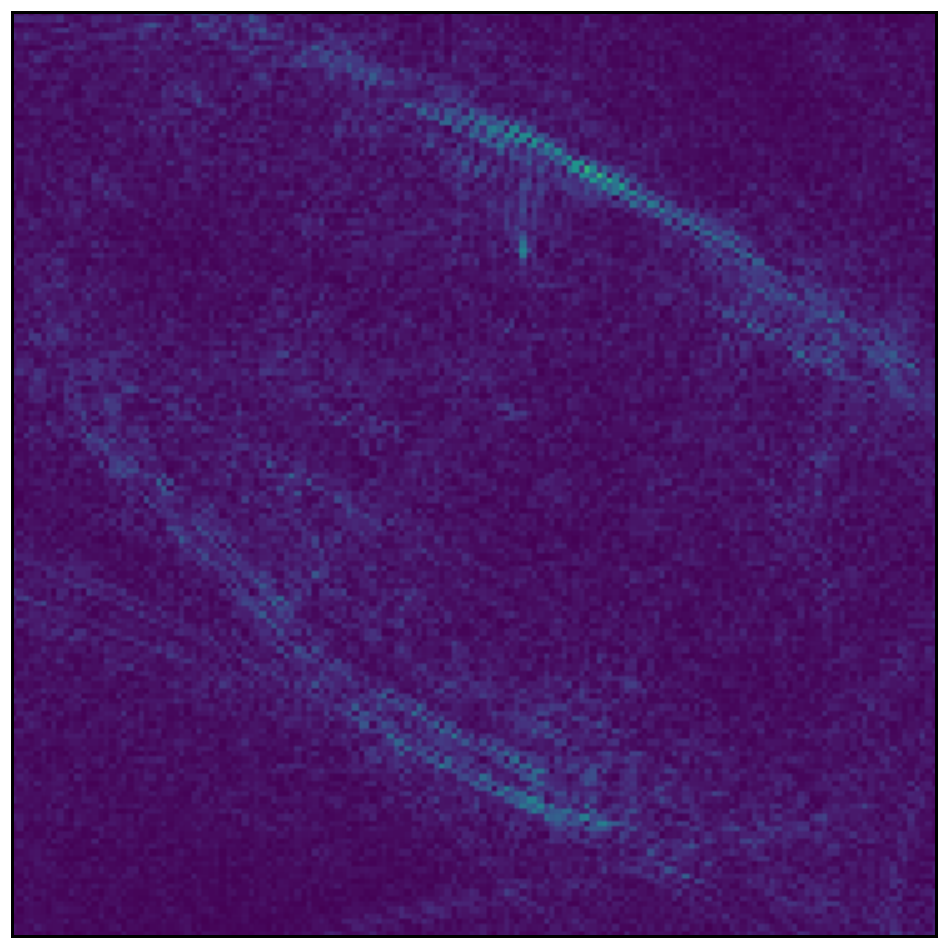}
\end{overpic}\hspace{-0.1cm}
}
\resizebox{\linewidth}{!}{
\includegraphics[height=3.4cm]{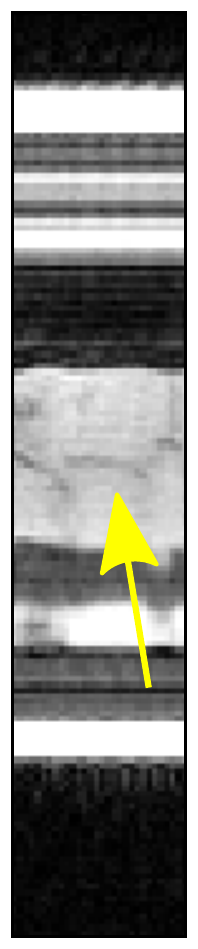}\hspace{-0.2cm}
\begin{overpic}[height=3.4cm,tics=10]{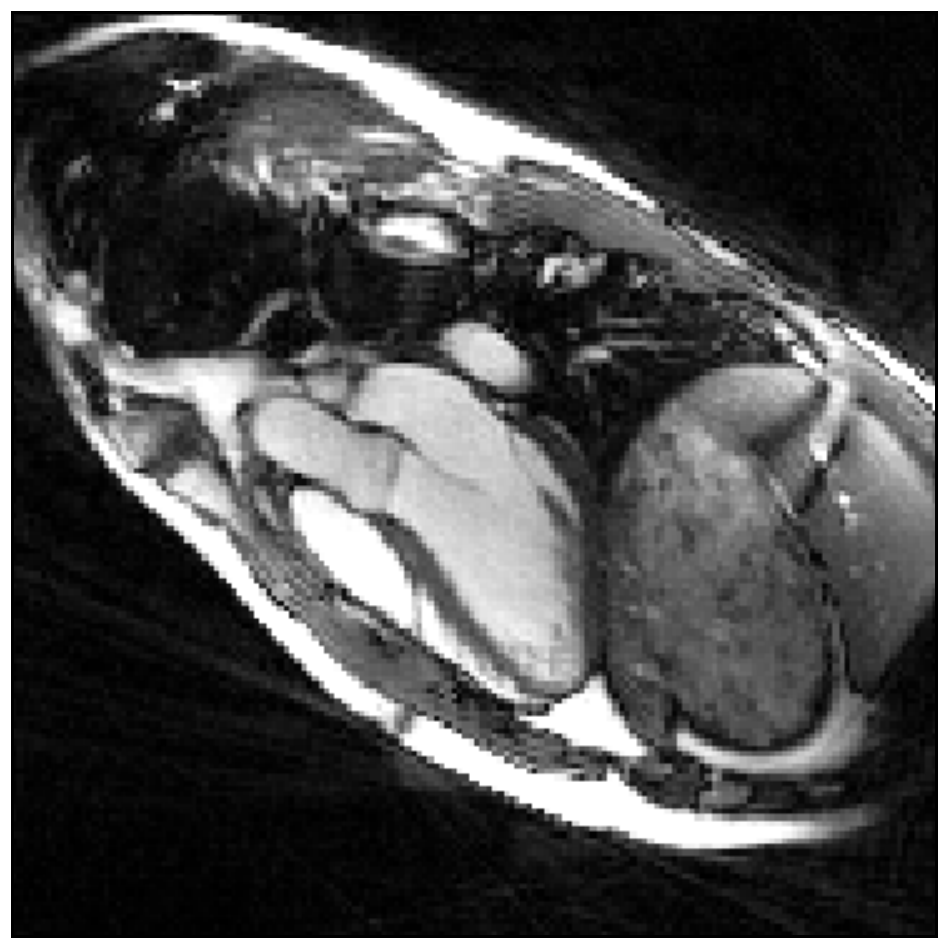}
 \put (82,82) {\Large\textcolor{white}{D}}\hspace{-0.1cm}
\end{overpic}
\includegraphics[height=3.4cm]{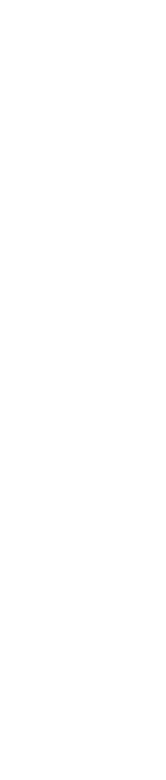}\hspace{-0.2cm}
\begin{overpic}[height=3.4cm,tics=10]{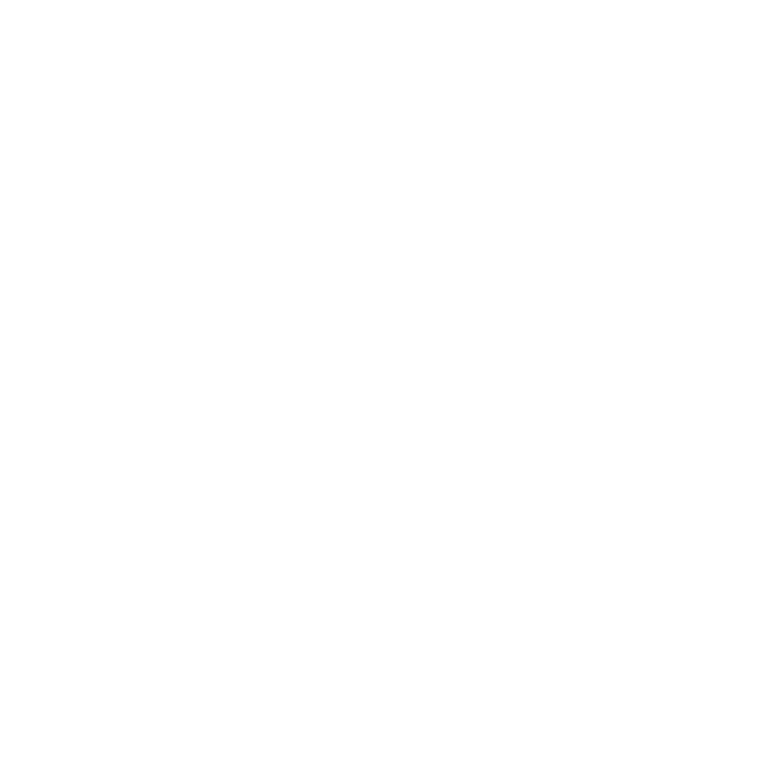}
\end{overpic}\hspace{-0.1cm}
}
\end{minipage}
\caption{Results for a healthy volunteer showing two slices with different orientations. A: Initial NUFFT-reconstruction $\Xu$ using $N_{\theta} =1130$ radial spokes, B: estimated output $\Xcnn$ using the spatio-temporal 2D XT,YT U-net, C: solution of the CNNs-based regularized functional $\XX_{\mathrm{REC}}$, D: ground truth image reconstruction with $kt$-SENSE and $N_{\theta} = 3400$ radial spokes. All images are displayed in the same scale. For better visibility, the point-wise error images are magnified by a factor of $\times 3$. The yellow arrows point at details which are smoothed out in the CNN-prior $\Xcnn$ but are visible again in the final reconstruction $\XX_{\mathrm{REC}}$.}\label{MRI_intermediate_results_figs}
\end{figure}

\begin{figure*}			
\centering
\begin{minipage}{0.495\linewidth}
\resizebox{\linewidth}{!}{
\includegraphics[height=3.4cm]{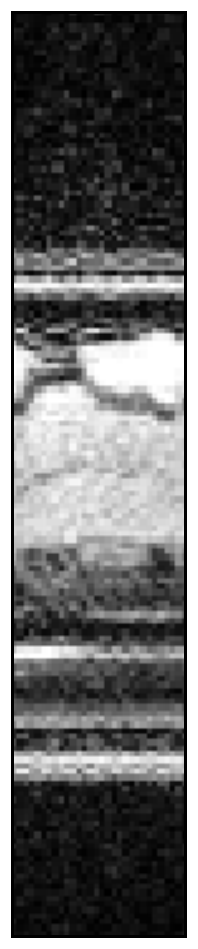}\hspace{-0.2cm}
\begin{overpic}[height=3.4cm,tics=10]{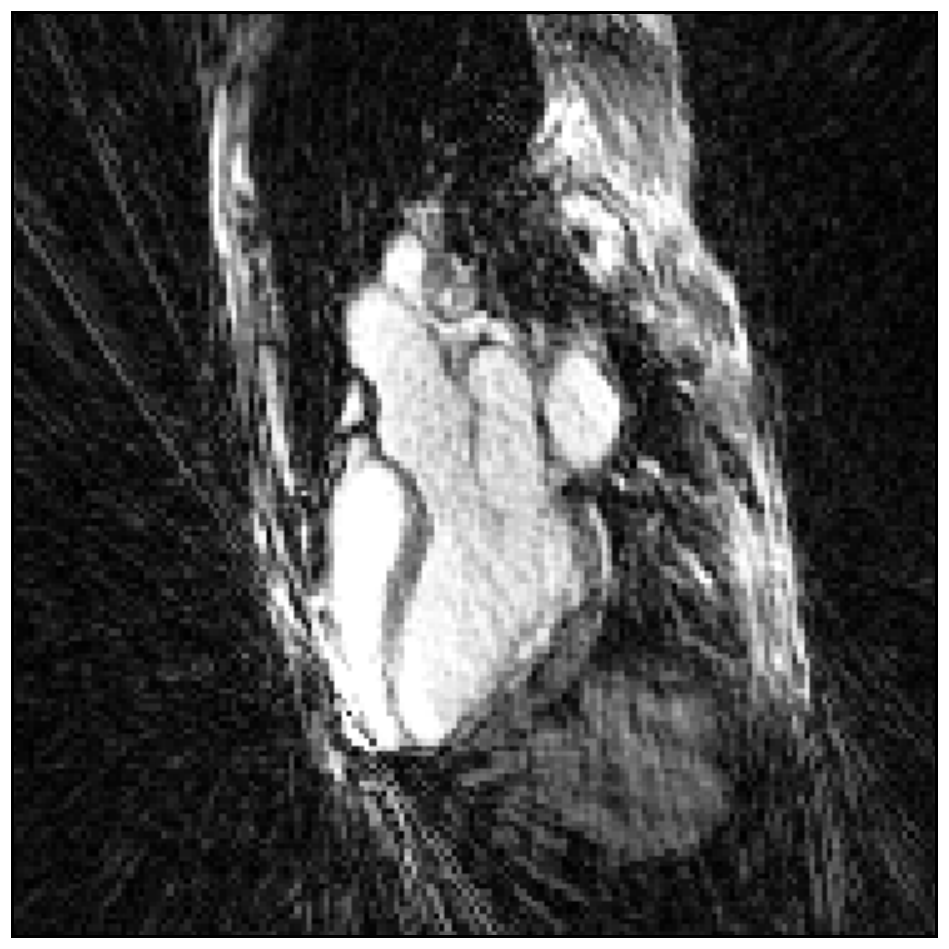}
 \put (84,84) {\Large\textcolor{white}{A}}
\end{overpic}\hspace{-0.1cm}
\includegraphics[height=3.4cm]{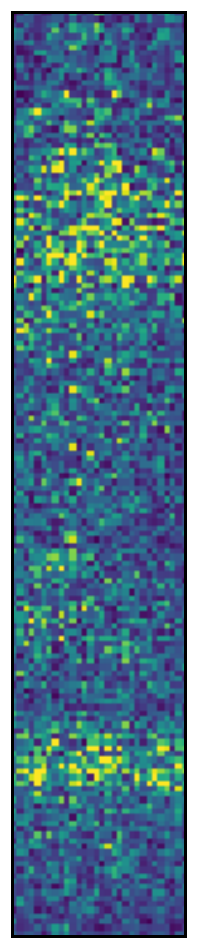}\hspace{-0.2cm}
\begin{overpic}[height=3.4cm,tics=10]{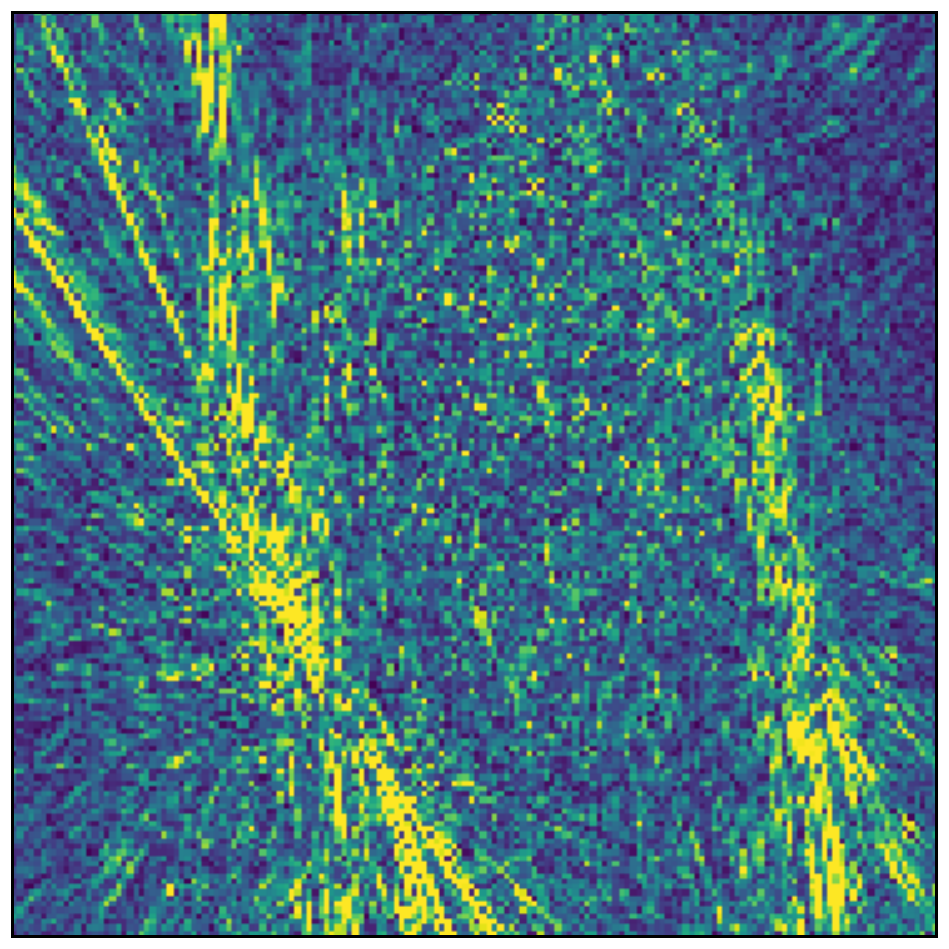}
\end{overpic}
}
\resizebox{\linewidth}{!}{
\includegraphics[height=3.4cm]{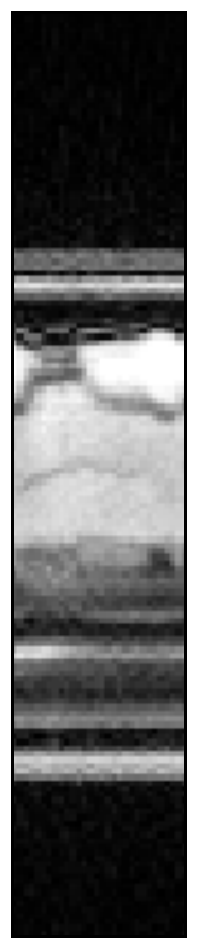}\hspace{-0.2cm}
\begin{overpic}[height=3.4cm,tics=10]{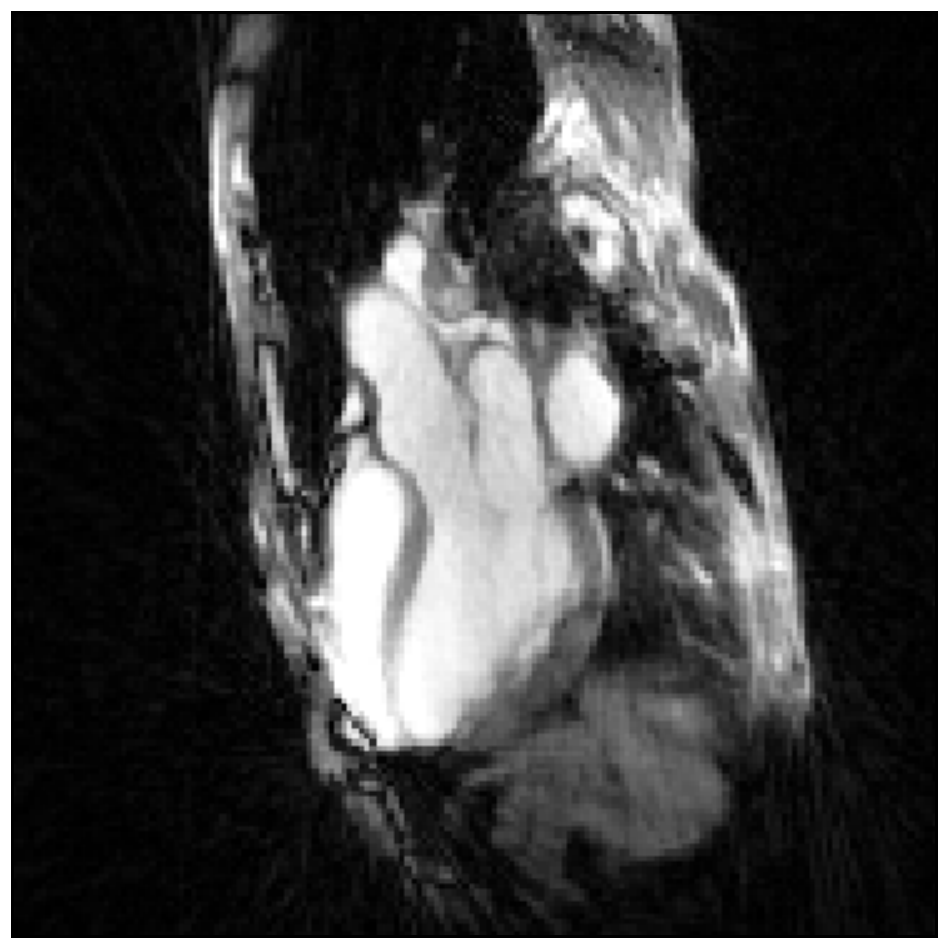}
 \put (84,84) {\Large\textcolor{white}{B}}
\end{overpic}\hspace{-0.1cm}
\includegraphics[height=3.4cm]{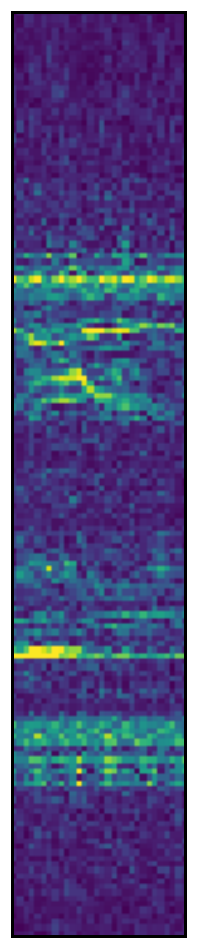}\hspace{-0.2cm}
\begin{overpic}[height=3.4cm,tics=10]{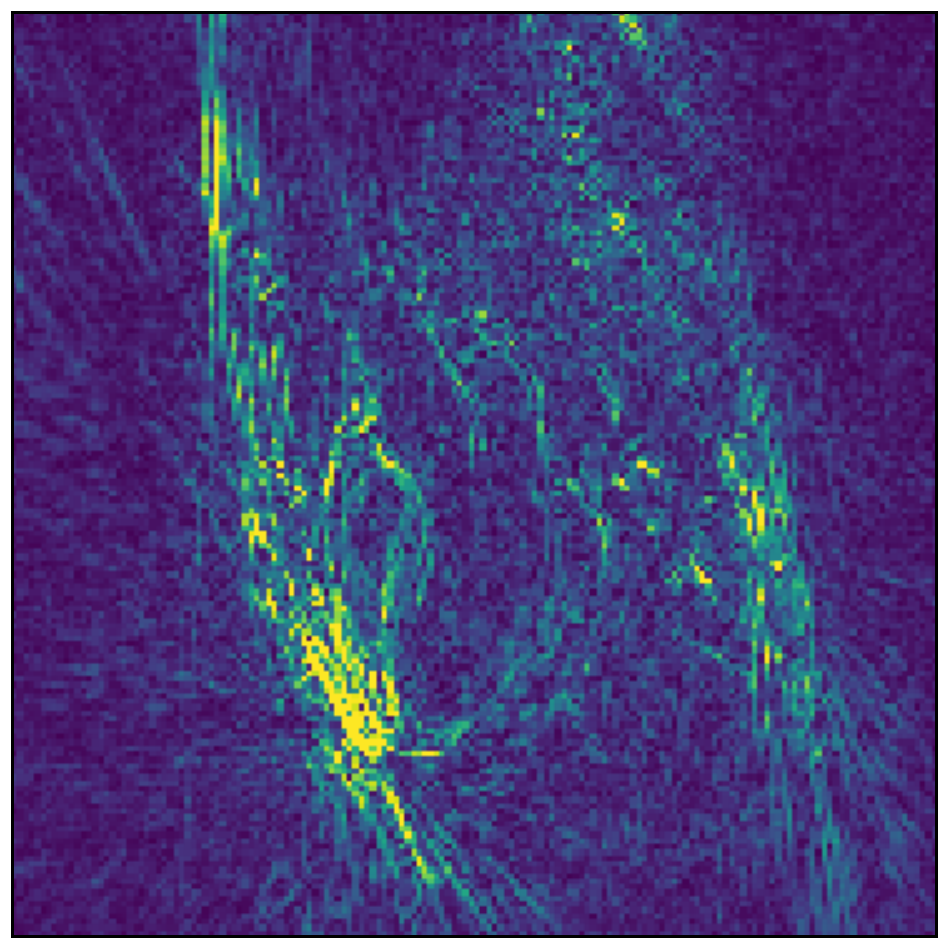}
\end{overpic}
}
\resizebox{\linewidth}{!}{
\includegraphics[height=3.4cm]{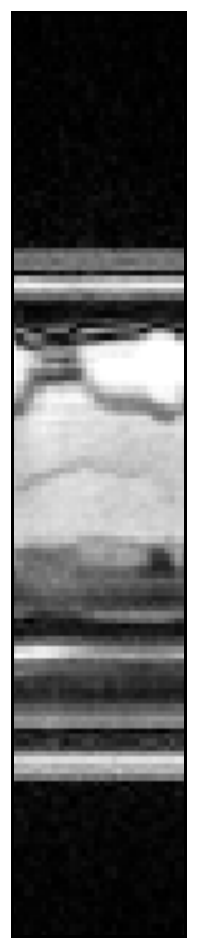}\hspace{-0.2cm}
\begin{overpic}[height=3.4cm,tics=10]{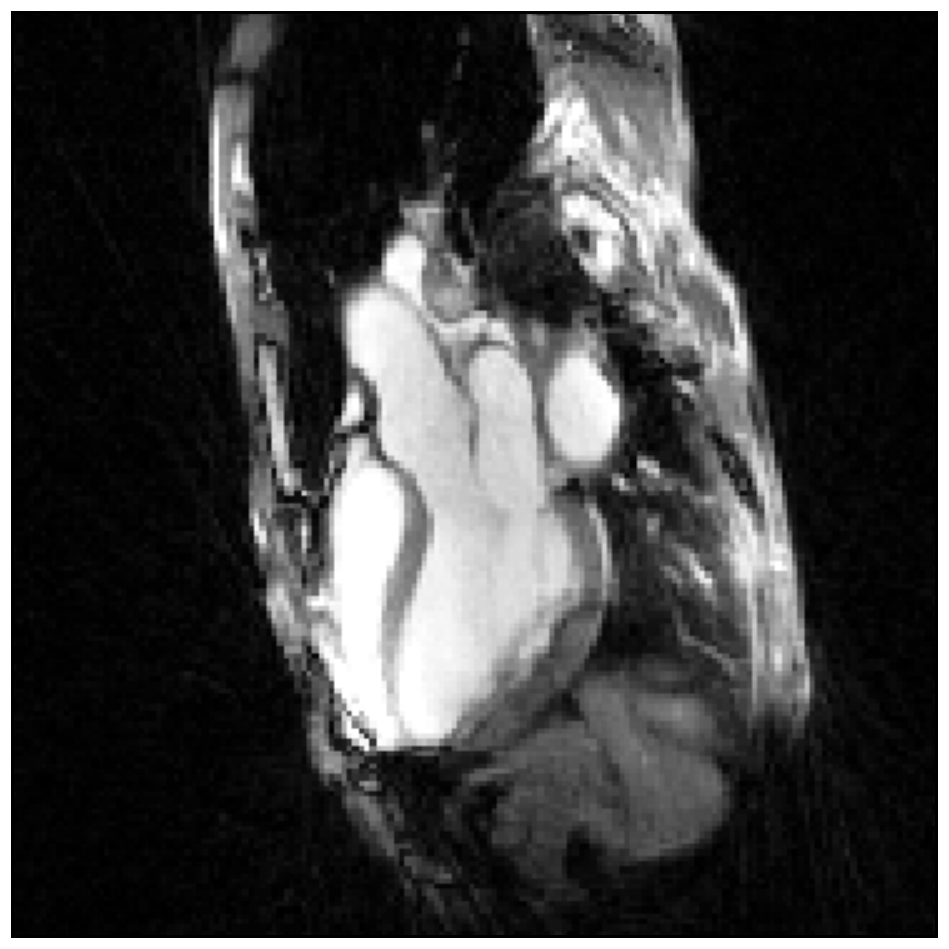}
 \put (84,84) {\Large\textcolor{white}{C}}
\end{overpic}\hspace{-0.1cm}
\includegraphics[height=3.4cm]{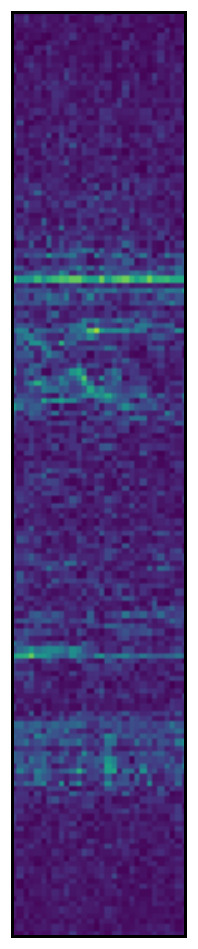}\hspace{-0.2cm}
\begin{overpic}[height=3.4cm,tics=10]{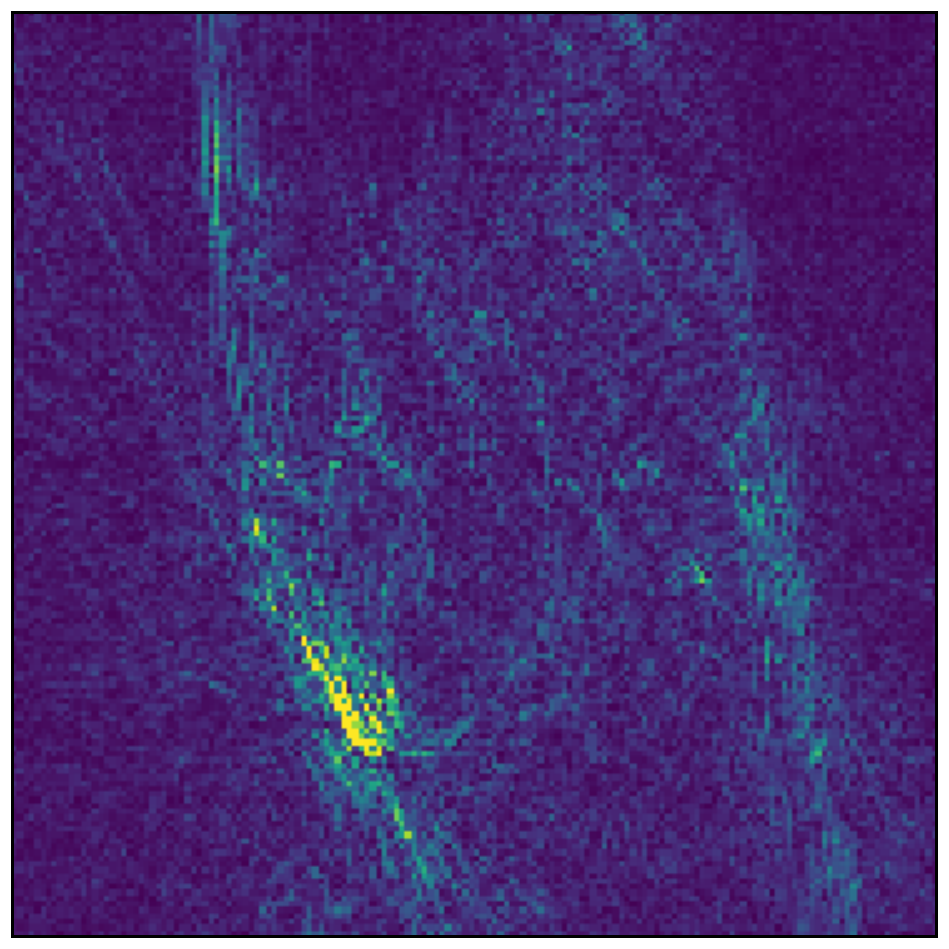}
\end{overpic}
}
\resizebox{\linewidth}{!}{
\includegraphics[height=3.4cm]{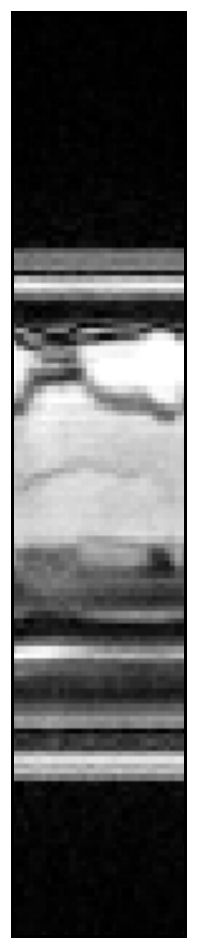}\hspace{-0.2cm}
\begin{overpic}[height=3.4cm,tics=10]{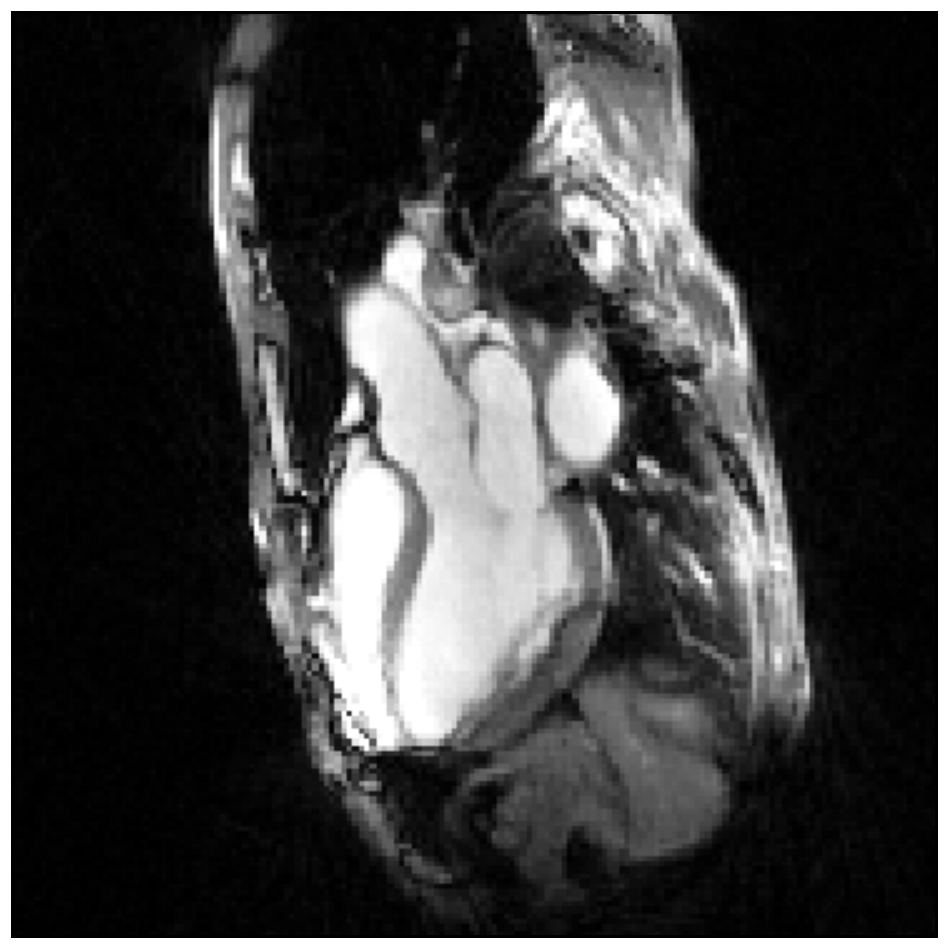}
 \put (84,84) {\Large\textcolor{white}{D}}
\end{overpic}\hspace{-0.1cm}
\includegraphics[height=3.4cm]{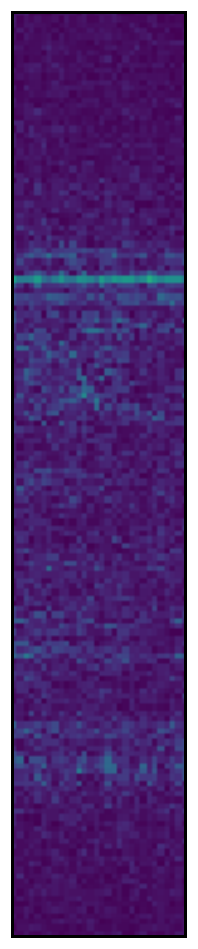}\hspace{-0.2cm}
\begin{overpic}[height=3.4cm,tics=10]{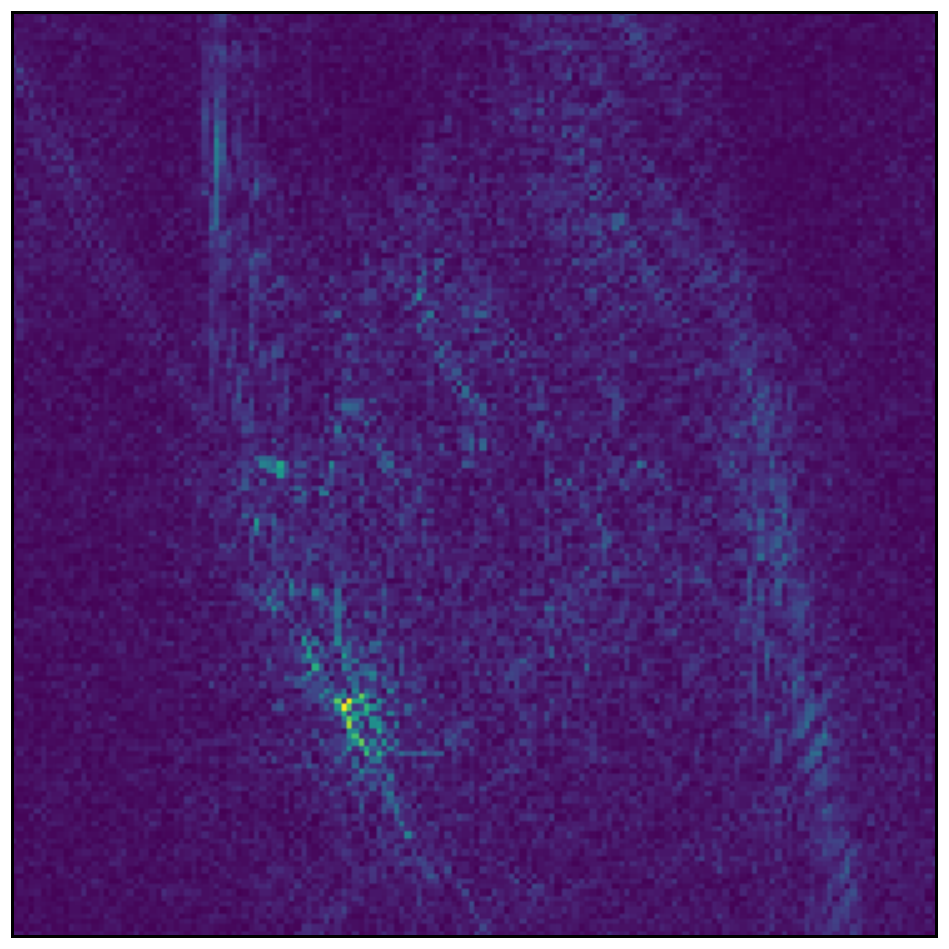}
\end{overpic}
}
\resizebox{\linewidth}{!}{
\includegraphics[height=3.4cm]{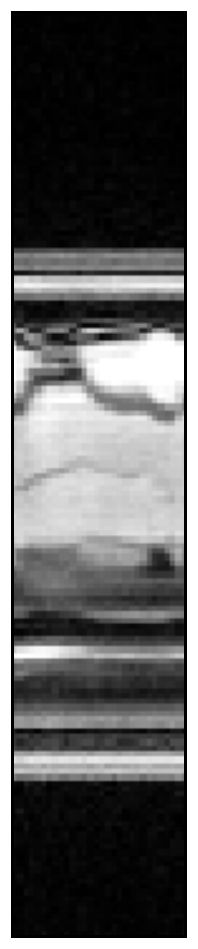}\hspace{-0.2cm}
\begin{overpic}[height=3.4cm,tics=10]{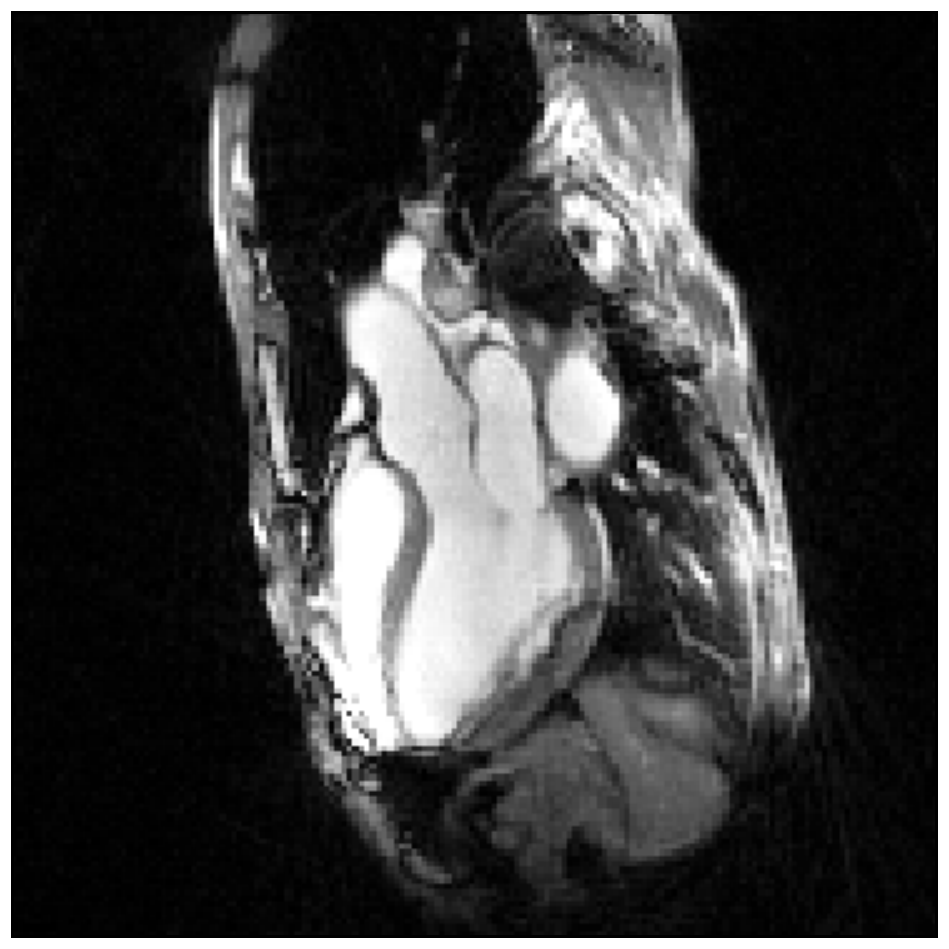}
 \put (84,84) {\Large\textcolor{white}{E}}
\end{overpic}
\includegraphics[height=3.4cm]{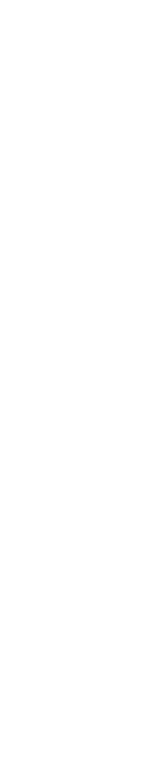}\hspace{-0.2cm}
\begin{overpic}[height=3.4cm,tics=10]{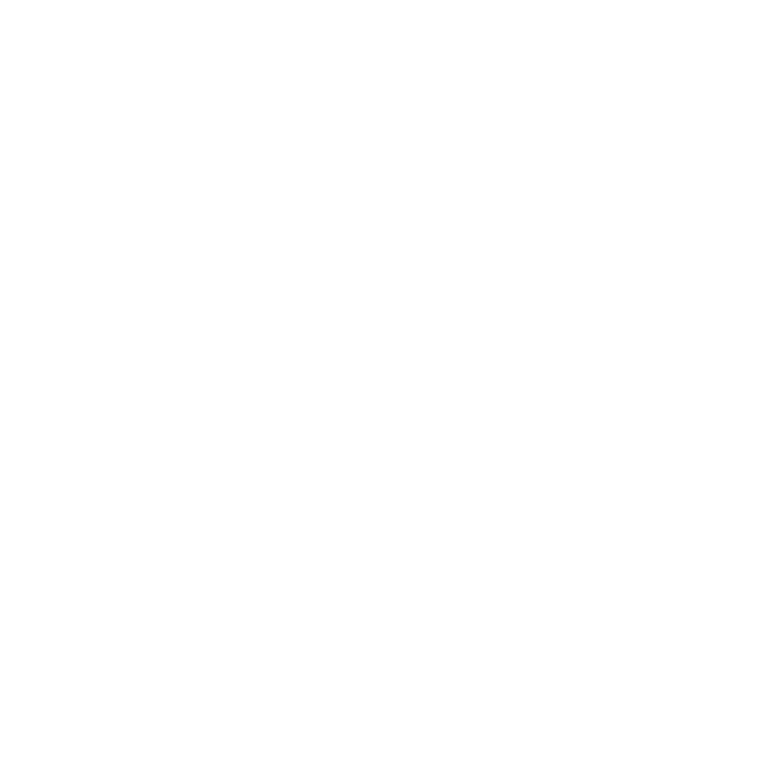}
\end{overpic}
}
\end{minipage}
\begin{minipage}{0.495\linewidth}
\resizebox{\linewidth}{!}{
\includegraphics[height=3.4cm]{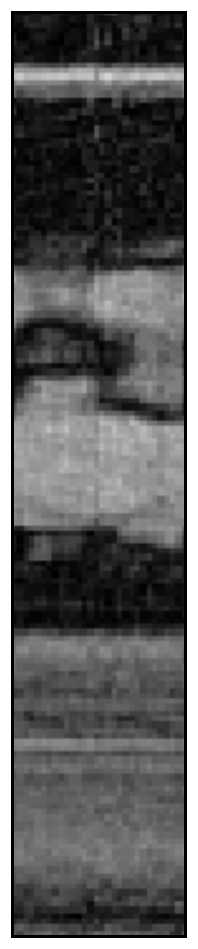}\hspace{-0.2cm}
\begin{overpic}[height=3.4cm,tics=10]{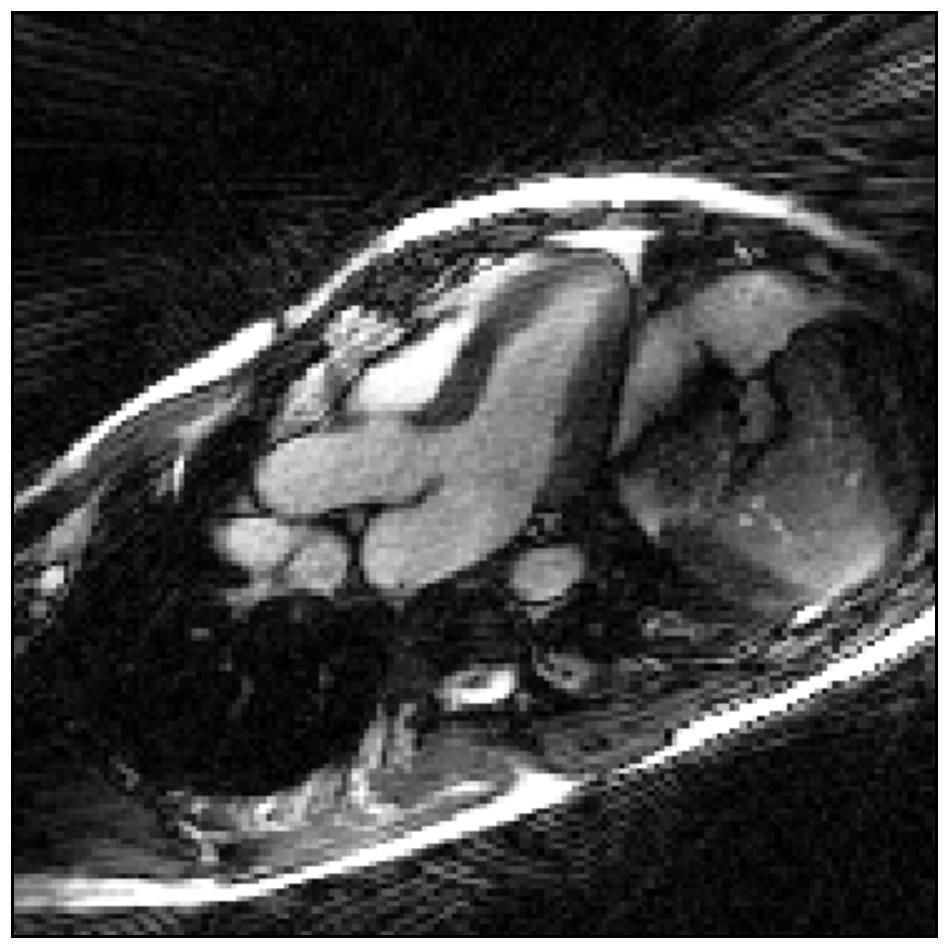}
 \put (84,84) {\Large\textcolor{white}{A}}
\end{overpic}\hspace{-0.1cm}
\includegraphics[height=3.4cm]{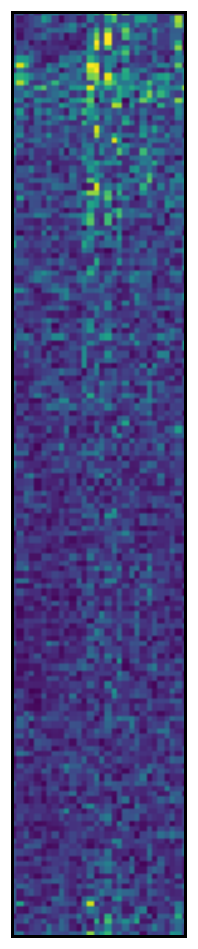}\hspace{-0.2cm}
\begin{overpic}[height=3.4cm,tics=10]{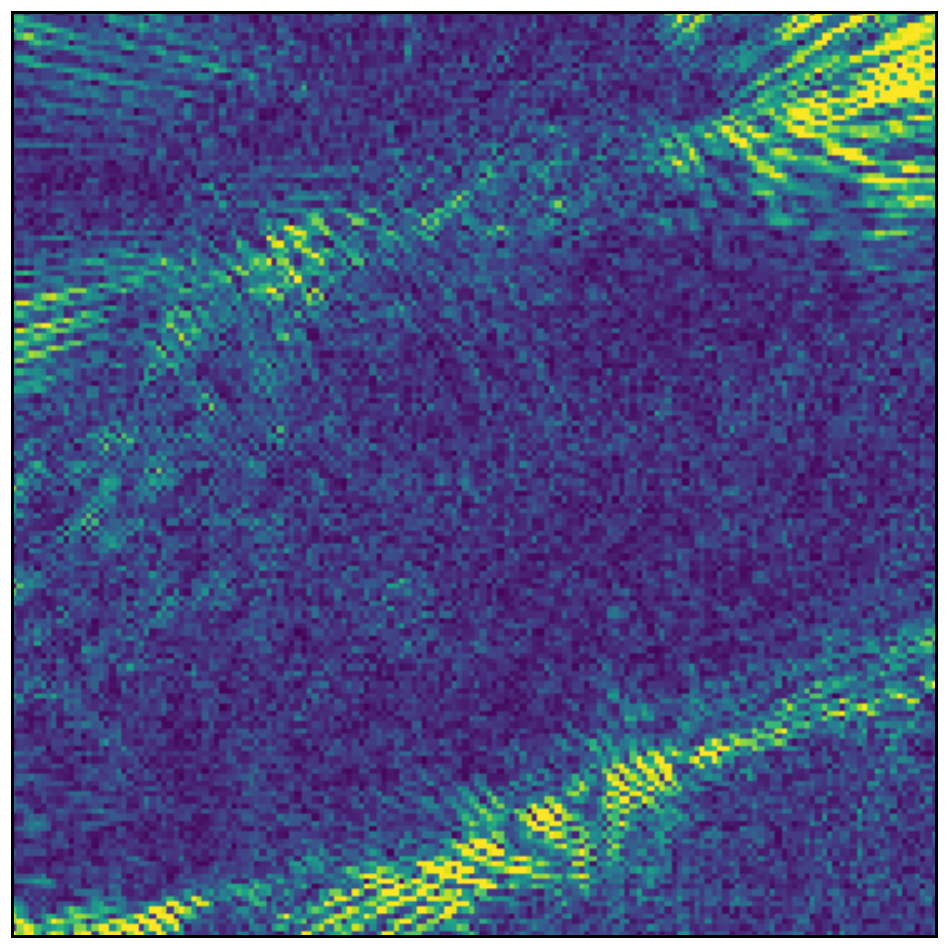}
\end{overpic}
}
\resizebox{\linewidth}{!}{
\includegraphics[height=3.4cm]{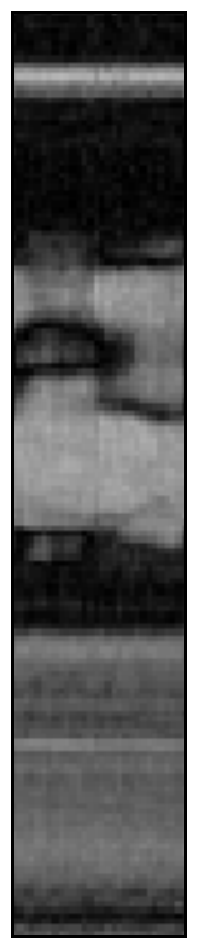}\hspace{-0.2cm}
\begin{overpic}[height=3.4cm,tics=10]{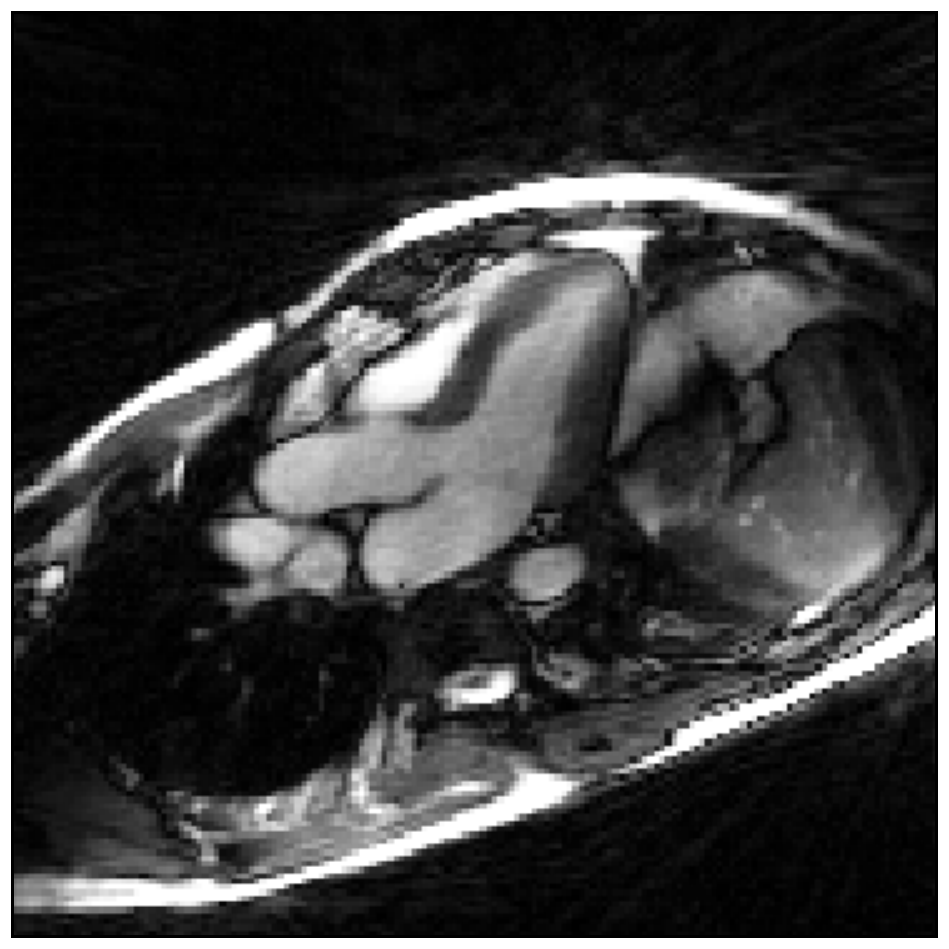}
 \put (84,84) {\Large\textcolor{white}{B}}
\end{overpic}\hspace{-0.1cm}
\includegraphics[height=3.4cm]{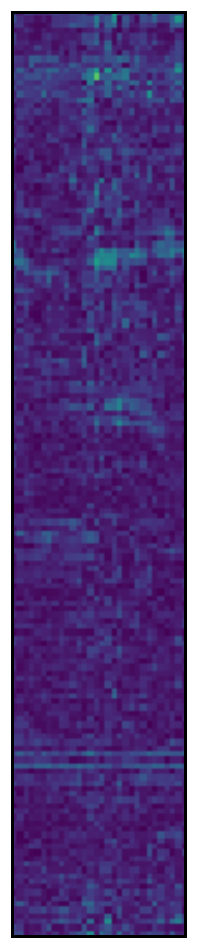}\hspace{-0.2cm}
\begin{overpic}[height=3.4cm,tics=10]{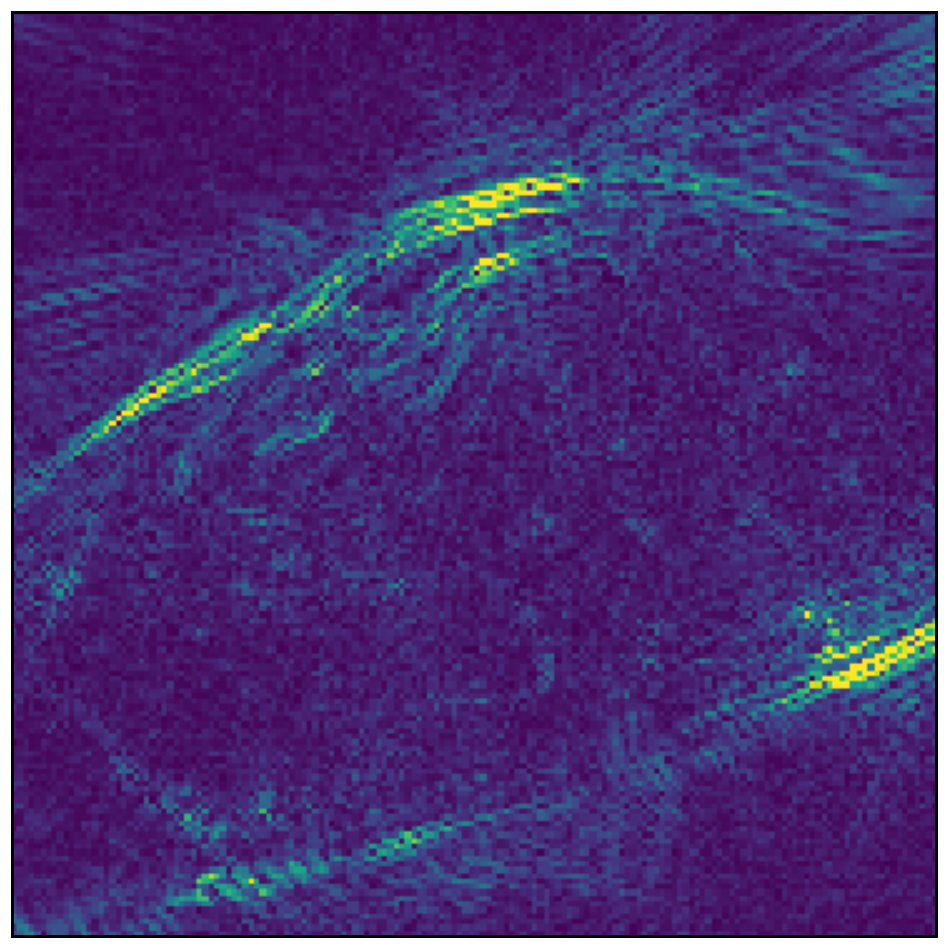}
\end{overpic}
}
\resizebox{\linewidth}{!}{
\includegraphics[height=3.4cm]{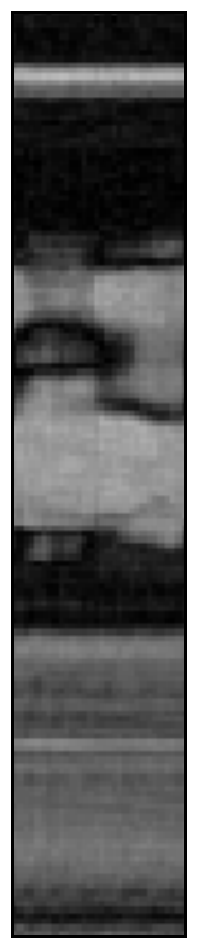}\hspace{-0.2cm}
\begin{overpic}[height=3.4cm,tics=10]{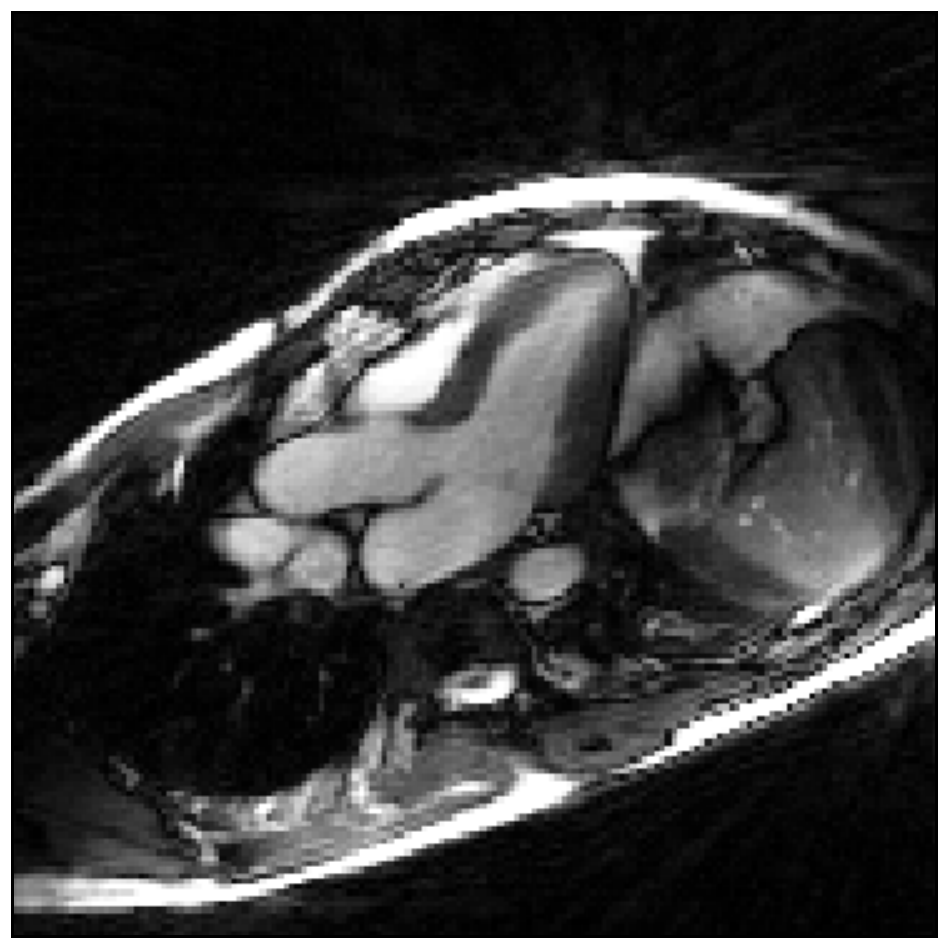}
 \put (84,84) {\Large\textcolor{white}{C}}
\end{overpic}\hspace{-0.1cm}
\includegraphics[height=3.4cm]{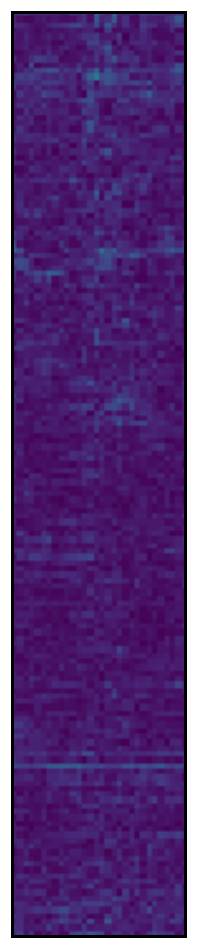}\hspace{-0.2cm}
\begin{overpic}[height=3.4cm,tics=10]{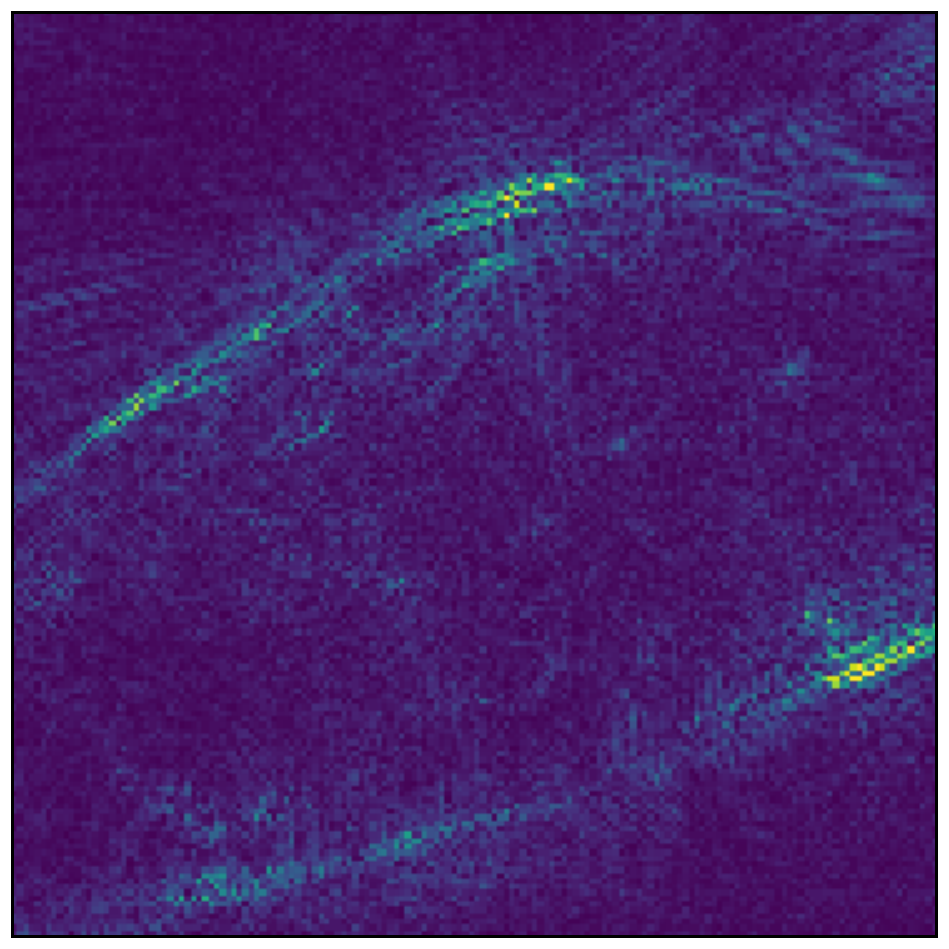}
\end{overpic}
}
\resizebox{\linewidth}{!}{
\includegraphics[height=3.4cm]{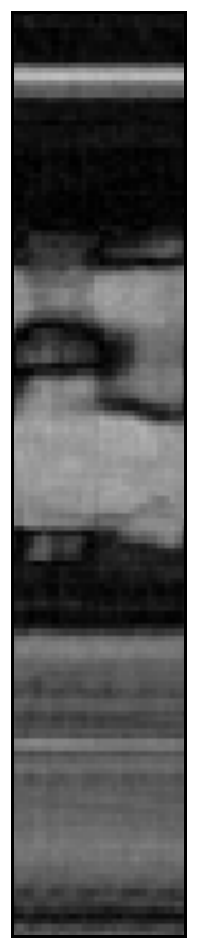}\hspace{-0.2cm}
\begin{overpic}[height=3.4cm,tics=10]{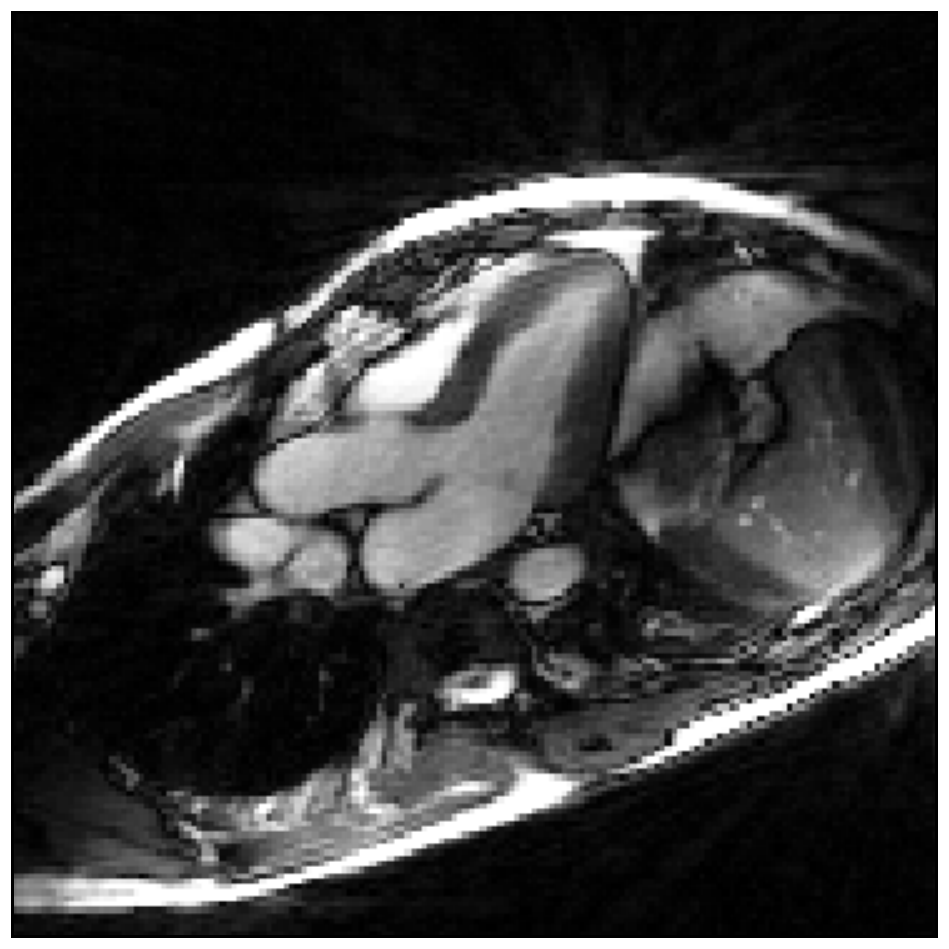}
 \put (84,84) {\Large\textcolor{white}{D}}
\end{overpic}\hspace{-0.1cm}
\includegraphics[height=3.4cm]{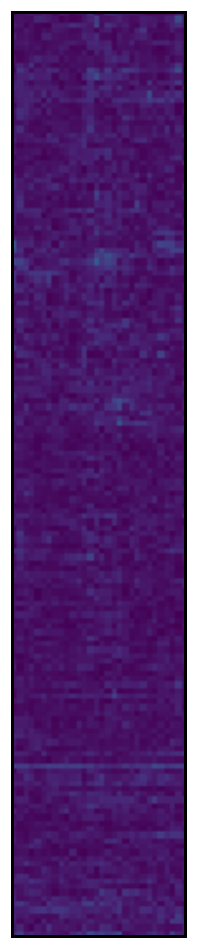}\hspace{-0.2cm}
\begin{overpic}[height=3.4cm,tics=10]{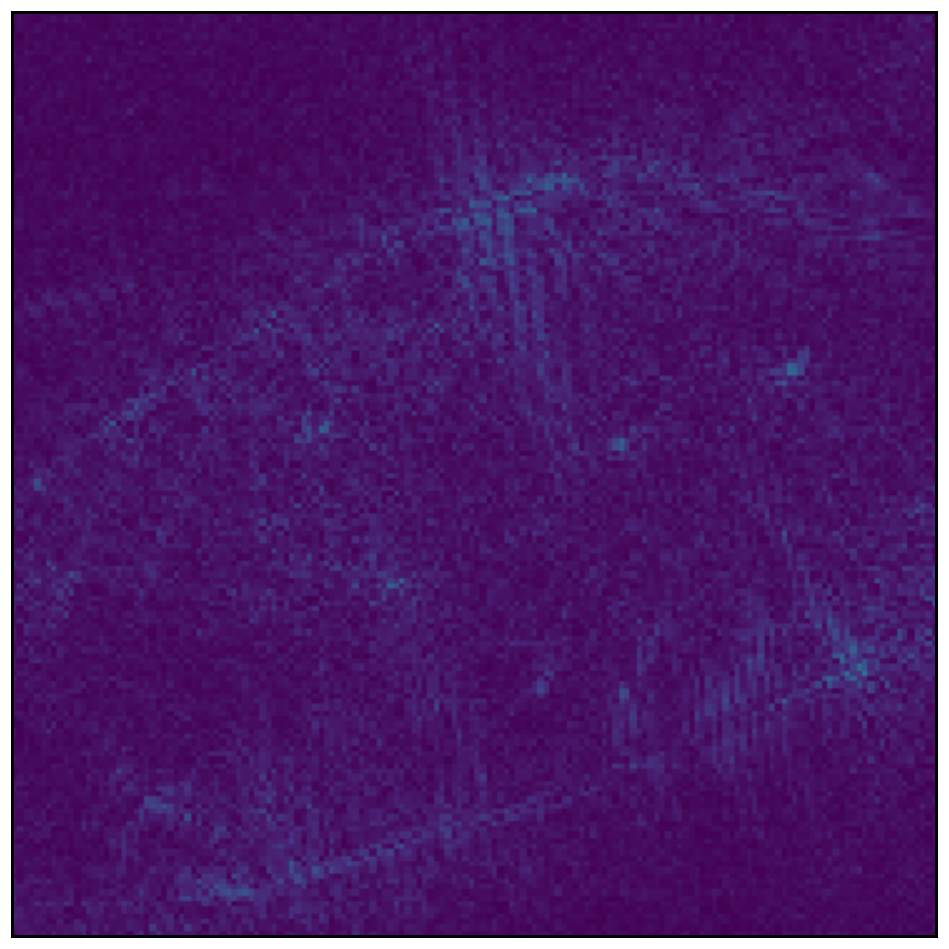}
\end{overpic}
}
\resizebox{\linewidth}{!}{
\includegraphics[height=3.4cm]{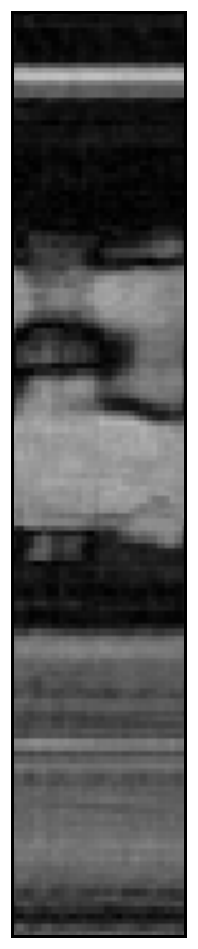}\hspace{-0.2cm}
\begin{overpic}[height=3.4cm,tics=10]{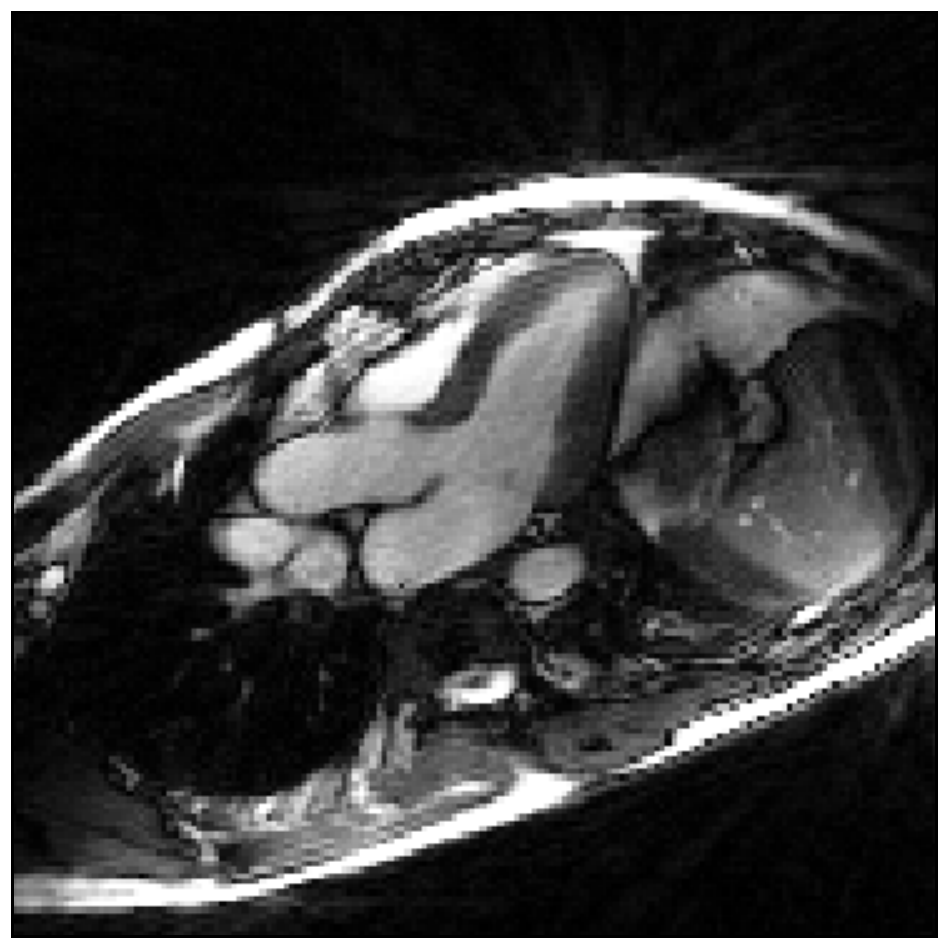}
 \put (84,84) {\Large\textcolor{white}{E}}
\end{overpic}
\includegraphics[height=3.4cm]{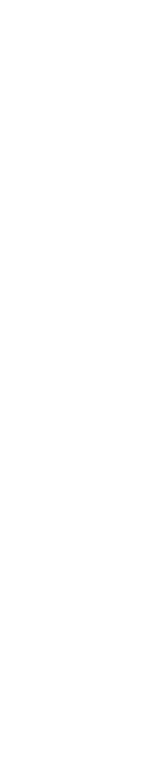}\hspace{-0.2cm}
\begin{overpic}[height=3.4cm,tics=10]{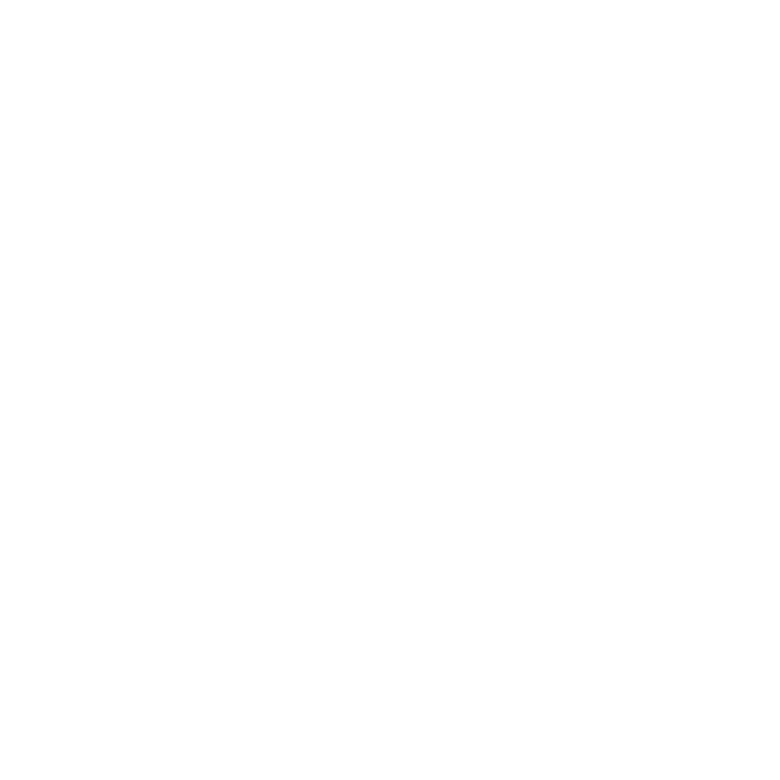}
\end{overpic}
}
\end{minipage}
\caption{Results for a patient (left panel) and a healthy volunteer (right panel). A: Initial NUFFT-reconstruction $\Xu$ using $N_{\theta} =1130$ radial spokes, B: solution of the TV-minimization approach (TV), C: dictionary learning-based regularization solution (DIC), D: CNN-regularized solution $\XX_{\mathrm{REC}}$, E: ground truth images obtained by $kt$-SENSE using $N_{\theta}=3400$ radial spokes.  All images are displayed in the same scale. For better visibility, the point-wise error images are magnified by a factor of $\times 5$. The point-wise error is the lowest for the reconstruction $\XX_{\mathrm{REC}}$.}\label{MRI_comparisons_results_figs}
\end{figure*}

Figure \ref{MRI_intermediate_results_figs} shows an example of the results obtained with our proposed method. Figure \ref{MRI_intermediate_results_figs}A shows the initial NUFFT-reconstruction $\Xu$ obtained from the undersampled $k$-space data $\Yu$. The CNN-prior $\Xcnn$ obtained by the XT,YT network can be seen in Figure \ref{MRI_intermediate_results_figs}B and shows a strong reduction of undersampling artefacts but also blurring of small structures as indicated  the yellow arrows. The CNN-prior $\Xcnn$ is then used as a prior in functional (\ref{DC_eq_MRI}) which is subsequently minimized in order to obtain the solution $\XX_{\mathrm{REC}}$ which can be seen in Figure \ref{MRI_intermediate_results_figs}C. Figure \ref{MRI_intermediate_results_figs}D shows the $kt$-SENSE reconstruction from the complete sampling pattern using $N_{\theta}=3400$ radial spokes for the acquisition. From the point-wise error images, we clearly see that the NRMSE is further reduced after performing the further iterations to minimize the CNN-prior-regularized functional. Further, fine details are recovered as can be seen from the yellow arrows in Figure \ref{MRI_intermediate_results_figs}C.
\begin{table}[h]
\renewcommand{\arraystretch}{1.3}
\centering
\caption{Quantitative measures for the 2D radial cine MRI example.  The measures are obtained as averages over the four different folds.}\label{MRI_results_table}
\begin{tabular}{@{}l|ccc|cc}
\toprule
 & \textbf{NUFFT} & $\Xcnn$ & $\XX_{\mathrm{REC}}$ & \textbf{TV} & \textbf{DIC}\\
\midrule
\textbf{PSNR} & 36.8023 & 42.5647 & 48.7752  & 41.6968 & 45.4743\\
\textbf{NRMSE} & 0.1228 & 0.0612 & 0.0302 & 0.0693 & 0.0442 \\ 
\textbf{SSIM} & 0.6649 & 0.7876 & 0.952 & 0.8635 & 0.9175 \\ 
\textbf{HPSI} & 0.9679 & 0.9910 & 0.9985 & 0.9878 & 0.9959\\ 
\bottomrule
\end{tabular}
\end{table}
Figure \ref{MRI_comparisons_results_figs} shows a comparison of all different reported methods. As can be seen from the point-wise error in Figure \ref{MRI_comparisons_results_figs}B, the TV-minimization \cite{block2007} method was  able to eliminate some artefacts but less accurately compared to both learning-based methods, see Figure \ref{MRI_comparisons_results_figs}C and Figure \ref{MRI_comparisons_results_figs}D. Table \ref{MRI_results_table} lists the obtained quantitative measures for all methods averaged over the 4 different folds. From Table \ref{MRI_results_table}, we see that the DIC method yielded better results than TV with respect to all reported measures. Our proposed solution $\XX_{\mathrm{REC}}$ further surpassed the dictionary learning-based method, by additionally increasing the PSNR and SSIM by approximately 3dB and 0.04, respectively. The difference with respect to HPSI, on the other hand, is relatively small. Our method also reduced the NRMSE by about 0.014 compared to the DIC method. In addition, from Table \ref{MRI_results_table}, we see that for this example, even though processing the initial NUFFT-reconstruction with a CNN improved image quality with respect to all reported measures, further iterations to minimize the CNN-prior regularized functional increased data-consistency and additionally improved the PSNR, SSIM, HPSI and NRMSE.  In fact, the statistics of the CNN-prior show that only post-processing the initial NUFFT-reconstruction leads to results which are inferior to the DIC method with respect to all reported measures.

\subsection{Results for 3D Low-Dose CT}\label{subsec_CT_results}
\begin{figure}[H]
\centering
\begin{minipage}{\linewidth} 
\begin{overpic}[width=0.495\linewidth,tics=10]{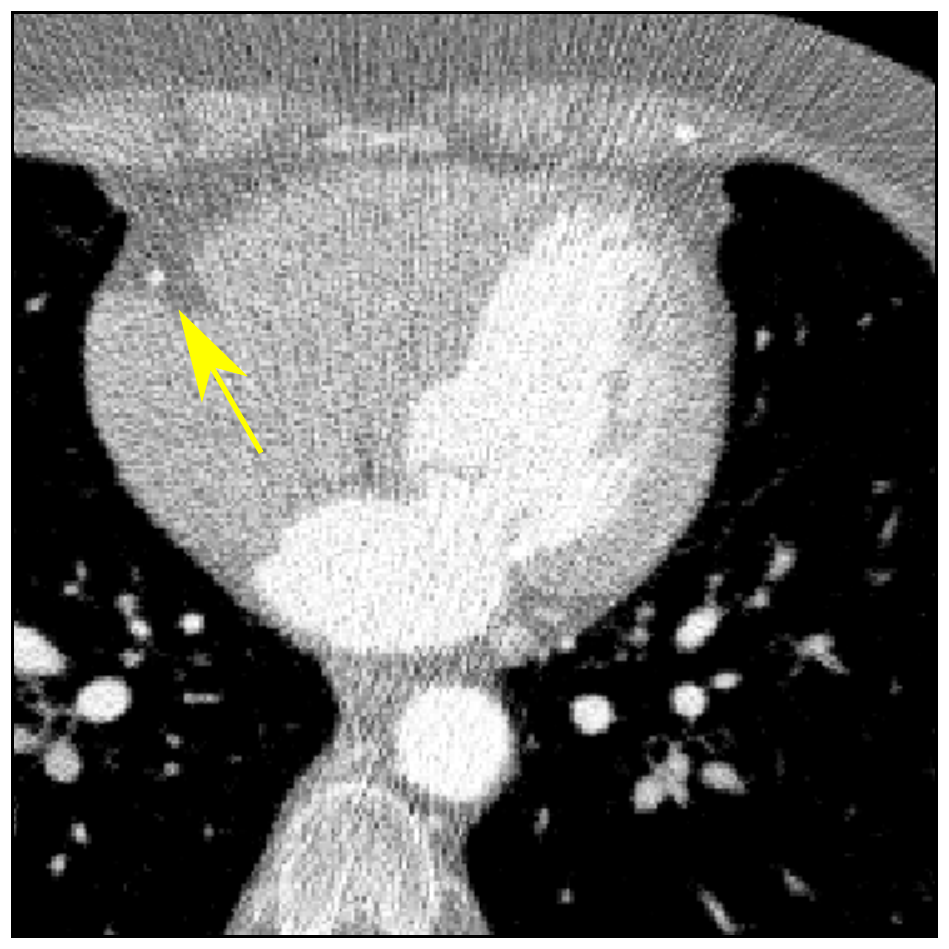}\put (84,84){\LARGE\textcolor{white}{A}}\end{overpic}\hspace{-0.1cm}
\includegraphics[width=0.495\linewidth]{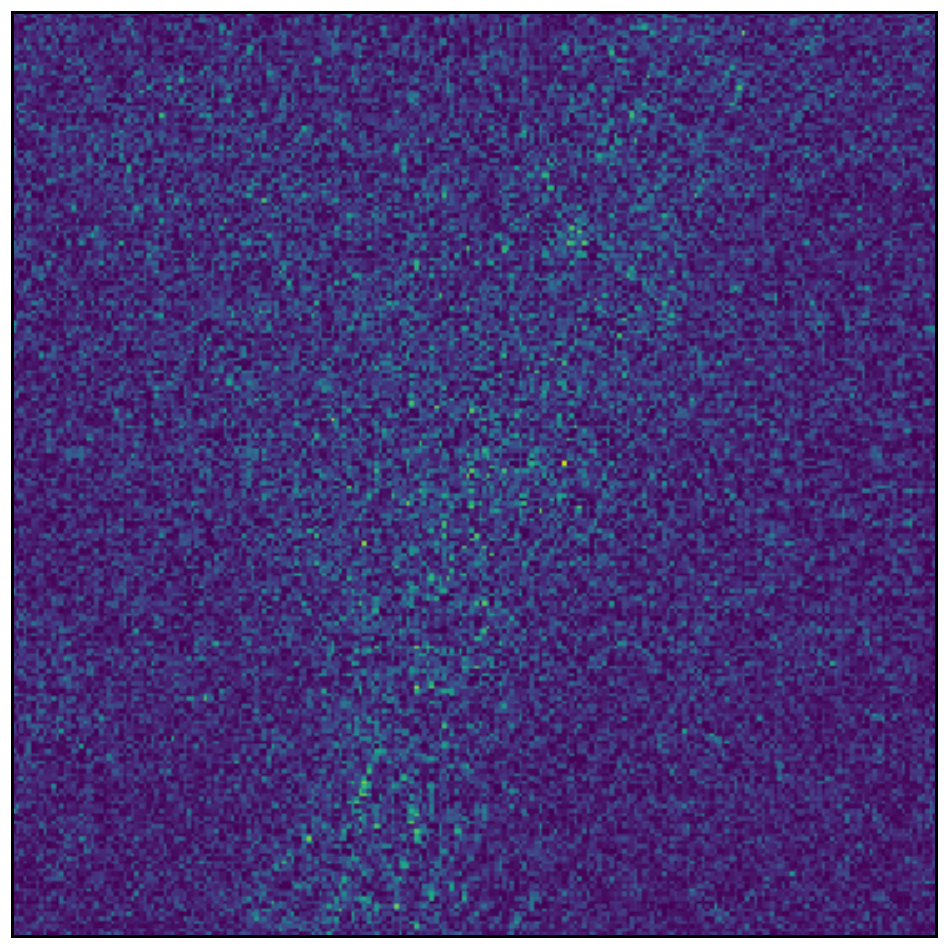}\\
\begin{overpic}[width=0.495\linewidth,tics=10]{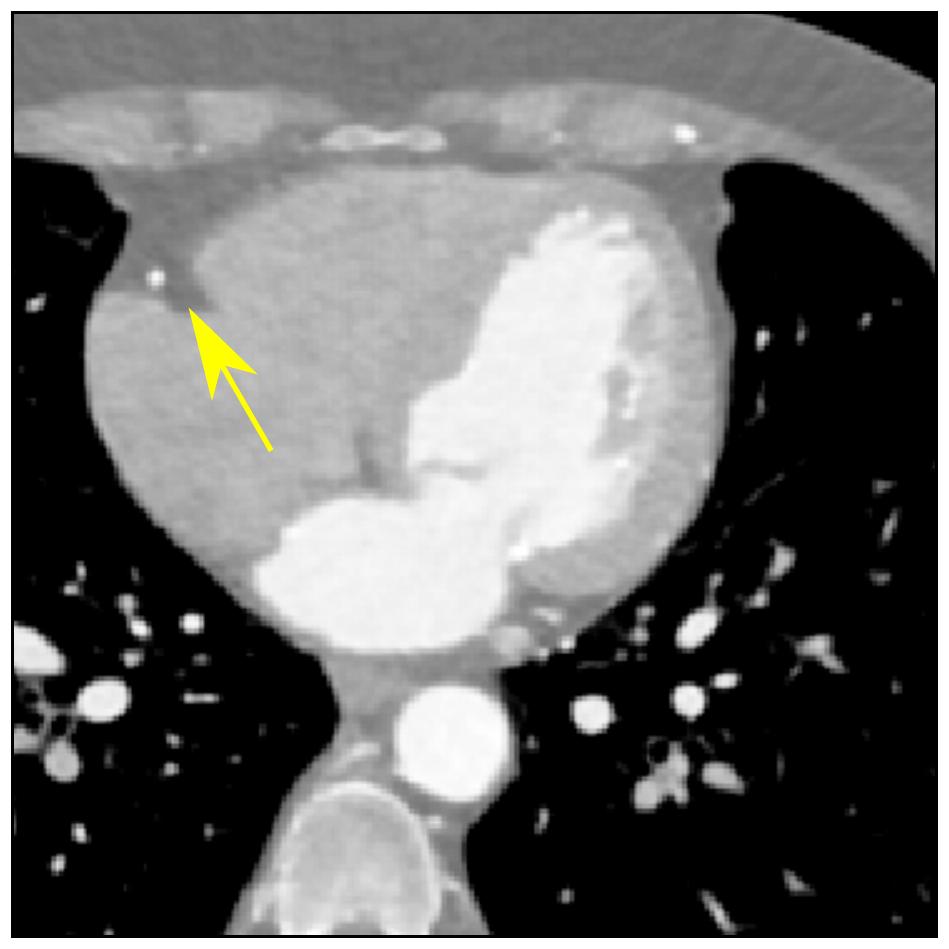}\put (84,84){\LARGE\textcolor{white}{B}}\end{overpic}\hspace{-0.1cm}
\includegraphics[width=0.495\linewidth]{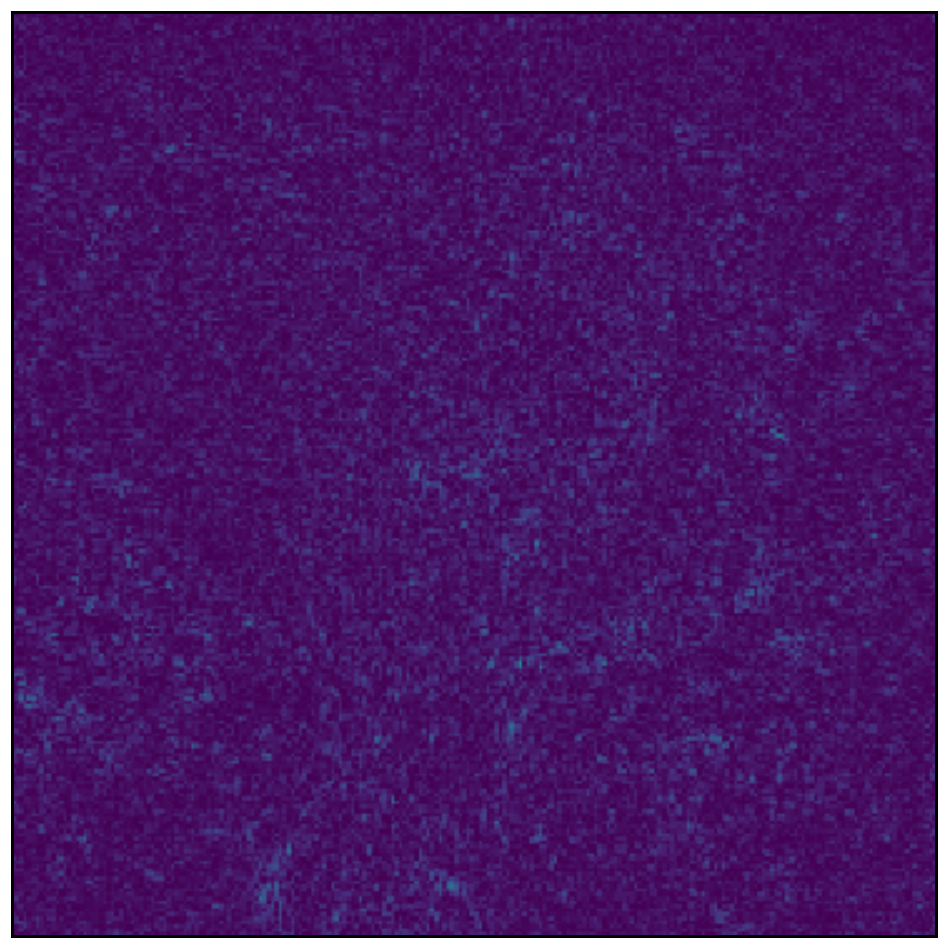}\\
\begin{overpic}[width=0.495\linewidth,tics=10]{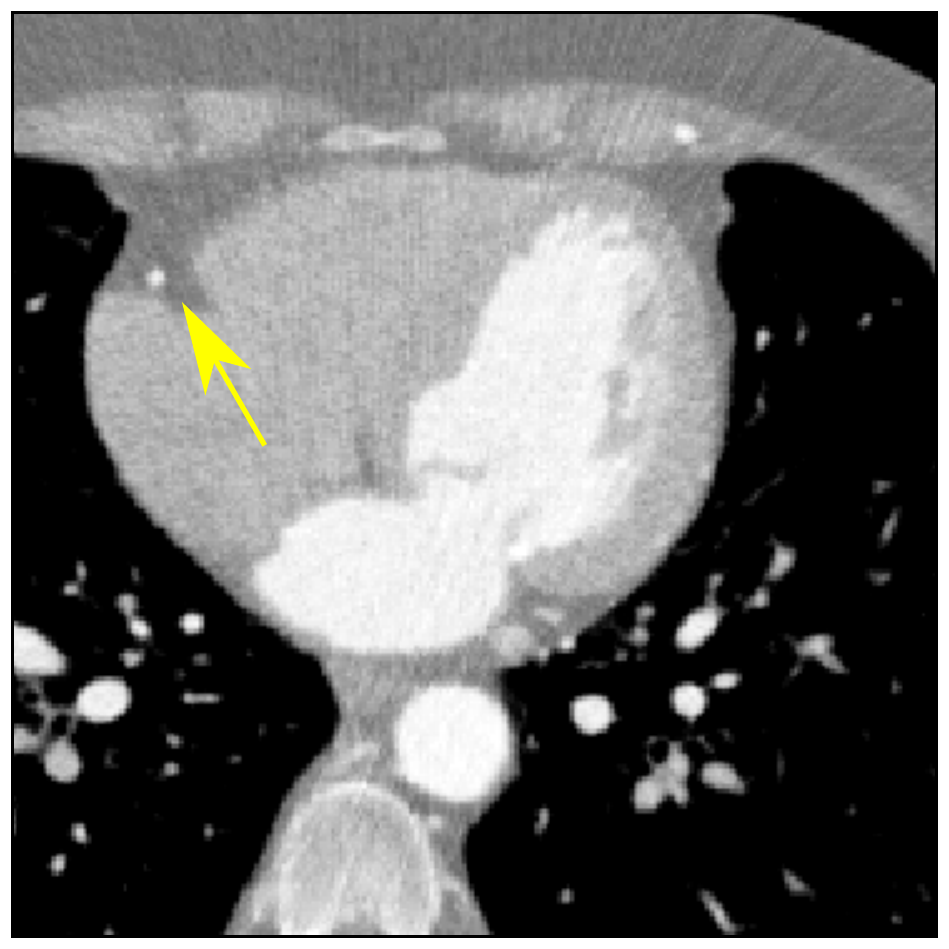}\put (84,84){\LARGE\textcolor{white}{C}}\end{overpic}\hspace{-0.1cm}
\includegraphics[width=0.495\linewidth]{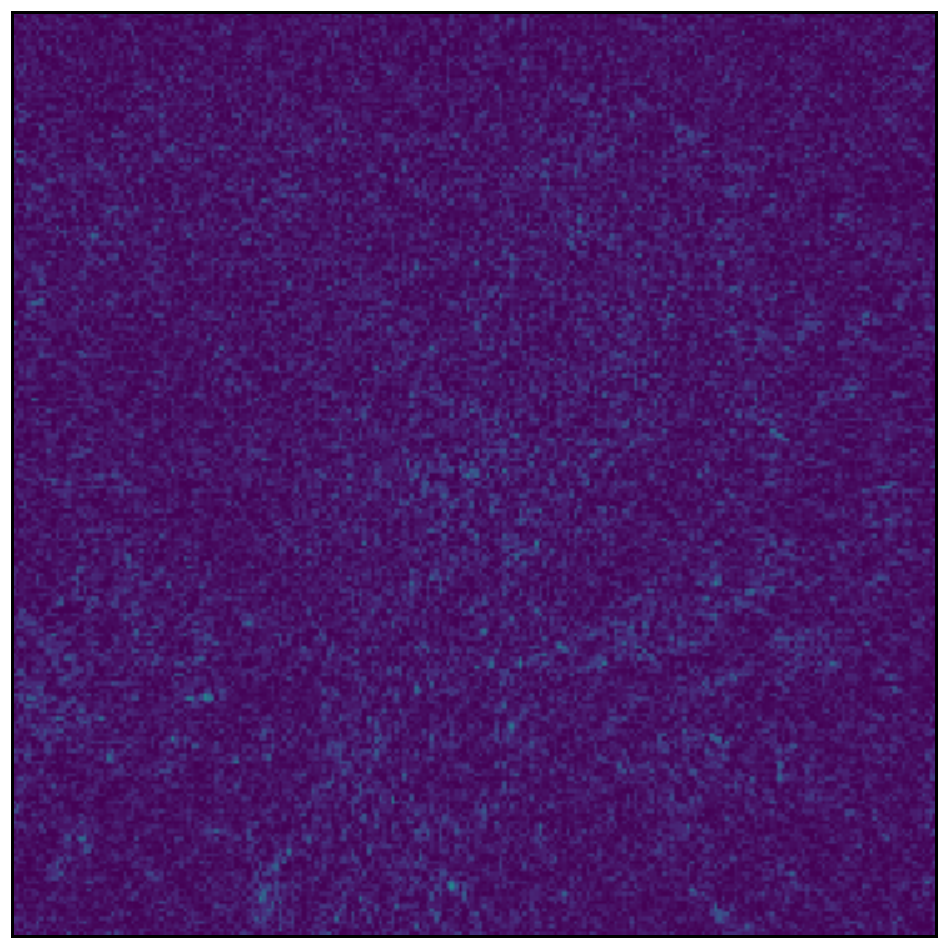}\\
\begin{overpic}[width=0.495\linewidth,tics=10]{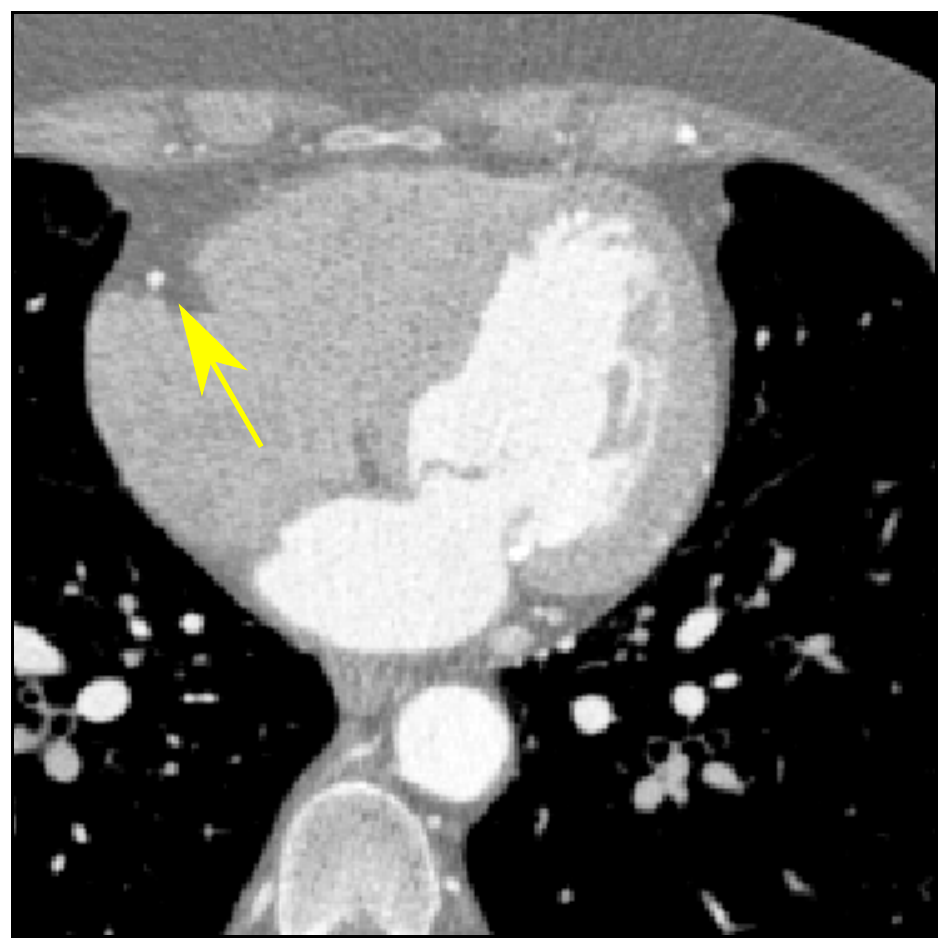}\put (84,84){\LARGE\textcolor{white}{D}}\end{overpic}\hspace{-0.1cm}
\includegraphics[width=0.495\linewidth]{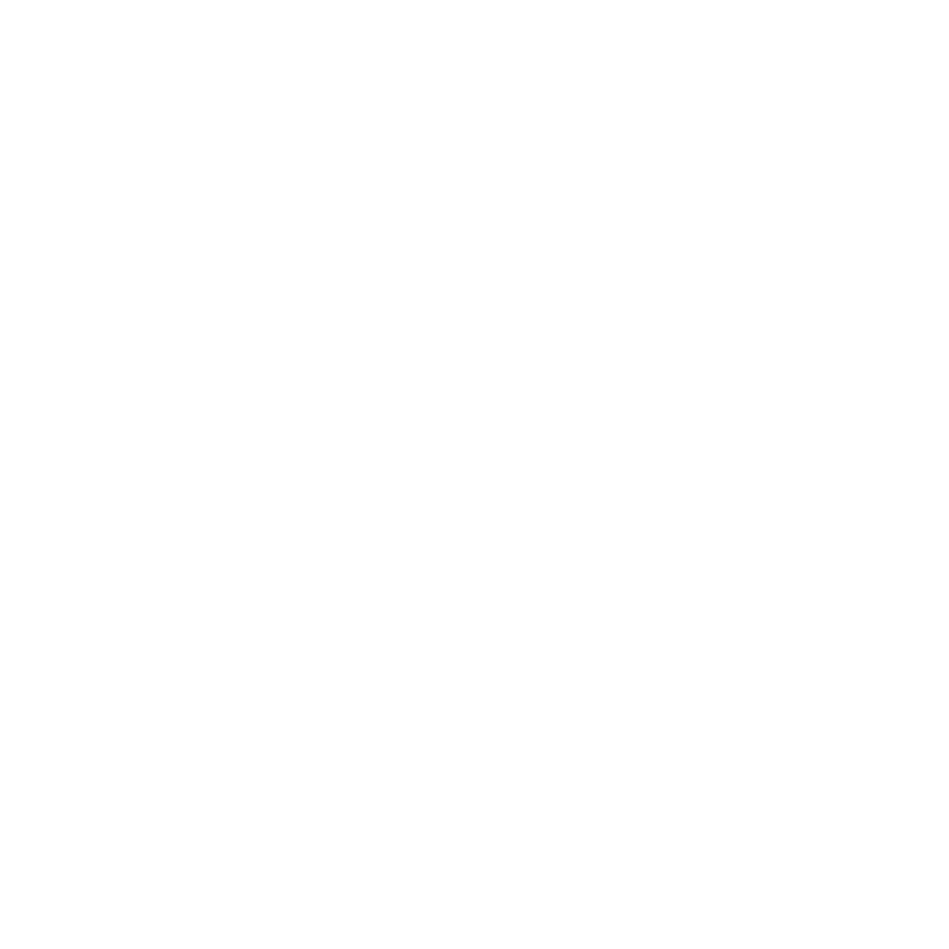}\\
\end{minipage}
\caption{Axial view of image reconstructions of low-dose 3D CT data of a 55 years old female patient.  A: Low-dose FBP-reconstruction $\Xn$, B: estimated output $\Xcnn$ using a 3D U-net, C: solution of the  CNNs-based regularized functional $\XX_{\mathrm{REC}}$, D: ground truth image.  The yellow arrow points at the right coronary artery, which is visible in the prior $\Xcnn$ as well as in the final reconstruction $\XX_{\mathrm{REC}}$. All images are windowed and displayed on the scale with $C=0$\,HU, $W=850$\,HU.}\label{CT_intermediate_results_figs}
\end{figure}

\begin{figure}
\centering
\begin{minipage}{0.95\linewidth} 
\begin{overpic}[width=0.49\linewidth,tics=10]{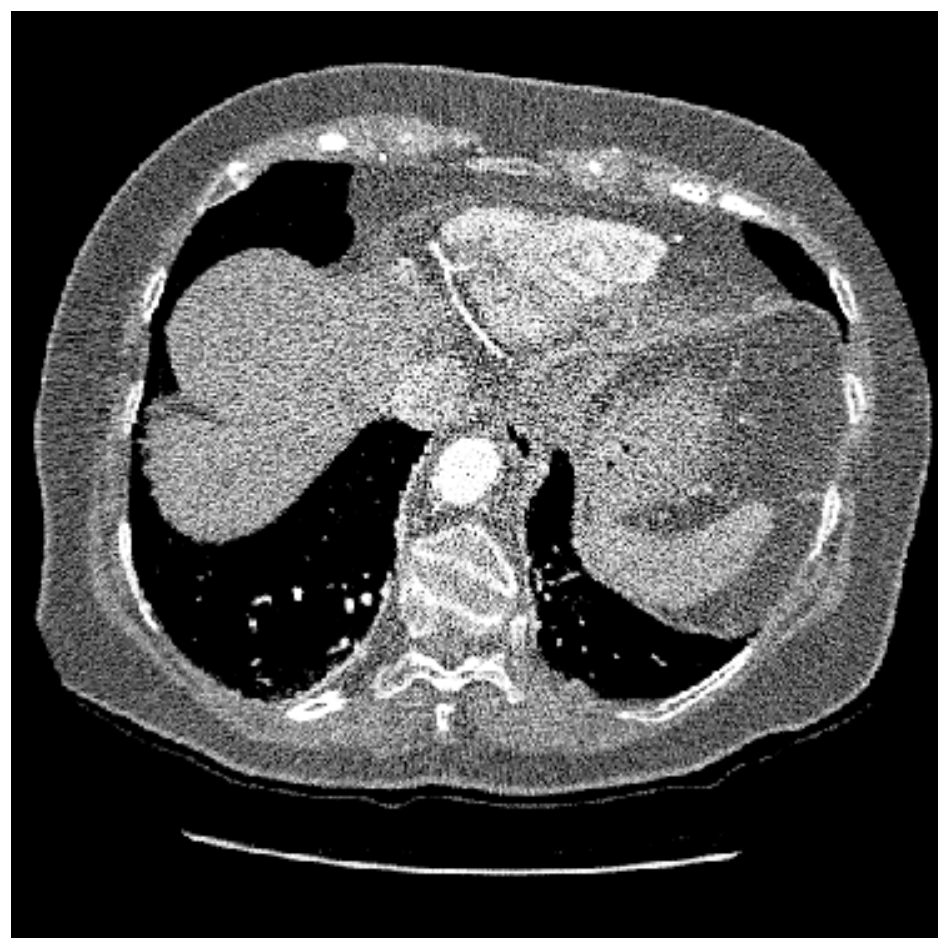}\put (84,84){\LARGE\textcolor{white}{A}}\end{overpic}\hspace{-0.1cm}
\includegraphics[width=0.49\linewidth]{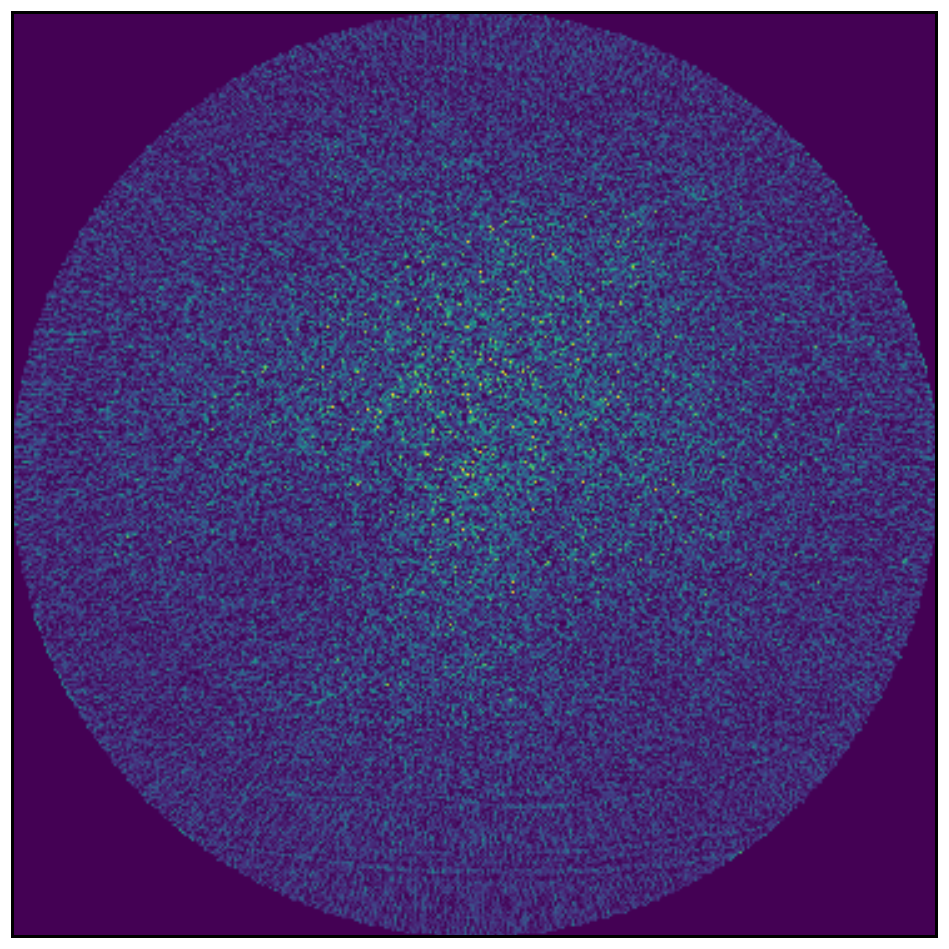}\\
\begin{overpic}[width=0.49\linewidth,tics=10]{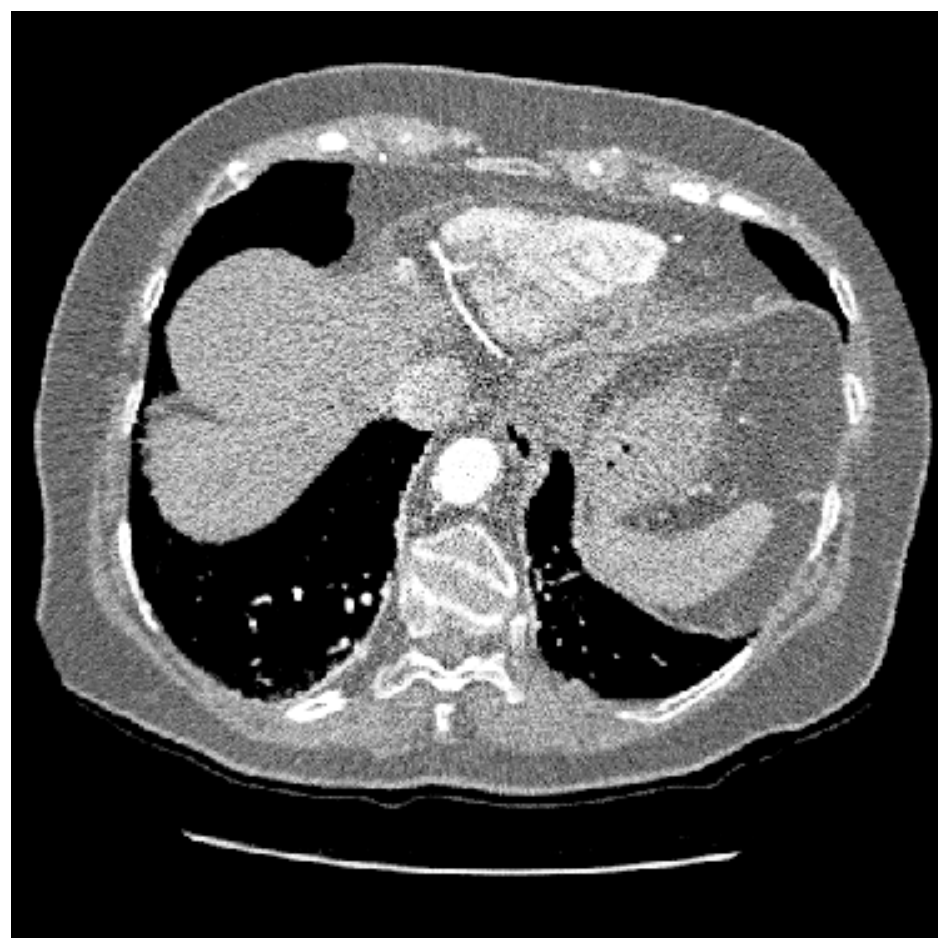}\put (84,84){\LARGE\textcolor{white}{B}}\end{overpic}\hspace{-0.1cm}
\includegraphics[width=0.49\linewidth]{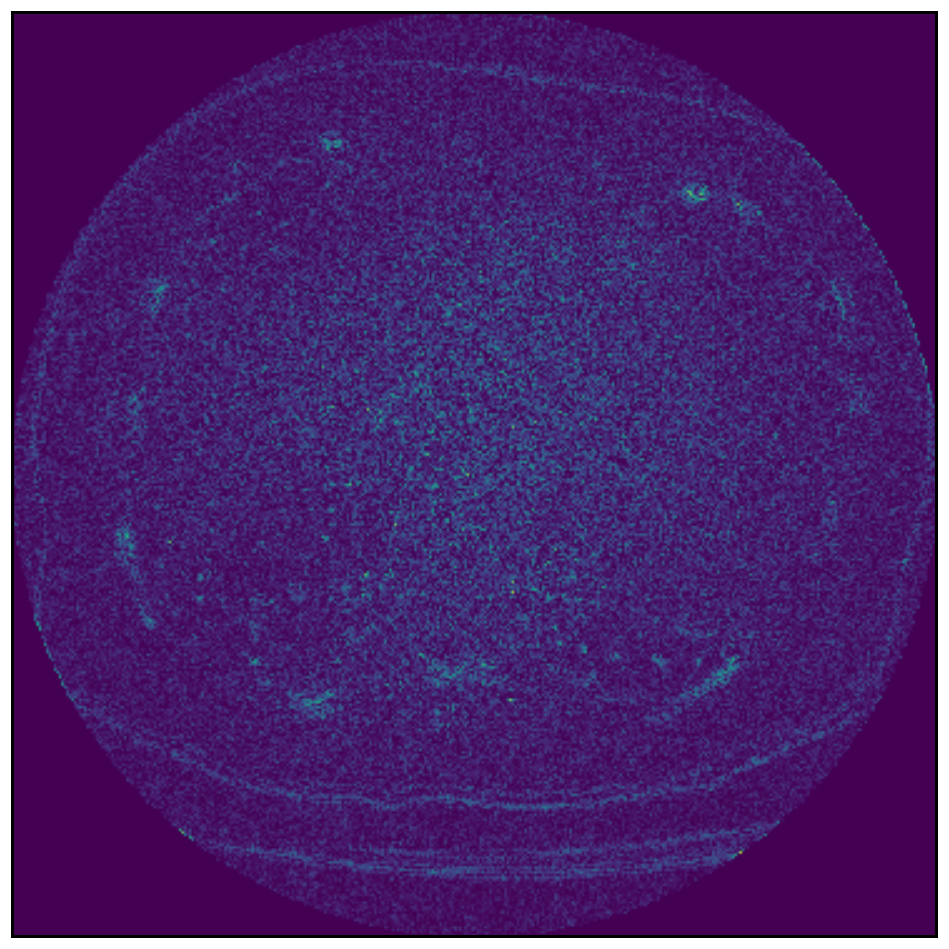}\\
\begin{overpic}[width=0.49\linewidth,tics=10]{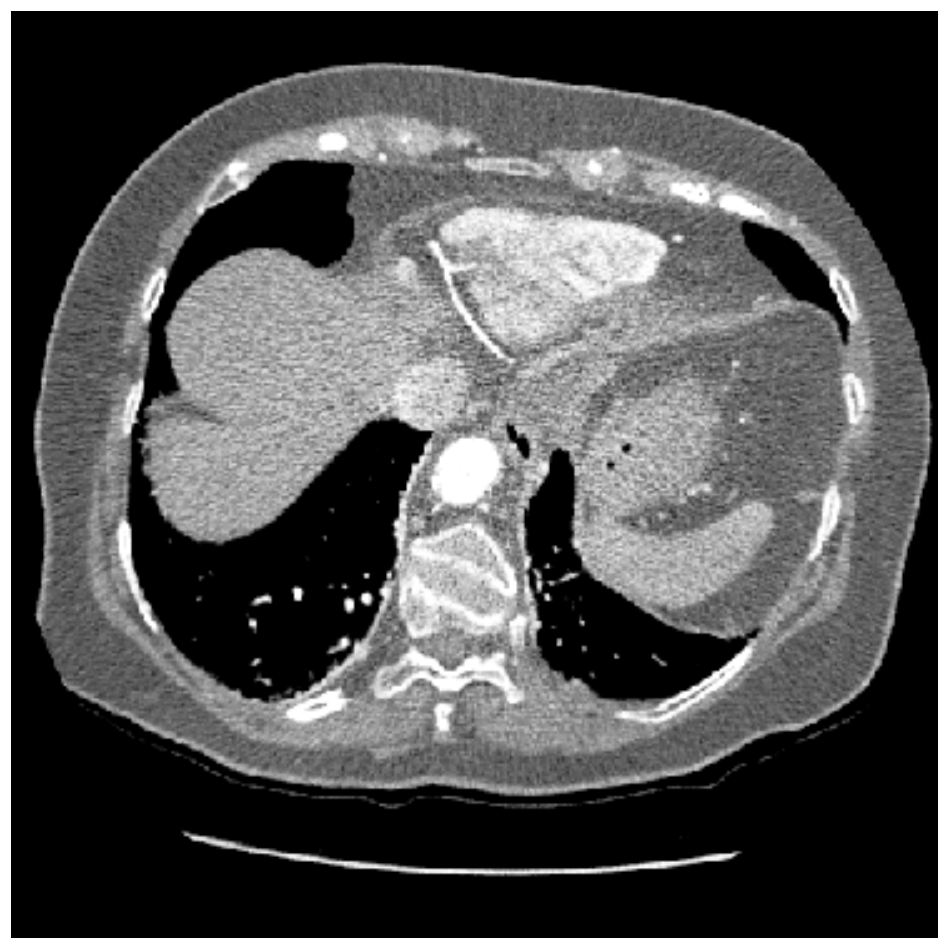}\put (84,84){\LARGE\textcolor{white}{C}}\end{overpic}\hspace{-0.1cm}
\includegraphics[width=0.49\linewidth]{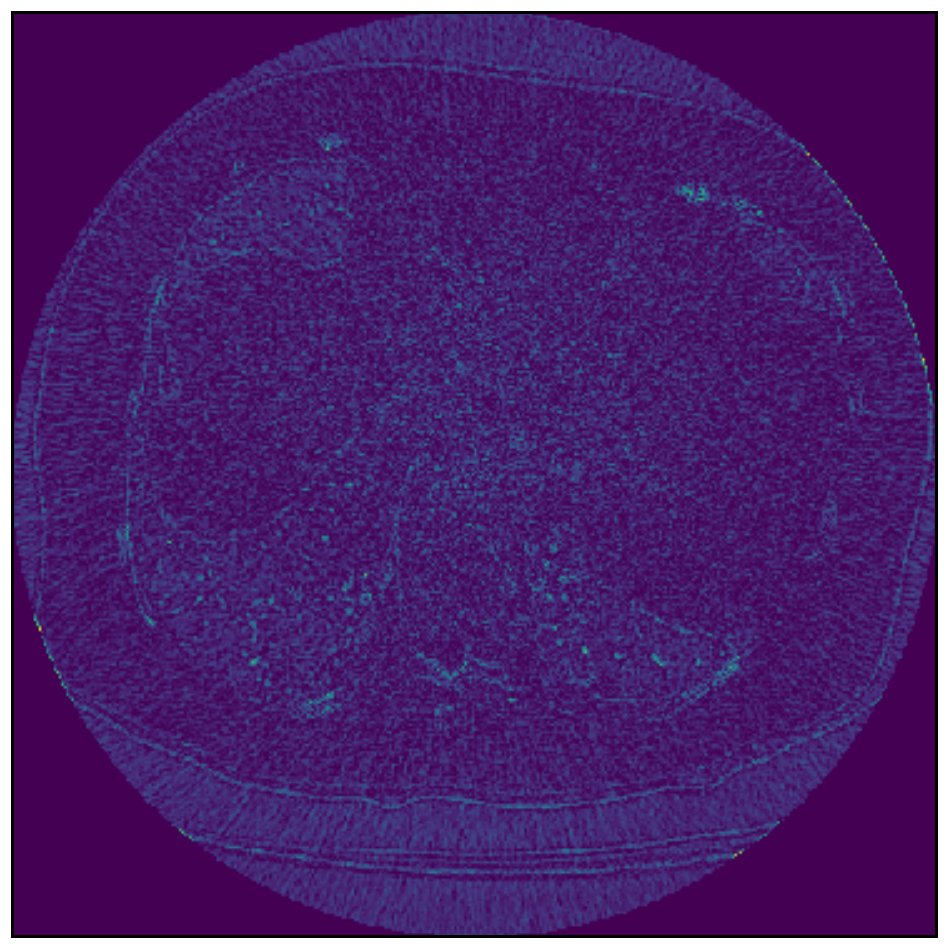}\\
\begin{overpic}[width=0.49\linewidth,tics=10]{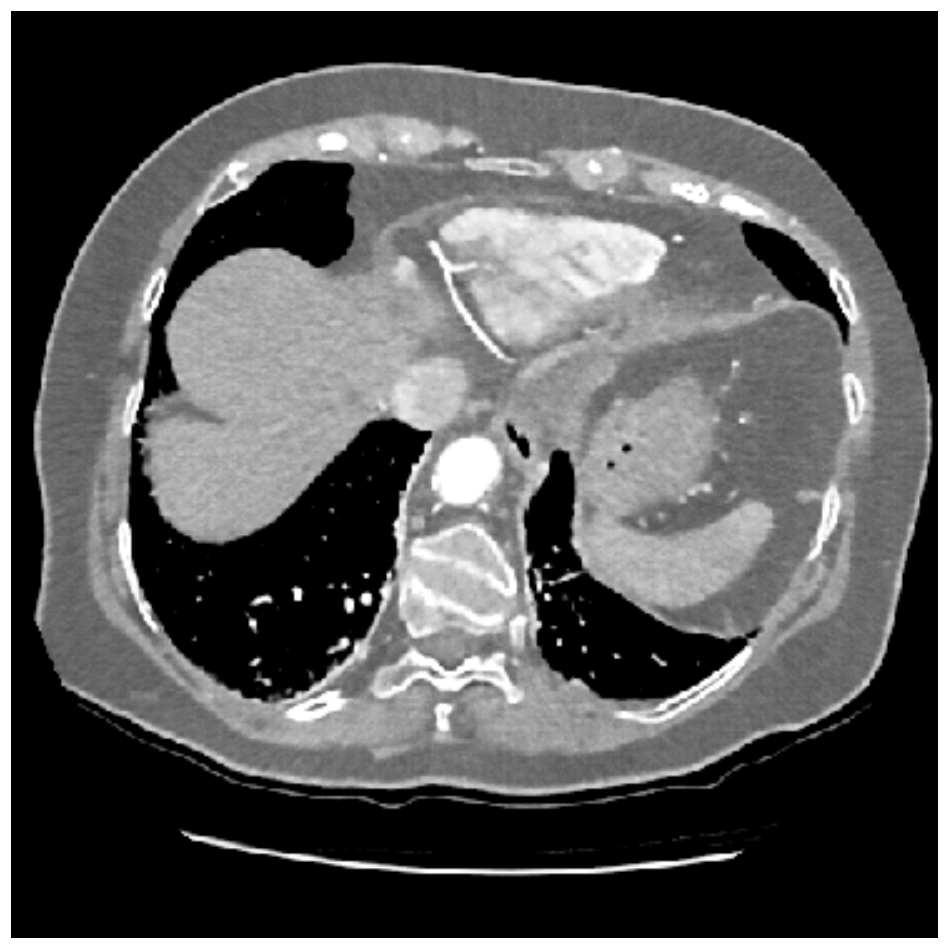}\put (84,84){\LARGE\textcolor{white}{D}}\end{overpic}\hspace{-0.1cm}
\includegraphics[width=0.49\linewidth]{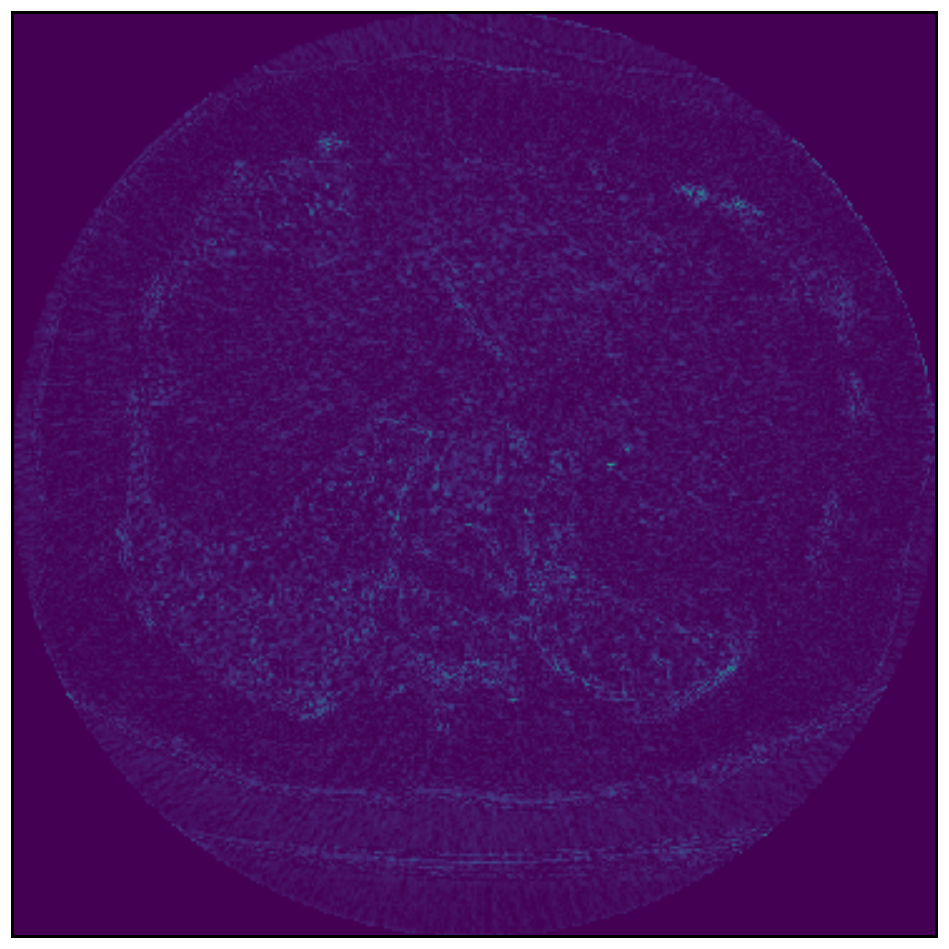}\\
\begin{overpic}[width=0.49\linewidth,tics=10]{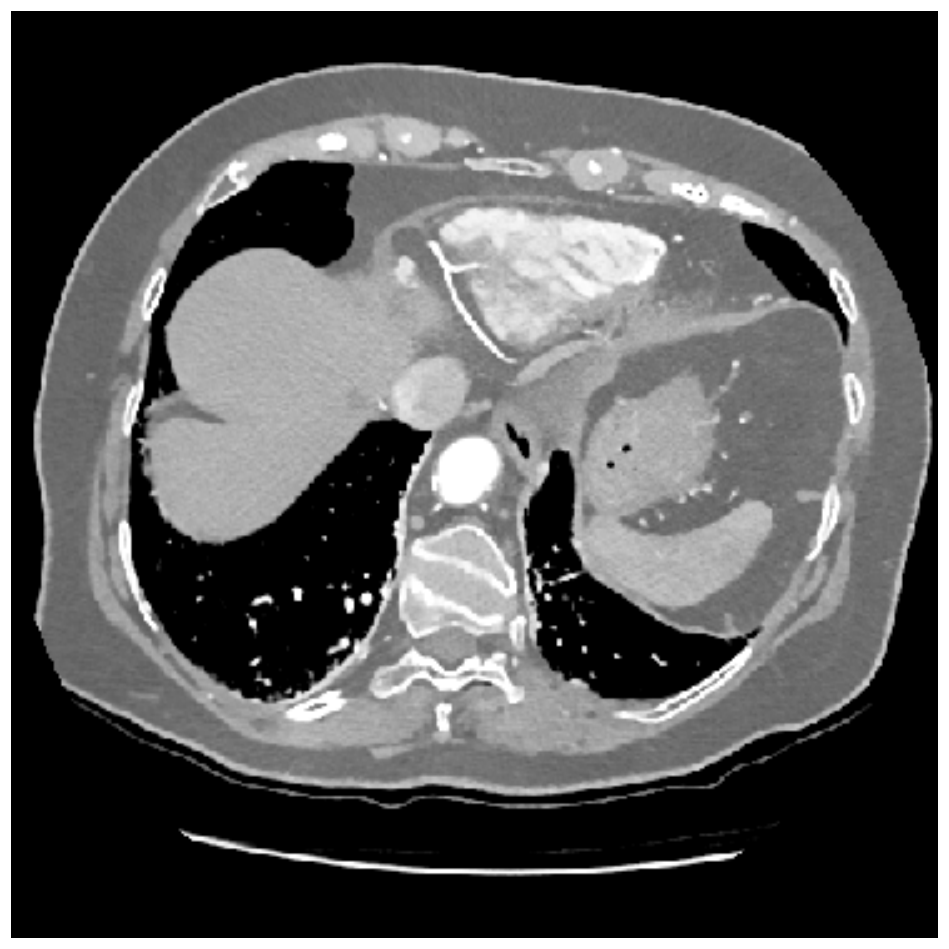}\put (84,84){\LARGE\textcolor{white}{E}}\end{overpic}\hspace{-0.1cm}
\includegraphics[width=0.49\linewidth]{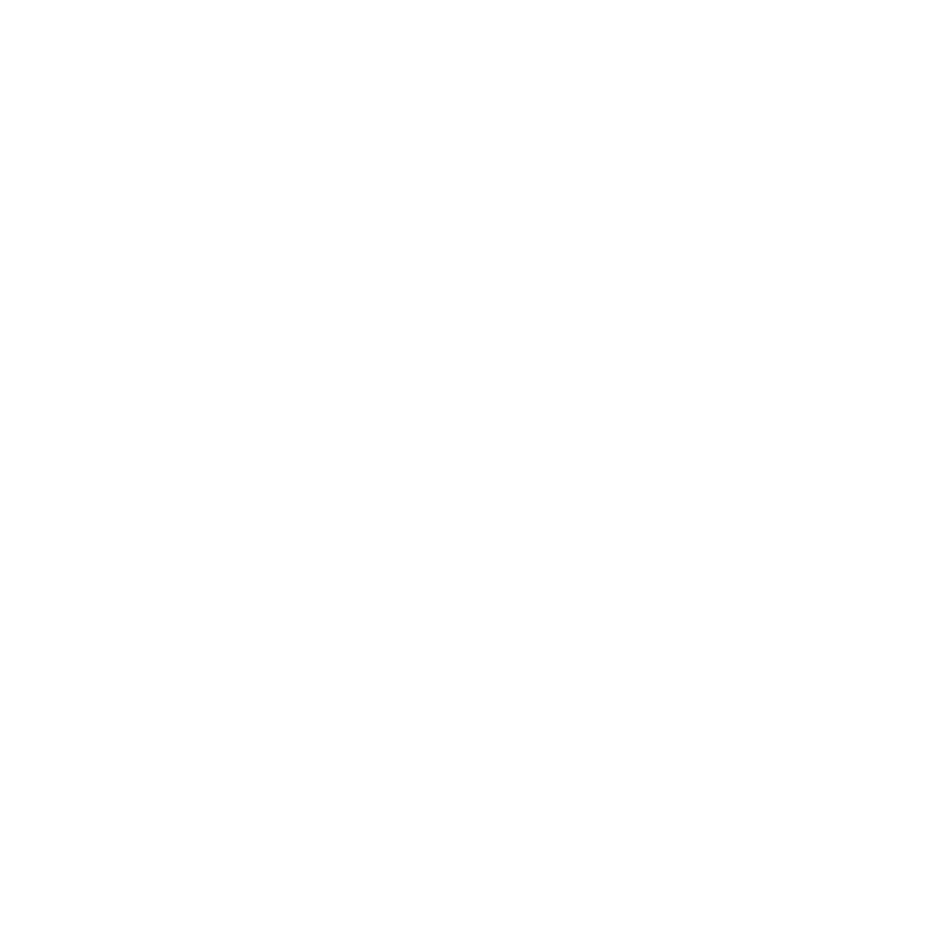}\\
\end{minipage}
\caption{Axial view of image reconstructions of low-dose 3D CT data of a 76 years old female patient. A: Low-dose FBP-reconstruction $\Xn$, B: TV-minimization based reconstruction (TV), C: DIC-regularization based reconstruction (DIC), D: CNN-regularization based reconstruction $\XX_{\mathrm{REC}}$, E: ground truth image. All images are windowed and displayed on the same scale with $C=0$\,HU, $W=800$\,HU.}\label{CT_comparisons_results_figs}
\end{figure}

Figure \ref{CT_intermediate_results_figs} shows all the intermediate results obtained with the proposed method.  Figure \ref{CT_intermediate_results_figs}A shows the initial FBP-reconstruction which is contaminated by noise. The FBP-reconstruction was then processed using the function $f_{\theta}$ described in (\ref{xcnn}) to obtain the prior $\Xcnn$ which can be seen in Figure \ref{CT_intermediate_results_figs}B. From the point-wise error, we see that patch-wise post-processing with the 3D U-net removed a large portion of the noise resulting from the low-dose acquisition. Solving problem (\ref{DC_eq_CT}) increases data-consistency since we make  use of the measured data $\Yn$. Note that in contrast to the previous example of undersampled radial MRI, the minimization of the functional increased data-consistency of the solution but also contaminated the solution with noise, since the measured data is noisy due to the simulated low-dose scan protocol.
Table \ref{CT_results_table} summarizes the obtained quantitative measures for all intermediate reconstructions of our approach as well as for the TV and the DIC method. In the first three columns of Table \ref{CT_results_table} we see the results obtained for all three intermediate reconstructions of our proposed scheme. The reconstruction metrics improved substantially from the FBP-reconstruction to the estimated prior $\Xcnn$. The difference in terms of PSNR was almost 10\,dB, while the NRMSE decreased by approximately 0.11. Further, the similarity measures SSIM and HPSI were increased by about $0.14$ and $0.04$, respectively.  Finally, the estimated solution given by $\XX_{\mathrm{REC}}$ which was obtained by performing $n_{\mathrm{iter}}=4$ iterations of Landweber to minimize (\ref{DC_eq_CT}) showed a slight decrease in PSNR and NRMSE which is related to the use of the noisy-measured data. However, fine diagnostic details as the coronary arteries are still visible in the prior $\Xcnn$ and in  the solution $\XX_{\mathrm{REC}}$ as indicated by the yellow arrows. SSIM slightly increased while HPSI stayed approximately the same.\\ 
Figure \ref{CT_comparisons_results_figs} shows a comparison of images obtained by the different reconstruction methods. In Figure \ref{CT_comparisons_results_figs}A, we see again the FBP-reconstruction obtained from the noisy data. Figure \ref{CT_comparisons_results_figs}B shows the result obtained by the TV-minimization method which removed some of the noise as can be taken from the point-wise error image. The result obtained by the DIC method can be seen in Figure \ref{CT_comparisons_results_figs}C which further reduced image noise compared to the TV method and surpasses TV with respect to the reported statistics, as can be seein in Table \ref{CT_results_table}. Finally, Figure \ref{CT_comparisons_results_figs}D shows the solution $\XX_{\mathrm{REC}}$ obtained with our proposed scheme and Figure \ref{CT_comparisons_results_figs}E shows the ground truth image. The reconstruction using the CNN output as a prior further increased the PSNR, SSIM and HPSI by also reducing the NRMSE as can be taken from Table \ref{CT_results_table}.

\begin{table}[h]
\centering
\caption{Quantitative measures for the 3D low-dose CT example. The measures are obtained as averages over the seven different folds. }\label{CT_results_table}
\begin{tabular}{@{}l|ccc|cc}
\toprule
 & \textbf{FBP} & $\Xcnn$ & $\XX_{\mathrm{REC}}$ & \textbf{TV} & \textbf{DIC}\\
\midrule
\textbf{PSNR} & 30.0052 & 40.3546 & 39.6264 & 33.946 & 34.7807\\
\textbf{NRMSE} & 0.1657 & 0.0498 & 0.0538 & 0.1051 & 0.0938 \\
\textbf{SSIM} & 0.425 & 0.5755 & 0.5813 & 0.4985 & 0.5465 \\
\textbf{HPSI} & 0.9394 & 0.9821  & 0.9819 & 0.9503 & 0.9581 \\
\bottomrule
\end{tabular}
\end{table} 

\subsection{Reconstruction Times}

Table \ref{reco_times} summarizes the times for the different components of the reconstructions using all different approaches for both examples. The abbreviations "SHRINK" and "LS1" stand for "shrinkage" and "linear system - one iteration" and denote the times which are needed to apply the iterative shrinkage method for the TV approach and to solve the sub-problems which are solved using iterative schemes, respectively.

\begin{table}[h]
\centering
\caption{Reconstruction and processing times for the different methods for one 3D CT volume and a 2D cine MR image sequence.}
\begin{tabular}{@{}ll|l|l}
\toprule
 & &  \textbf{3D Low Dose CT} & \textbf{2D Radial Cine MRI}\\
\midrule
$\Ad^{\dagger} \Yn$ &  & $\approx 23$ s (FBP)  & $\approx 11$ s  (NUFFT)  \\
\midrule
\textbf{TV} & $\mathrm{SHRINK}$ & $\ll 1$ s & $\ll 1$ s   \\
& LS1 & $\approx 40$ s & $\approx 1:20$m   \\
& Total & $\approx 11$ m & $\approx 42$ m \\
\midrule
\textbf{DIC} & $\XX_{\mathrm{DIC}}$ & $\approx$ \textbf{1:24 h}   & $\approx$ \textbf{7 m}  \\
	& LS1 & $\approx 40$ s & $\approx$ 1:20 m   \\
	& Total & $\approx$ 1:28 h & $\approx 28$ m\\
\midrule
\textbf{Proposed} & $\Xcnn$  & $\approx$ \textbf{4 m}  & $\approx $ \textbf{ 5 s}  \\
& LS1  & $\approx 40$ s  &  $\approx $ 1:20 m  \\
& Total & $\approx 8$ m & $\approx 21$ m \\
\bottomrule
\end{tabular}\label{reco_times}
\end{table}

Obviously, in terms of achieved image quality, the advantage of the DIC- and the CNN-based Tikhonov regularization are given by obtaining stronger priors which allow to use a smaller number of iterations to regularize the solution. The advantage of our proposed approach compared to the dictionary learning-based is the highly reduced time to compute the prior which is used for regularization. The reason lies in the fact that the DIC-based method requires to solve problem (\ref{sparse_coding}) to obtain the prior $\XX_{\mathrm{DIC}}$, while in our method a CNN is used to obtain the prior $\Xcnn$. Since problem (\ref{sparse_coding}) is separable, OMP is applied for each image/volume patch which is prohibitive as the number of overlapping patches in a 3D volume is  in the order of $\mathcal{O}(N_x\cdot N_y \cdot N_z)$ or $\mathcal{O}(N_x\cdot N_y \cdot N_t)$, respectively. Obtaining $\Xcnn$, on the other hand, does not involve the solution of any minimization problem but only requires the application of the network $u_{\theta}$ to the different patches. As this corresponds to matrix-vector multiplications with sparse matrices, its computational cost is lower and the calculations are further highly accelerated by performing the computations on a GPU.

\section{Discussion}\label{discussion_section}
The proposed three-steps reconstruction scheme provides a general framework for solving large-scale inverse problems. The method is motivated by the observations stated in the ablation study \cite{kofler2018u}, where the performance of cascades of CNNs  with different numbers of intercepting data-consistency layers but approximately fixed number of trainable parameters was studied. First, it was noted that the  replacement of simple blocks of convolutional layers by multi-scale CNNs given by U-nets had a visually positive impact on the obtained results. Further, it was empirically shown that the results obtained by cascades of U-nets of different length but with approximately the same number of trainable parameters were all visually and quantitatively comparable in terms of all reported measures. This suggests that, for large-scale problems, where the construction of cascaded networks might be infeasible, investing the same computational effort and expressive power in terms of number of trainable parameters in one single network might be similarly beneficial to intercepting several smaller sub-networks by data-consistency layers as for example in \cite{schlemper2017deep}, \cite{qin2018convolutional}.\\
Due to the large sizes of the considered objects of interest, the prior $\Xcnn$ is obtained by processing patches of the images. Training the network on patches or slices of the images further has the advantage of reducing the computational overhead while naturally enlarging the available training data and therefore being able to successfully train neural networks even with datasets coming from a  relatively small number of subjects. Further, as demonstrated in \cite{kofler2019}, for the case of 2D radial MRI, one can also exploit the low topological complexity of 2D spatio-temporal slices for training the network $u_{\theta}$. This allows to reduce the network complexity by using 2D- instead of 3D-convolutional layers and still exploiting spatio-temporal correlations and therefore to prevent overfitting. Note that the network architectures we are considering are CNNs and, since they mainly consist of convolutional and max-pooling layers, we can expect the networks to be translation-equivariant and therefore, patch-related artefacts arising from the re-composition of the processed overlapping patches are unlikely to occur in the CNN-prior.\\
We have tested and evaluated our method on two examples of large-scale inverse problems given by 2D undersampled radial MRI and 3D low-dose CT. For both examples, our method outperformed the TV-minimization method and the dictionary learning-based method with respect to all reported quantitative measures. 
For the case of 2D undersampled radial cine MRI, using the CNN-prior as a regularizer in the subsequent iterative reconstruction increased the achieved image quality with respect to all reported measures, as can be taken from Table \ref{MRI_results_table}. For the CT example, due to the inherent presence of noise in the measured data, the quantitative measures of the final reconstruction are only similar to the ones obtained by post-processing the FBP-reconstruction. However, performing a few iterations to minimize functional (\ref{DC_eq_CT}), increased data-consistency of the obtained solution and resulted in a slight re-enhancement of  the edges and gave back the CT images their characteristic texture. Future work to qualitatively assess the achieved image quality with respect to clinically relevant features, e.g.\ the visibility of coronary arteries for the assessment of coronary artery disease in cardiac CT, is already planned.\\
Using the CNN for obtaining a learning-based prior is faster by several orders of magnitude compared to the dictionary learning-based approach. This is because obtaining the prior with a CNN reduces to a forward pass of all patches, i.e.\ to multiplications of vectors with sparse matrices, where instead, the sparse coding of all patches involves the solution of an optimization problem for each patch. Further, the time needed for OMP is dependent on the sparsity level and the number of atoms of the dictionary, see \cite{sturm2012comparison}.
In our comparison, for the 2D radial MRI example, the total reconstruction times of our proposed method and the DIC-based regularization method mainly differ in the step of obtaining the priors $\XX_{\mathrm{DIC}}$ and $\Xcnn$. Note that, in contrast to \cite{wang2014compressed} and \cite{caballero2014dictionary}, in our comparison, the  prior $\XX_{\mathrm{DIC}}$ was only calculated once. In the original works, however, the proposed reconstruction algorithms use an alternating direction method of multipliers (ADMM) which alternates between first training the dictionary $\mathbf{D}$ and sparse coding with OMP and then updating the image estimate. Therefore, the realistic time needed to reconstruct the 2D cine MR images according to \cite{wang2004image} and \cite{caballero2014dictionary} is given by the product of the seven minutes needed for one sparse approximation and the number of iterations in the ADMM algorithm and the total time used for PCG for solving the obtained linear systems. Note that for the 3D low-dose CT example, even one patch-wise sparse approximation of the whole volume already takes about one hour and therefore, applying an ADMM type of reconstruction method is computationally prohibitive. 
Also, note that, even if the size of the image sequences for the MRI example is smaller than the one of the 3D CT volumes, the reconstruction of the 2D cine MR images takes relatively long compared to the CT example due to the fact that we use two different iterative methods (Landweber and PCG) for two different systems with different operators. Further, the number of iterations for the CT example is on purpose smaller than for the MR example, as the measurement data is noisy and early stopping of the iteration can already be thought of as a proper  regularization method, see for example \cite{strand1974theory}. Also, the operators used for the CT examples were implemented by using the operators provided by the \texttt{ODL} library and are therefore optimized for performing calculations on the GPU. On the other hand, for the MRI example, we used our own implementation of a radial encoding operator $\Ed$ which could be further improved and accelerated.\\
Clearly, one difficulty of the proposed method is the one shared by all iterative reconstruction schemes with regularization: the need to choose the hyper-parameter $\lambda$ which balances the contribution of the regularization and the data-fidelity term can highly affect the achieved image quality, especially when the data is contaminated by noise. In cascaded networks, the parameter $\lambda$ can on the other hand be learned as well during training. Further, some other hyper-parameters as the number of iterations to minimize Tikhonov functional have to be chosen as well. In this work, we empirically chose $\lambda$ but point out that an exhaustive parameter search might yield superior results.  \\
The proposed method is related to the ones presented in \cite{schlemper2017deep}, \cite{qin2018convolutional}, \cite{kofler2018u} in the sense that  steps 2 and 3 in Algorithm 1 are iterated in a cascaded network which represents the different iterations. However, in \cite{schlemper2017deep} and \cite{qin2018convolutional}, the encoding operator is given by a Fourier transform sampled on a Cartesian grid and therefore is an isometry. Thus, assuming a single-coil data-acquistion, given $\Xcnn$, the  solution of (\ref{NN_reg_inv_problem})  has a closed-form solution which is also fast and cheap to compute since it corresponds to performing a linear combination of the acquired $k$-space data and the one estimated from the CNN outputs and subsequently  applying the inverse Fourier transform. In the case where the operator $\Ad$ is not an isometry, one usually needs to either solve a system of linear equations in order to obtain a solution which matches the measured data or, alternatively, rely on another formulation of the functional (\ref{NN_reg_inv_problem}) which is suitable for more general, also non-orthogonal operators \cite{kofler2018u}. However, if the operator $\Ad$ and its adjoint $\Ad^\herm$ are computationally demanding to apply as in the case of radial multi-coil MRI, or if the objects of interest are high-dimensional, e.g.\ 3D volumes in low-dose CT, the construction of cascaded or iterative networks is prohibitive with nowadays available hardware. In contrast, in the proposed approach, since the regularization is separated from the data-consistency step, large-scale problems can be tackled as well. 
Hence, by decoupling the regularization from further iteration of the reconstruction, one can also choose to employ more complex and sophisticated neural networks to obtain the prior $\Xcnn$ as it is typically the case for cascaded or iterative networks.  For example, in \cite{schlemper2017deep} or \cite{adler2017solving}, the CNNs were given by simple blocks of fully convolutional neural networks with residual connection. In contrast, in \cite{kofler2018u}, the CNNs were replaced by more sophisticated U-nets \cite{ronneberger2015u}, \cite{jin2017deep}. However, the examples in \cite{kofler2018u}, \cite{adler2017solving} or \cite{adler2018learned} all use two-dimensional CT geometries, which do not correspond to the ones used in clinical practice. Therefore, particularly for large-scale  inverse problems where the construction of iterative networks is  infeasible, our method represents a valid alternative to obtain accurate reconstructions. \\
While in this work we used a relatively simple neural network architecture given by a plain U-net as in \cite{jin2017deep}, further focus could be put on the choice of the network $u_{\theta}$, also by using more sophisticated approaches, e.g.\ improved versions of the U-net \cite{han2018framing} or generative adversarial networks for obtaining a more accurate prior to be further used in the proposed reconstruction scheme. 

\section{Conclusion}\label{conclusion_section}

We have presented a general framework for solving large-scale ill-posed inverse problems in medical image reconstruction. The strategy consists in decoupling the regularization of the solution from ensuring data-consistency by solving the problem in three stages. First, an initial guess of the solution is obtained by the direct reconstruction from the measured data. As a second step, the initial solution is patch-wise processed by a previously trained CNN in order to obtain a prior which is then used in a Tikhonov-regularized functional to obtain the final reconstruction in a third step.
The decoupling of the steps of obtaining a CNN-prior and minimizing a Tikhonov-functional allows to tackle large-scale problems. 
For both shown examples of 2D undersampled radial MRI and 3D low-dose CT, the proposed method outperformed the total variation-minimization method and the dictionary learning-based approach with respect to all reported quantitative measures. Since the reconstruction scheme is a general one, we expect the proposed method to be successfully applicable to other imaging modalities as well.\\

\section*{Acknowledgements}
A.  Kofler and M. Dewey acknowledge  the  support  of  the German  Research  Foundation  (DFG),  project  number  GRK2260, BIOQIC. We thank Dr. V. Wieske for providing clinically relevant cases for the 3D low-dose CT experiments.

\bibliographystyle{IEEEtran}
\bibliography{IEEEabrv,references.bib}

\begin{thebibliography}{10}
\providecommand{\url}[1]{#1}
\csname url@samestyle\endcsname
\providecommand{\newblock}{\relax}
\providecommand{\bibinfo}[2]{#2}
\providecommand{\BIBentrySTDinterwordspacing}{\spaceskip=0pt\relax}
\providecommand{\BIBentryALTinterwordstretchfactor}{4}
\providecommand{\BIBentryALTinterwordspacing}{\spaceskip=\fontdimen2\font plus
\BIBentryALTinterwordstretchfactor\fontdimen3\font minus
  \fontdimen4\font\relax}
\providecommand{\BIBforeignlanguage}[2]{{%
\expandafter\ifx\csname l@#1\endcsname\relax
\typeout{** WARNING: IEEEtran.bst: No hyphenation pattern has been}%
\typeout{** loaded for the language `#1'. Using the pattern for}%
\typeout{** the default language instead.}%
\else
\language=\csname l@#1\endcsname
\fi
#2}}
\providecommand{\BIBdecl}{\relax}
\BIBdecl

\bibitem{Lustig2008}
M.~Lustig, D.~L. Donoho, J.~M. Santos, and J.~M. Pauly, ``Compressed sensing
  mri,'' \emph{IEEE signal processing magazine}, vol.~25, no.~2, p.~72, 2008.

\bibitem{block2007}
K.~T. Block, M.~Uecker, and J.~Frahm, ``Undersampled radial mri with multiple
  coils. iterative image reconstruction using a total variation constraint,''
  \emph{Magnetic Resonance in Medicine}, vol.~57, no.~6, pp. 1086--1098, 2007.

\bibitem{wang2014compressed}
Y.~Wang and L.~Ying, ``Compressed sensing dynamic cardiac cine mri using
  learned spatiotemporal dictionary,'' \emph{IEEE Transactions on Biomedical
  Engineering}, vol.~61, no.~4, pp. 1109--1120, 2014.

\bibitem{xu2012low}
Q.~Xu, H.~Yu, X.~Mou, L.~Zhang, J.~Hsieh, and G.~Wang, ``Low-dose x-ray ct
  reconstruction via dictionary learning,'' \emph{IEEE Transactions on Medical
  Imaging}, vol.~31, no.~9, pp. 1682--1697, 2012.

\bibitem{zhu2018image}
B.~Zhu, J.~Z. Liu, S.~F. Cauley, B.~R. Rosen, and M.~S. Rosen, ``Image
  reconstruction by domain-transform manifold learning,'' \emph{Nature}, vol.
  555, no. 7697, p. 487, 2018.

\bibitem{jin2017deep}
K.~H. Jin, M.~T. McCann, E.~Froustey, and M.~Unser, ``Deep convolutional neural
  network for inverse problems in imaging,'' \emph{IEEE Transactions on Image
  Processing}, vol.~26, no.~9, pp. 4509--4522, 2017.

\bibitem{schwab2018deep}
J.~Schwab, S.~Antholzer, and M.~Haltmeier, ``Deep null space learning for
  inverse problems: convergence analysis and rates,'' \emph{Inverse Problems},
  2018.

\bibitem{han2018framing}
Y.~Han and J.~C. Ye, ``Framing u-net via deep convolutional framelets:
  Application to sparse-view ct,'' \emph{IEEE Transactions on Medical Imaging},
  vol.~37, no.~6, pp. 1418--1429, 2018.

\bibitem{adler2017solving}
J.~Adler and O.~{\"O}ktem, ``Solving ill-posed inverse problems using iterative
  deep neural networks,'' \emph{Inverse Problems}, vol.~33, no.~12, p. 124007,
  2017.

\bibitem{adler2018learned}
------, ``Learned primal-dual reconstruction,'' \emph{IEEE Transactions on
  Medical Imaging}, vol.~37, no.~6, pp. 1322--1332, 2018.

\bibitem{schlemper2017deep}
J.~Schlemper, J.~Caballero, J.~V. Hajnal, A.~N. Price, and D.~Rueckert, ``A
  deep cascade of convolutional neural networks for dynamic mr image
  reconstruction,'' \emph{IEEE Transactions on Medical Imaging}, vol.~37,
  no.~2, pp. 491--503, 2018.

\bibitem{kofler2018u}
A.~Kofler, M.~Haltmeier, C.~Kolbitsch, M.~Kachelrie{\ss}, and M.~Dewey, ``{A
  U-nets cascade for sparse view computed tomography},'' in \emph{Lecture Notes
  in Computer Science (including subseries Lecture Notes in Artificial
  Intelligence and Lecture Notes in Bioinformatics)}, 2018.

\bibitem{li2018nett}
H.~Li, J.~Schwab, S.~Antholzer, and M.~Haltmeier, ``{NETT}: Solving inverse
  problems with deep neural networks,'' \emph{arXiv preprint arXiv:1803.00092},
  2018.

\bibitem{aggarwal2018modl}
H.~K. Aggarwal, M.~P. Mani, and M.~Jacob, ``Modl: Model-based deep learning
  architecture for inverse problems,'' \emph{IEEE Transactions on Medical
  Imaging}, vol.~38, no.~2, pp. 394--405, 2018.

\bibitem{qin2018convolutional}
C.~Qin, J.~Schlemper, J.~Caballero, A.~N. Price, J.~V. Hajnal, and D.~Rueckert,
  ``Convolutional recurrent neural networks for dynamic mr image
  reconstruction,'' \emph{IEEE Transactions on Medical Imaging}, vol.~38,
  no.~1, pp. 280--290, 2018.

\bibitem{yang2018low}
Q.~Yang, P.~Yan, Y.~Zhang, H.~Yu, Y.~Shi, X.~Mou, M.~K. Kalra, Y.~Zhang,
  L.~Sun, and G.~Wang, ``Low-dose ct image denoising using a generative
  adversarial network with wasserstein distance and perceptual loss,''
  \emph{IEEE Transactions on Medical Imaging}, vol.~37, no.~6, pp. 1348--1357,
  2018.

\bibitem{hammernik2018learning}
K.~Hammernik, T.~Klatzer, E.~Kobler, M.~P. Recht, D.~K. Sodickson, T.~Pock, and
  F.~Knoll, ``Learning a variational network for reconstruction of accelerated
  mri data,'' \emph{Magnetic Resonance in Medicine}, vol.~79, no.~6, pp.
  3055--3071, 2018.

\bibitem{hauptmann2018model}
A.~Hauptmann, F.~Lucka, M.~Betcke, N.~Huynh, J.~Adler, B.~Cox, P.~Beard,
  S.~Ourselin, and S.~Arridge, ``Model-based learning for accelerated,
  limited-view 3-d photoacoustic tomography,'' \emph{IEEE Transactions on
  Medical Imaging}, vol.~37, no.~6, pp. 1382--1393, 2018.

\bibitem{strand1974theory}
O.~N. Strand, ``Theory and methods related to the singular-function expansion
  and {Landweber’s} iteration for integral equations of the first kind,''
  \emph{SIAM Journal on Numerical Analysis}, vol.~11, no.~4, pp. 798--825,
  1974.

\bibitem{engl1996regularization}
H.~W. Engl, M.~Hanke, and A.~Neubauer, \emph{Regularization of inverse
  problems}.\hskip 1em plus 0.5em minus 0.4em\relax Springer Science \&
  Business Media, 1996, vol. 375.

\bibitem{scherzer2009variational}
O.~Scherzer, M.~Grasmair, H.~Grossauer, M.~Haltmeier, and F.~Lenzen,
  \emph{Variational methods in imaging}.\hskip 1em plus 0.5em minus 0.4em\relax
  Springer, 2009.

\bibitem{grasmair2010generalized}
M.~Grasmair, ``Generalized bregman distances and convergence rates for
  non-convex regularization methods,'' \emph{Inverse Problems}, vol.~26,
  no.~11, p. 115014, 2010.

\bibitem{tian2011low}
Z.~Tian, X.~Jia, K.~Yuan, T.~Pan, and S.~B. Jiang, ``Low-dose ct reconstruction
  via edge-preserving total variation regularization,'' \emph{Physics in
  Medicine \& Biology}, vol.~56, no.~18, p. 5949, 2011.

\bibitem{winkelmann2006optimal}
S.~Winkelmann, T.~Schaeffter, T.~Koehler, H.~Eggers, and O.~Doessel, ``An
  optimal radial profile order based on the golden ratio for time-resolved
  mri,'' \emph{IEEE Transactions on Medical Imaging}, vol.~26, no.~1, pp.
  68--76, 2006.

\bibitem{rasche1999resampling}
V.~Rasche, R.~Proksa, R.~Sinkus, P.~Bornert, and H.~Eggers, ``Resampling of
  data between arbitrary grids using convolution interpolation,'' \emph{IEEE
  Transactions on Medical Imaging}, vol.~18, no.~5, pp. 385--392, 1999.

\bibitem{smith2019trajectory}
D.~S. Smith, S.~Sengupta, S.~A. Smith, and E.~Brian~Welch, ``Trajectory
  optimized {NUFFT}: Faster non-cartesian mri reconstruction through prior
  knowledge and parallel architectures,'' \emph{Magnetic Resonance in
  Medicine}, vol.~81, no.~3, pp. 2064--2071, 2019.

\bibitem{han2019k}
Y.~{Han}, L.~{Sunwoo}, and J.~C. {Ye}, ``k-space deep learning for accelerated
  mri,'' \emph{IEEE Transactions on Medical Imaging}, p.
  {DOI:}10.1109/TMI.2019.2927101, 2019.

\bibitem{kofler2019}
A.~{Kofler}, M.~{Dewey}, T.~{Schaeffter}, C.~{Wald}, and C.~{Kolbitsch},
  ``Spatio-temporal deep learning-based undersampling artefact reduction for 2d
  radial cine mri with limited training data,'' \emph{IEEE Transactions on
  Medical Imaging}, no. {DOI:} 10.1109/TMI.2019.2930318, 2019.

\bibitem{adler_github}
J.~Adler, H.~Kohr, and O.~Oktem, ``Operator discretization library,''
  \emph{https://github. com/odlgroup/odl}, 2017.

\bibitem{Adam}
D.~P. Kingma and J.~Ba, ``Adam: A method for stochastic optimization,'' 2014,
  arxiv:1412.6980 Published as a conference paper at the 3rd International
  Conference for Learning Representations, San Diego, 2015.

\bibitem{dewey2009noninvasive}
M.~Dewey, E.~Zimmermann, F.~Deissenrieder, M.~Laule, H.-P. D{\"u}bel,
  P.~Schlattmann, F.~Knebel, W.~Rutsch, and B.~Hamm, ``Noninvasive coronary
  angiography by 320-row computed tomography with lower radiation exposure and
  maintained diagnostic accuracy: comparison of results with cardiac
  catheterization in a head-to-head pilot investigation.'' \emph{Circulation},
  vol. 120, no.~10, pp. 867--875, 2009.

\bibitem{Hauptmann2019}
A.~Hauptmann, S.~Arridge, F.~Lucka, V.~Muthurangu, and J.~A. Steeden,
  ``Real-time cardiovascular mr with spatio-temporal artifact suppression using
  deep learning--proof of concept in congenital heart disease,'' \emph{Magnetic
  resonance in medicine}, vol.~81, no.~2, pp. 1143--1156, 2019.

\bibitem{napp2017computed}
A.~E. Napp, R.~Haase, M.~Laule, G.~M. Schuetz, M.~Rief, H.~Dreger,
  G.~Feuchtner, G.~Friedrich, M.~{\v{S}}pa{\v{c}}ek, V.~Such{\'a}nek
  \emph{et~al.}, ``Computed tomography versus invasive coronary angiography:
  design and methods of the pragmatic randomised multicentre discharge trial,''
  \emph{European radiology}, vol.~27, no.~7, pp. 2957--2968, 2017.

\bibitem{chambolle2005total}
A.~Chambolle, ``Total variation minimization and a class of binary mrf
  models,'' in \emph{International Workshop on Energy Minimization Methods in
  Computer Vision and Pattern Recognition}.\hskip 1em plus 0.5em minus
  0.4em\relax Springer, 2005, pp. 136--152.

\bibitem{schnass2018convergence}
K.~Schnass, ``Convergence radius and sample complexity of itkm algorithms for
  dictionary learning,'' \emph{Applied and Computational Harmonic Analysis},
  vol.~45, no.~1, pp. 22--58, 2018.

\bibitem{tropp2007signal}
J.~A. Tropp and A.~C. Gilbert, ``Signal recovery from random measurements via
  orthogonal matching pursuit,'' \emph{IEEE Trans. on information theory},
  vol.~53, no.~12, pp. 4655--4666, 2007.

\bibitem{wang2004image}
Z.~Wang, A.~C. Bovik, H.~R. Sheikh, E.~P. Simoncelli \emph{et~al.}, ``Image
  quality assessment: from error visibility to structural similarity,''
  \emph{IEEE Transactions on Image Processing}, vol.~13, no.~4, pp. 600--612,
  2004.

\bibitem{caballero2014dictionary}
J.~Caballero, A.~N. Price, D.~Rueckert, and J.~V. Hajnal, ``Dictionary learning
  and time sparsity for dynamic mr data reconstruction,'' \emph{IEEE
  Transactions on Medical Imaging}, vol.~33, no.~4, pp. 979--994, 2014.

\bibitem{reisenhofer2018haar}
R.~Reisenhofer, S.~Bosse, G.~Kutyniok, and T.~Wiegand, ``{A Haar wavelet-based
  perceptual similarity index for image quality assessment},'' \emph{Signal
  Processing: Image Communication}, 2018.

\bibitem{sturm2012comparison}
B.~L. Sturm and M.~G. Christensen, ``Comparison of orthogonal matching pursuit
  implementations,'' in \emph{2012 Proceedings of the 20th European Signal
  Processing Conference (EUSIPCO)}.\hskip 1em plus 0.5em minus 0.4em\relax
  IEEE, 2012, pp. 220--224.

\bibitem{ronneberger2015u}
O.~Ronneberger, P.~Fischer, and T.~Brox, ``U-net: Convolutional networks for
  biomedical image segmentation,'' in \emph{International Conference on Medical
  image computing and computer-assisted intervention}.\hskip 1em plus 0.5em
  minus 0.4em\relax Springer, 2015, pp. 234--241.

\end{thebibliography}

\end{document}